\documentclass[12pt]{article}

\usepackage{textcomp}
\usepackage{chetnew}
\usepackage{mathtools}
\usepackage{amssymb,amsmath,amsfonts,mathrsfs,braket}
\usepackage[utf8]{inputenc}
\usepackage{tikz,tikz-cd,url}
\usepackage{xcolor}
\usepackage{tcolorbox}
\usepackage{booktabs,caption}

\usetikzlibrary{arrows,decorations.markings,shapes.geometric,decorations.pathmorphing}
\tikzset{snake it/.style={decorate, decoration=snake}}

\usepackage{bm}

\usepackage[colorinlistoftodos,textsize=tiny]{todonotes}

\renewcommand{\d}[1]{\ensuremath{\operatorname{d}\!{#1}}}

\def\one{{\,\hbox{1\kern-.8mm l}}}

\allowdisplaybreaks

\def\makeatletter{\catcode`\@=11}% 11:letter
\makeatletter
\def\mathbox#1{\hbox{$\m@th#1$}}%
\def\math@ccstyles#1#2#3#4#5#6#7{{\leavevmode
      \setbox0\mathbox{#6#7}%
      \setbox2\mathbox{#4#5}%
      \dimen@ #3%
      \baselineskip\z@\lineskiplimit#1\lineskip\z@
      \vbox{\ialign{##\crcr
             \hfil \kern #2\box2 \hfil\crcr
             \noalign{\kern\dimen@}%
             \hfil\box0\hfil\crcr}}}}
\def\mathaccstyles{\math@ccstyles\maxdimen}
\def\maththroughstyles{\math@ccstyles{-\maxdimen}}
\def\unity%
 {\maththroughstyles{.45\ht0}\z@\displaystyle {\mathchar"006C}\displaystyle 1}

\def\AA{{\cal A}}

\def\FF{{\cal F}}

\def\II{{\cal I}}

\def\KK{{\cal K}}
\def\LL{{\cal L}}

\def\NN{{\cal N}}
\def\OO{{\cal O}}

\def\XX{{\cal X}}

\def\IR{{\mathbb R}}
\def\IT{{\mathbb T}}

\def\IN{{\mathbb N}}

\def\d{{\partial}}

\def\beq{\begin{equation}}
\def\eeq{\end{equation}}
\newcommand{\bea}{\begin{eqnarray}}
\newcommand{\eea}{\end{eqnarray}}
\def\bal{\begin{align}}
\def\eal{\end{align}}

\def\disc{{\rm Disc}}

\usepackage{wrapfig}
\usepackage{subcaption}
\usepackage{dsfont}

\preprint{CCTP-2025-8 \hfill QMUL-PH-25-16 \\ ITCP-2025/8}

\title{Deep Finite Temperature Bootstrap} 
\author{
V.~Niarchos\;$^{a,\bigstar}$, 
C.~Papageorgakis\;$^{b,\blacklozenge}$,
A.~Stratoudakis\;$^{a,\spadesuit}$,
M.~Woolley$^{b,\clubsuit}$}

\affiliation{
$^a$ ITCP \& CCTP, Department of Physics,\\
University of Crete, 71003 Heraklion, Greece \\
$^b$ Centre for Theoretical Physics, Department of Physics and Astronomy\\ Queen Mary University of London, London E1 4NS, UK \vspace{0.3cm} $ $ \\
\vspace{0.3cm} $ $\\

\vspace{0.0cm}
{\tt \small
$^\bigstar$niarchos@physics.uoc.gr,
$^\blacklozenge$c.papageorgakis@qmul.ac.uk, 
$^\spadesuit$astratoudakis@physics.uoc.gr,
$^\clubsuit$mitchell.woolley@qmul.ac.uk}
}

\abstract{\centerline{\bf Abstract}

\vspace{0.5cm}
\noindent
We introduce a novel method to bootstrap crossing equations in Conformal Field Theory and apply it to finite temperature theories on $S^1\times \IR^{d-1}$. The proposed approach does not rely on positivity constraints and does not employ uncontrolled truncation schemes. Instead, we capture the contribution of an infinite number of operators in conformal block expansions using suitable functions, which are bootstrapped (numerically) together with a finite number of exposed CFT data. Our approach at finite temperature employs three key ingredients: $(i)$ the Kubo-Martin-Schwinger (KMS) condition, $(ii)$ thermal dispersion relations and $(iii)$ Neural Networks that model spin-dependent tail functions within the conformal block expansions. We test the efficiency of the new method in the case of Generalized Free Fields and use it to perform a preliminary bootstrap analysis of double-twist thermal data in holographic CFTs.    
}

\date{\today}

\begin{document}

\maketitle

\hypersetup{pageanchor=true}

\setcounter{tocdepth}{3}

\toc
\newpage
%%%%%%%%%%%%%%%%%%%%%%%%%%%%%%%%%%%%%%

\section{Introduction and summary}
\label{intro}

\subsection{The need for an efficient primal bootstrap}

Some of the most successful techniques in the modern conformal bootstrap program rely on the study of feasibility conditions, reformulated suitably as convex (semi-definite programming) optimization problems. Most notably, in the analysis of the crossing equations of 4-point correlation functions in unitary Conformal Field Theories (CFTs), positivity constraints are used to exclude assumptions and obtain allowed regions for selected CFT data. This approach has been remarkably useful, because it cleverly bypasses the seemingly impossible problem of solving explicitly the crossing equations, which involves dealing with continuous families of equations for an infinite set of data.  

There are, however, several reasons why one would like to go beyond the feasibility analysis, and solve the \textit{primal bootstrap} problem that aims to reconstruct full solutions of the bootstrap equations. For example, if there is an island of allowed parameters, it would be useful to know how consistent correlation functions look like in the bulk of the island in order to gain better intuition about potential assumptions and guided searches for theories of interest. In addition, there are many physically relevant contexts, where the positivity conditions, necessary for the feasibility analysis, are altogether absent. Notable cases include: finite-temperature theories, theories with defects or boundaries, higher-point crossing equations etc.

To make the direct (numerical) solution of the bootstrap equations more tractable, one typically truncates the sum over an infinite number of contributions to a finite subset of low-energy CFT data, which (one hopes) are the dominant contributions to the equations of interest. In a drastic truncation, only the low-lying CFT data are considered. In more nuanced, \textit{soft} truncations, an approximation scheme of the contribution of the truncated high-energy data is also employed. Although in the past such schemes have produced results with notable accuracy in the context of specific examples, so far there has been no general, systematic methodology for truncations in the conformal bootstrap program, and the pertinent difficulties of such methodologies are well known (see e.g. \cite{Gliozzi:2013ysa,Gliozzi:2014jsa,Gliozzi:2015qsa,Gliozzi:2016cmg,Esterlis:2016psv,Li:2017ukc,Li:2017agi,Leclair:2018trn,Kantor:2021jpz,Niarchos:2023lot} and the review \cite{Poland:2018epd} for related discussions).

To address this situation, we introduce a new framework for the primal bootstrap, which is not based on traditional hard or soft truncations, and involves a more systematic control over the employed approximations, compared to previous schemes: No positivity conditions are assumed, the approach is general and applies to many different contexts. In this work we center the presentation exclusively around the analysis of thermal 2-point functions of scalar operators in finite-temperature CFTs on $S^1\times \IR^{d-1}$. The case of 4-point correlation functions at zero temperature will be discussed in a companion paper \cite{NPS}.

\subsection{The strategy}
\label{mainidea}

A typical conformal bootstrap problem involves a vector of sum rules of the form
\beq
\label{introaa}
\sum_{J} \sum_\Delta \vec \XX_{\Delta, J} = 0
\eeq
that contain an infinite number of contributions labeled by the
scaling dimension $\Delta$ and spin $J$ of the corresponding CFT operators. Our strategy for solving \eqref{introaa} involves 3 steps:
\begin{itemize}
    \item[(1)] We make an arbitrary choice of a spin cutoff $J_*$ and express the `high'-spin\footnote{'High' here refers to any spin above the arbitrary $J_*$, which does not have to be a large number.} part of Eq.\ \eqref{introaa}
    \beq
    \label{introab}
    \sum_{J>J_*}\sum_\Delta \vec \XX_{\Delta, J}
    \eeq
    using dispersion relations in terms of a discontinuity. This is the only place where we need to perform an approximation in our numerical implementations, and the corresponding error can be systematically suppressed by increasing the arbitrary value of $J_*$.
    
    \item[(2)] The remaining, `low'-spin part of Eq.\ \eqref{introaa}
    \beq
    \label{introab}
    \sum_{J\leq J_*} \sum_\Delta \vec \XX_{\Delta, J}
    \eeq
    involves sums over a finite number of spins, but an infinite number of scaling dimensions. At each spin, we freely choose to expose a finite number of contributions from specific operators at low scaling dimension and the remaining combined contribution of all the other operators (with arbitrarily high scaling dimension) is expressed in terms of suitable `tail' functions.\footnote{The precise realization of these contributions, as one-dimensional functions of a single radial space coordinate, will be explained in the context of the thermal bootstrap in the main part of the paper.} In summary, this part contains a finite number of explicit CFT data, call them $\tt D$, and a finite number of unknown functions, call them $\tt F$. In this part, there is no approximation.  
    
    \item[(3)] We formulate a non-convex optimization problem that varies over the unknowns $({\tt D},{\tt F})$ to determine solutions of \eqref{introaa}. To efficiently vary over the space of unknown functions ${\tt F}$ we use a representation based on Multi-Layer-Perceptron (MLP) Neural Networks. More specifically, we use an architecture known as Multi-Branch MLPs.
\end{itemize}

The crucial new element of this approach is that the traditionally obscure tails of contributions from operators with arbitrarily high scaling dimension are not dropped, or approximated in some fixed ad hoc manner. In the low-spin component of the bootstrap equations the tails become part of the data we are bootstrapping dynamically. In the high-spin part we employ dispersion relations. These features enable the method to remove critical limitations of previous truncation schemes in the literature. The method additionally allows a lot of freedom on the choice of assumptions one would like to implement, or the specific choice of the exposed CFT data, the value of which one would like to compute explicitly. The latter does not require any prior knowledge of the structure of the solution or what constitutes a significant low-energy CFT datum. The exposed data are arbitrary. 

This paper establishes the above approach in a specific physical context, demonstrates its implementation through a proof-of-concept example compared against known analytic results, and identifies error sources and potential technical challenges to guide future developments.

\subsection{The challenge of finite temperature}

At finite temperature, the 1-point functions of many primary operators are non-vanishing. The goal of the finite-temperature bootstrap is to determine (or constrain) these thermal data assuming knowledge of the zero-temperature CFT data (scaling dimensions and three-point function coefficients). It has been suggested \cite{El-Showk:2011yvt} that this might be achievable by studying general consistency conditions on thermal 2-point functions. One of these conditions is the periodicity of 2-point correlation functions on the periodic, thermal coordinate in Euclidean space, also known as the Kubo--Martin--Schwinger (KMS) condition. This is a non-trivial condition, because the Operator Product Expansion (OPE) of the 2-point functions is not automatically periodic and that imposes consistency conditions on the unknown thermal 1-point functions that appear in the OPE. 

At infinite spatial volume, the KMS condition is superficially very similar to the crossing equations of 4-point correlation functions at zero-temperature, but the analysis exhibits special complications due to the following features:
\begin{itemize}
    \item In the OPE expressions of the KMS condition, there are no positivity requirements on the OPE coefficients. As a result, it is unclear how to reformulate the analysis in terms of a convex-optimization, semi-definite programming problem for feasibility conditions.\footnote{It is worth noting, however, that in finite-temperature Quantum Mechanics positivity constraints were recently used in Ref.\ \cite{Cho:2024kxn}.}
    
    \item The OPE of thermal 2-point functions does not share the familiar, quick convergence properties of conformal block expansions for 4-point functions at zero-temperature (see e.g.\  \cite{Marchetto:2023xap} for a relevant discussion). That complicates attempts to perform hard truncations.

    \item Soft truncations---such as those estimating tail contributions from high-dimension operators using Tauberian theorems \cite{Marchetto:2023xap,Barrat:2025wbi}---are unsatisfactory for the following reason: They reduce the problem to convex linear regression for finitely many unknown thermal OPE coefficients, yielding a unique solution even when the KMS condition admits multiple consistent solutions.\footnote{Holographic CFTs provide explicit examples where the KMS condition exhibits continuous families of solutions.} As a result, linear regression-based soft truncation schemes cannot detect multiple solutions and become intractable in generic problems, requiring large numbers of operators with explicit spectral knowledge. These difficulties motivate the analysis of the KMS condition without explicit truncation, using methods that are able to detect multiple solutions and can treat the high-energy CFT spectrum more economically. 
    
\end{itemize}

\subsection{Key results}

In the main part of the paper, we explain how the general approach outlined in Section \ref{mainidea} can be used to bypass the above-mentioned issues. We focus on the finite-temperature case at infinite spatial volume, as a problem of intrinsic interest, but also as a special concrete case for an explicit illustration of the capabilities and potential challenges of the proposed approach. The most significant contributions of the present paper can be summarized as follows:

\begin{list}{}{\leftmargin=0pt \itemindent=0pt}

\item \textbf{Approximate KMS Formulation}: We formulate an approximate KMS condition, using suitable spin-cutoff-dependent thermal dispersion relations and consider thermal 2-point functions at finite spatial separation, obtaining constraints on spin-dependent thermal OPE coefficients  $a_{\Delta,J}$. In contrast, several recent approaches \cite{Marchetto:2023xap,Barrat:2024aoa,Barrat:2025wbi,Barrat:2025nvu,Buric:2025anb} focus on the properties of thermal 2-point functions---and the associated KMS condition---at zero spatial separation that are sensitive only to spin-independent thermal OPE data $a_{\Delta}$. More importantly, our work shifts the focus away from the exclusive computation of individual CFT data, to suitable one-dimensional functions of a radial space coordinate $r$ that repackage the information of an infinite tower of operators at increasing scaling dimension, but fixed spin. Under suitable conditions, we argue that these functions exhibit a universal behavior in the vicinity of $r=\beta$ (where $\beta$ is the inverse temperature). Knowledge of these functions leads to an approximate reconstruction of the thermal 2-point function in the OPE convergence region.

\item \textbf{Generalized Free Field testing}: We perform explicit tests of the approach in the case of Generalized Free Field (GFF) theories, where thermal 1-point functions can be computed analytically. We also use this case, to exhibit the systematic error introduced by the approximations in the high-spin part and demonstrate how one can systematically approach the exact KMS condition by increasing the arbitrary spin cutoff $J_*$.

\item \textbf{Tail contribution bootstrap}: By analyzing specific theories (GFF and holographic CFTs) we provide a proof-of-concept that tail contributions can indeed be bootstrapped, avoiding the aforementioned truncation issues as well as issues inherent in linear regression schemes.

\item \textbf{Neural network architecture}: Unknown optimizable functions are modeled using suitable Neural Networks (NNs). We present a specific sparse architecture (Multi-Branch MLPs) that involves separate subnets for different output functions.

\item \textbf{Loss function analysis}: We explore, and compare, several different loss functions, designed to handle non-convex optimization in diverse situations, with or without exposed CFT data. In addition, we discuss practical features of the searches we perform, and of the outputs of optimization while also identifying potential difficulties.

\item \textbf{Holographic CFT applications}: As a non-trivial application, we discuss the bootstrap of holographic CFTs, perform preliminary computations of explicit double-twist CFT data and compare with the limited results in the literature. We emphasize that in this context, our approach has access to quantitatively new, spin-dependent, thermal double-twist data, which are not accessible to other methods based on the study of thermal 2-point functions at zero-spatial separation.

\item \textbf{Future $\mathcal{N}=4$ SYM extension}: We dedicate a section to summarizing key elements needed for future applications of our approach to $\mathcal{N}=4$ SYM holography. We emphasize the role of operator-mixing and the ensuing logarithmic contributions in the thermal OPE, and set up several necessary elements for the implementation of our approach in this context.
\end{list}

\subsection{Plan of the paper}

In Section \ref{kms1} we set the notation and introduce the necessary background material for the analysis of the main text. This includes a summary of the relevant conformal block expansions of thermal 2-point functions and the KMS condition. In Section \ref{dispersion} we collect useful material about the specific form of the finite-temperature dispersion relations that are employed in our approach, and relegate several pertinent technical details to Appendix\ \ref{dispApp}. In Section \ref{kms2} we formulate an approximate version of the KMS condition, the solution of which is the main target of the paper. The Neural Network representation of the tail functions is introduced in this section, with the technical details deferred to Appendix\ \ref{nn}. Another important element introduced in Section \ref{kms2} are three different versions of loss functions that are used throughout the paper to find approximate solutions to the KMS condition. 

The main applications of the proposed formalism appear in Sections \ref{GFF}--\ref{sym}. Section \ref{GFF} contains tests in the context of the Generalized Free Field Theories and Sections \ref{holo1}, \ref{holo2} applications to holographic CFTs. Section \ref{sym} summarizes necessary material for future applications to the bootstrap of thermal 2-point functions of half-BPS operators in the supergravity limit of 4d $\NN=4$ SYM theory. Further useful material for these sections appears in Appendices \ref{apptails} and \ref{largec}--\ref{addplots}.

We conclude in Section \ref{outlook} with a brief summary of interesting open problems and promising directions for future research.

\vspace{0.5cm}
\noindent
{\bf Note added:} As we were finalizing the draft, we became aware of the paper \cite{Buric:2025fye}, which partially overlaps with our results.

\section{Thermal block expansions, KMS and all that}
\label{kms1}

We study $d$-dimensional CFTs on $S^1\times \IR^{d-1}$. We parametrize the space with coordinates $x=(\tau, \vec x)$, where $\tau$ describes the thermal circle with period $\beta$ and $\vec x$ the spatial $\IR^{d-1}$. This setup captures thermal physics at inverse temperature $\beta$ in the infinite spatial volume limit. It can also be viewed as the high temperature limit of the theory on $S^1 \times S^{d-1}$.

In this section, we collect useful well-known facts about this context and set up a significant part of the notation we will be using in the main text.

\subsection{2-point functions and the KMS condition}
\label{kms12point}

For concreteness, we will focus on 2-point functions of identical scalar operators $\phi$
\beq
\label{notaaa}
g(\tau, |x|) := \langle \phi(x) \phi(0) \rangle_\beta
~.
\eeq
These functions depend separately on the Euclidean time $\tau$ and
\beq
\label{notaab}
|x| = \sqrt{\tau^2 + \vec x^2}
\eeq
because of the reduced $SO(d-1)$ symmetry of the background. It will be convenient to use this symmetry to set $\vec x = (\sigma, 0, \ldots, 0)$. In this frame, we define the following set of variables that will be used interchangeably in the main text
\beq
\label{notaac}
z := \tau + i \sigma ~, ~~ \bar z := \tau - i \sigma
~,
\eeq
and
\beq
\label{notaad}
z = r w~, ~~ \bar z = r w^{-1}
~.
\eeq
In this definition, $w$ is a pure phase, but later it will be continued to the whole complex plane. To summarize, $|x|=r$ and the 2-point function $g$ depends on $(\tau, r)$ or equivalently on $(z,\bar z)$.

The Kubo-Martin-Schwinger (KMS) condition states that the 2-point function is invariant under the transformation $\tau \to \beta - \tau$,
\beq
\label{notaae}
g(\tau, r) = g(\beta - \tau, r)
~.
\eeq
Henceforth, we will set $\beta=1$ (without loss of generality). By using the invariance of the 2-point function under the parity transformation $\sigma \to -\sigma$, we can further recast the KMS condition as a `crossing' equation
\beq
\label{notaaf}
g(z, \bar z) = g(1-z, 1-\bar z)
~.
\eeq
The importance of this equation as a non-trivial consistency condition on thermal CFT data was first recognized by El-Showk and Papadodimas in \cite{El-Showk:2011yvt}.

\subsection{Operator product expansions}
\label{kms1OPE}

In a 2-point function $\langle \phi(x) \phi(0) \rangle_\beta$, we can use the Operator Product Expansion (OPE) to express the correlation function as a series over thermal 1-point functions. For an operator with scaling dimension $\Delta$, the 1-point function is proportional to the temperature raised to the power $\Delta$, namely it behaves as $\beta^{-\Delta}$.\footnote{As expected, the 1-point functions of primary operators, other than the identity, vanish at zero temperature ($\beta\to \infty$) in agreement with the conformal Ward identities.} Moreover, since conformal descendants have vanishing thermal 1-point functions, only the primaries in a conformal family contribute. This fact simplifies the form of the thermal conformal blocks. Putting everything together (and setting $\beta=1$), one ends up with the following conformal block expansion of thermal 2-point functions of identical scalar operators \cite{Iliesiu:2018fao}
\beq
\label{opeaa}
g(rw, rw^{-1}) = \sum_{\OO_{\Delta,J} \in \phi\times \phi} a_{\OO_{\Delta,J}} \, C_{J}^{(\nu)}\left(\frac{1}{2}(w+w^{-1})\right) r^{\Delta-2\Delta_\phi}
~,
\eeq
where
\beq
\label{opeab}
a_\OO := \frac{f_{\phi\phi \OO}b_\OO}{c_\OO} \frac{J !}{2^{J} (\nu)_{J}}
~,~~
\nu := \frac{d-2}{2}
\eeq
and $C_J^{(\nu)}(\eta)$ are Gegenbauer polynomials. The coefficients $f_{\phi\phi\OO}$ are 3-point function coefficients in the zero-temperature theory, and $b_\OO$ are the thermal 1-point coefficients. $\Delta, J$ represent the scaling dimension and spin of each contributing operator $\OO$ in the OPE. This expansion is convergent for $r<1$. Earlier discussions of thermal 2-point functions and OPEs have appeared in \cite{Katz:2014rla,Witczak-Krempa:2015pia}. 

We note in passing that for later purposes, it will be convenient to write \eqref{opeaa} as a power-series in $z,\bar z$. We can achieve that by expanding the Gegenbauer polynomials using the identity
\beq
\label{appab}
C_{J}^{(\nu)}\left(\frac{1}{2}\left( \sqrt{\frac{z}{\bar z}} + \sqrt{\frac{\bar z}{z}} \right)\right) = 
\sum_{s=0}^J p_s(J) z^{-\frac{J}{2}+s} \bar z^{\frac{J}{2}-s}
~,
\eeq
where
\beq
\label{appac}
p_s(J) = \frac{\Gamma(J-s+\nu)\Gamma(s+\nu)}{\Gamma(J-s+1)\Gamma(s+1)}\frac{1}{\Gamma(\nu)^2}
~.
\eeq
In this manner,
\beq
\label{appad}
g(z,\bar z) = \sum_{\OO_{\Delta,J} \in \phi\times \phi} a_{\OO_{\Delta,J}}
\sum_{s=0}^J p_s(J) z^{h-\Delta_\phi+s} \bar z^{\bar h -\Delta_\phi -s}
~.
\eeq
where
\beq
\label{appae}
h := \frac{\Delta -J}{2}~, ~~
\bar h := \frac{\Delta +J}{2}
~.
\eeq

Substituting the OPE \eqref{opeaa} into the KMS condition \eqref{notaaf} yields an equation of the form 
\beq
\label{appaf}
\sum_{\OO_{\Delta,J} \in \phi \times \phi} a_{\OO_{\Delta,J}} \, \bigg[ C_J^{(\nu)}\left( \frac{1}{2}(w+w^{-1})\right) r^{\Delta - 2\Delta_\phi} - C_J^{(\nu)}\left( \frac{1}{2}(\tilde w+ \tilde w^{-1})\right) \tilde r^{\Delta - 2\Delta_\phi} \bigg] = 0
~,
\eeq
where we used the parametrization $z=rw$, $1-z=\tilde r \tilde w$. This equation requires both OPEs to be valid, namely 
\beq
\label{appag}
r < 1~, ~~ \tilde r < 1
~.
\eeq
Following familiar nomenclature from the analysis of 4-point functions at zero temperature, we will sometimes call the OPE expansion around $r=0$ the $s$-channel, and the OPE expansion around $\tilde r=0$ the $t$-channel.

Eq.\ \eqref{appaf} can be viewed as an infinite set of non-trivial sum rules for the 1-point function coefficients $b_\OO$ (or, equivalently, the coefficients $a_\OO$). These sum rules can be reformulated in many different ways, e.g.\ by acting on the LHS of \eqref{appaf} with arbitrary linear functionals. In the numerical implementations of later sections, we will employ sum rules that arise from the point-wise evaluation of the LHS on a grid of different points on the allowed region of OPE-convergence on the $z$-plane.

The goal of the thermal bootstrap program is to flesh out explicit constraints on individual 1-point coefficients assuming knowledge about the zero-temperature CFT data (scaling dimensions and 3-point function coefficients). As we discussed in the introduction, this task is complicated by the fact that the solutions to the KMS condition with a fixed spectrum of scaling dimensions is not, in general, unique and by the fact that hard truncations (even with some universal fixed approximation of the high-scaling dimension contributions) cannot recover this ambiguity. A proper analysis of the KMS condition requires the inclusion of the {\it full}, infinite set of contributions to the OPE expansions. In the next section, we discuss the first step towards the implementation of such an analysis.

\section{OPE tails and thermal dispersion relations}
\label{dispersion}

Dispersion relations for scalar 2-point functions on $S^1\times \IR^{d-1}$ were discussed for the first time, to our knowledge, in Ref.\ \cite{Alday:2020eua}. In Appendix \ref{dispApp} we discuss a variety of dispersion relations that can be deduced either with a straightforward use of Cauchy's theorem, or with a use of a thermal Lorentzian inversion formula \cite{Iliesiu:2018fao}. In this section, we present without detailed explanations, a subtracted dispersion relation that will be particularly useful in the following sections. For an explicit derivation of this dispersion relation we refer the reader to Appendix \ref{dispApp}.

A 2-point function $g(z,\bar z)$ of identical scalars $\phi$ at finite temperature, \eqref{notaaa}, can be expressed in terms of its discontinuity
\beq
\label{dispaa}
\disc[g(z,\bar z)] = - i \bigg( g(z+i\epsilon, \bar z) - g(z-i\epsilon, \bar z) \bigg)
\eeq
across $z\in (-\infty, -1) \cup (1,\infty)$ as follows\footnote{For notational convenience, here and in the rest of the text we are changing notation from $a_{\OO_{\Delta,J}} \to a_{\Delta,J}$.}
\bea
\label{dispab}
g(rw,rw^{-1}) &=& \sum_{J=0}^{J_*} \sum_{\Delta}  a_{\Delta,J}\, C_J^{(\nu)}\left(\frac{1}{2}(w+w^{-1})\right) r^{\Delta-2\Delta_\phi} 
\nonumber\\
&&+ 2 \left( \int_{-\infty}^{-r^{-1}} +  \int^{\infty}_{r^{-1}} \right)
dw' \, \KK_{J_*}(w,w')\,
{\rm Disc}\left[ g(rw', r{w'}^{-1} ) \right]
~,
\eea
where $J_*\geq 0$ is an arbitrary spin cutoff. As detailed in Appendix\ \ref{dispApp}, $J_*$ needs to be selected above some number $J_0$ (related to the Regge behavior of the 2-point function), otherwise Eq.\ \eqref{dispab} includes an extra arc contribution (see Eq.\ \eqref{subdia} for details). In what follows, we assume $J_*\geq J_0$. The $J_*$-dependent kernel $\KK_{J_*}(w,w')$ has a universal contribution (independent of the spacetime dimension) $\KK(w,w')$, and a spacetime-dependent part
\bea
\label{dispac}
\KK_{J_*}(w,w') := \frac{1}{2} \KK(w,w') 
- w'^{-1}(w'-w'^{-1})^{2\nu} \left[ \sum_{J=0}^{J_*} K_J\, C_J^{(\nu)} \left( \frac{1}{2}(w+w^{-1})\right) F_J(w'^{-1})\right]
\, ,
\eea
where
\beq
\label{dispad}
\KK(w,w') := \frac{1}{2\pi w'} \frac{w'^2 - 1}{(w'-w)(w'-w^{-1})}
~,
\eeq
\beq
\label{dispae}
K_J := \frac{\Gamma(J+1)\Gamma(\nu)}{4\pi \Gamma(J+\nu)}
~,
\eeq
\beq
\label{dispaf}
F_J(w) = w^{J+d-2} ~_2F_1\left( J+d-2, \frac{d}{2}-1, J + \frac{d}{2}, w^2 \right)
~.
\eeq

In Eq.\ \eqref{dispab} the integrated discontinuity term on the second line is capturing the full contribution of operators with spin $J>J_*$. The contributions of the operators with $J\leq J_*$ are captured by the sum in the first term on the RHS of \eqref{dispab}. This sum is a truncated version of the thermal OPE in spin. It still involves an infinite number of contibutions with arbitrarily high scaling dimension $\Delta$ at spins at or below $J_*$.

In these expressions $J_*$ is a free integer parameter. As an extreme choice, we can remove completely the truncated OPE term and allow the full 2-point function $g(z,\bar z)$ to be captured by the integrated discontinuity with the universal kernel $\frac{1}{2}\KK(w,w')$ and a potential arc contribution that is detailed in Appendix \ref{dispApp}. On the other end of extreme choices, we can increase $J_*$ to an arbitrarily high value in order to include a very large number of spins in the truncated OPE term. As we do that, the contribution of the integrated discontinuity decreases, and (as one can easily check from the explicit definition \eqref{dispac}) the kernel $\KK_{J*}(w,w')$ suppresses more and more the integral away from the branch-cut endpoints at $z=-1, 1$.

To summarize, a subtracted thermal dispersion relation, \eqref{dispab}, allows us to capture all the `large'-spin\footnote{To avoid potential confusions, we repeat that `large'-spin in this context refers to spins $J>J_*$. Similarly, `low'-spin refers to spin $J\leq J_*$. The value of the cutoff $J_*$ remains arbitrary in this language.} contributions to the thermal block expansion in terms of the discontinuity of the 2-point function. In the next section, we will use approximations of the discontinuity in order to approximate the `large'-spin contributions to the OPE. These will be the only approximations we implement to the KMS condition. 

Now that we have a way to deal with the `large'-spin part, let us elaborate on the `low'-spin part of Eq.\ \eqref{dispab}. Using a spin-dependent cutoff $\Delta_*(J)$ we can recast the OPE term on the RHS of Eq.\ \eqref{dispab} as
\bea
\label{dispag}
\sum_{J=0}^{J_*} \sum_{\Delta}  a_{\Delta,J}\, C_J^{(\nu)}\left(\frac{1}{2}(w+w^{-1})\right) r^{\Delta-2\Delta_\phi} 
&=& \sum_{J=0}^{J_*} \sum_{\Delta\leq \Delta_*(J)}  a_{\Delta,J}\, C_J^{(\nu)}\left(\frac{1}{2}(w+w^{-1})\right) r^{\Delta-2\Delta_\phi} 
\nonumber\\
&&\hspace{-1.5cm}+\sum_{J=0}^{J_*} \sum_{\Delta > \Delta_*(J)}  a_{\Delta,J}\, C_J^{(\nu)}\left(\frac{1}{2}(w+w^{-1})\right) r^{\Delta-2\Delta_\phi} 
~.
\eea
The first term on the RHS of this expression is a truly truncated OPE that involves a finite number of terms. The second term involves an infinite number of contributions from operators with arbitrarily high scaling dimension. We recast it in the form
\beq
\label{dispai}
\sum_{J=0}^{J_*} \sum_{\Delta > \Delta_*(J)}  a_{\Delta,J}\, C_J^{(\nu)}\left(\frac{1}{2}(w+w^{-1})\right) r^{\Delta-2\Delta_\phi} 
= \sum_{J=0}^{J_*} A_{\Delta_*(J),J}(r) \, C_J^{(\nu)}\left(\frac{1}{2}(w+w^{-1})\right) 
~,
\eeq
in terms of a finite new set of one-dimensional (tail) functions
\beq
\label{dispaj}
A_{\Delta_*(J),J}(r) := \sum_{\Delta > \Delta_*(J)} a_{\Delta,J}\, r^{\Delta -  2\Delta_\phi}
~.
\eeq
Notice that up to a subtraction of a finite number of contributions with $\Delta\leq \Delta_*(J)$, the tail functions $A_{\Delta_*(J),J}(r)$ are simply a Gegenbauer projection of the full thermal 2-point functions to a specific spin (similar to Eq.\ \eqref{asac}). In that sense, these functions are always well-defined and the sums in \eqref{dispaj} are suitably convergent.

Putting everything together, we can write
\begin{tcolorbox}
\vspace{-0.5cm}
\bea
\label{dispak}
g(rw,rw^{-1}) &=& \sum_{J=0}^{J_*} \sum_{\Delta\leq \Delta_*(J)}  a_{\Delta,J}\, C_J^{(\nu)}\left(\frac{1}{2}(w+w^{-1})\right) r^{\Delta-2\Delta_\phi} 
\nonumber\\
&&+ \sum_{J=0}^{J_*} A_{\Delta_*(J),J}(r) \, C_J^{(\nu)}\left(\frac{1}{2}(w+w^{-1})\right)
\\
&&+ 2 \left( \int_{-\infty}^{-r^{-1}} +  \int^{\infty}_{r^{-1}} \right)
dw' \, \KK_{J_*}(w,w')\,
{\rm Disc}\left[ g(rw', r{w'}^{-1} ) \right]
~.
\nonumber
\eea
\end{tcolorbox}
\noindent
We emphasize that this is an exact relation with freely tunable scaling-dimension and spin cutoffs $\Delta_*(J)$, $J_*$, respectively. This relation forms the basis of the bootstrap approach proposed in this paper.

\section{KMS condition 2.0}
\label{kms2}

We can use Eq.\ \eqref{dispak} to reformulate the KMS condition \eqref{notaaf} in terms of a finite number of CFT data, a finite number of tail functions and the discontinuity. Our ultimate goal is to approximate the contribution of the discontinuity and use the resulting approximate crossing equation to bootstrap the explicit low-lying CFT data, as well as the corresponding tail functions. To our knowledge, this is the first general attempt to dynamically determine entire OPE tails (at fixed spin) using bootstrap methods.

\subsection{A reformulation of the exact KMS condition}
\label{exactcross}

Employing the decomposition \eqref{dispak}, the {\it exact} KMS condition 
\beq
\label{kms2aa}
g(z,\bar z)=g(1-z,1-\bar z)
\eeq
becomes (adopting the notation $z=rw$, $1-z=\tilde r \tilde w$)
\bea
\label{kms2ab}
&&\sum_{J=0}^{J_*} \sum_{\Delta\leq \Delta_*(J)}  a_{\Delta,J}\, \Bigg[
C_J^{(\nu)}\left(\frac{1}{2}(w+w^{-1})\right) r^{\Delta-2\Delta_\phi} 
- C_J^{(\nu)}\left(\frac{1}{2}(\tilde w+ \tilde w^{-1})\right) \tilde r^{\Delta-2\Delta_\phi} \bigg]
\nonumber\\
&&+\sum_{J=0}^{J_*} \Bigg[ A_{\Delta_*(J),J}(r) \, C_J^{(\nu)}\left(\frac{1}{2}(w+w^{-1})\right) 
- A_{\Delta_*(J),J}(\tilde r) \, C_J^{(\nu)}\left(\frac{1}{2}(\tilde w+\tilde w^{-1})\right) 
\Bigg]
\\
&&+ \IT_\disc[J_*;rw,rw^{-1}] - \IT_\disc[J_*;\tilde r \tilde w, \tilde r \tilde w^{-1}] = 0
~,\nonumber
\eea
where
\beq
\label{kms2ac}
\IT_\disc[J_*;rw,rw^{-1}] := 2 \left( \int_{-\infty}^{-r^{-1}} +  \int^{\infty}_{r^{-1}} \right) dw' \, \KK_{J_*}(w,w')\, {\rm Disc}\left[ g(rw', r w'^{-1} ) \right]
\eeq 
is the contribution of the integrated discontinuity.
Equation \eqref{kms2ab} imposes a necessary condition on $g(z,\bar z)$, but is not practically useful unless we find a sensible way to approximate the discontinuity and the tail functions $A_{\Delta_*(J),J}(r)$. This is our next goal.

\subsection{Approximations}
\label{approxcross}

\subsubsection{Capturing the discontinuity}

A common way to approximate discontinuities of correlation functions is through the use of the first few terms of the OPE in the crossed channel. In several examples, and in different contexts, it has been observed that such approximations can work well \cite{Alday:2017vkk,Alday:2017zzv,Lemos:2017vnx,Liu:2018jhs,Caron-Huot:2018kta,Cardona:2018qrt,Li:2019dix,Albayrak:2019gnz,Carmi:2019cub,Alday:2019clp,Iliesiu:2018fao,Lemos:2021azv}. A positive aspect of our setup is that the systematic error introduced by such an ad hoc approximation can be reduced by increasing $J_*$. Indeed, the discontinuity appears inside an integral with kernel $\KK_{J_*}$, which naturally suppresses the contributions of high-twist operators in the crossed-channel expansion of the discontinuity (as we will see more clearly in a moment). The suppressing effects of $\KK_{J_*}$ can also ameliorate another problem.

At finite temperature, a well-known difficulty in the above-mentioned approximation scheme for the discontinuity stems from the need to use the OPE simultaneously in both the $s$ and $t$-channels. The convergence properties of the corresponding OPEs require both $r<1$ and $\tilde r<1$. As a result, when we integrate the discontinuity along $z\in(-\infty,-1)\cup(1,\infty)$, a large part of the integrals in \eqref{kms2ab} lies outside the region of convergence of the crossed channel in the approximation of $\disc[g]$. Specifically, it is not possible to use the crossed-channel OPE to approximate the discontinuity for $z\in (-\infty,-2)\cup(2,\infty)$.\footnote{We note in passing that this problem does not arise for similar schemes in zero-temperature 4-point crossing equations that employ dispersion relations with integrated double discontinuities \cite{NPS}.} This difficulty was also noted in the context of the Lorentzian OPE inversion of thermal 2-point functions in \cite{Iliesiu:2018fao}. 

To address this problem, we can restrict by fiat the integrals in \eqref{dispak} over $rw'\in (-\infty,-1)\cup(1,\infty)$ (or equivalently $w'\in (-\infty,-r^{-1})\cup(r^{-1},\infty)$) to the subregion of common $s$-  and $t$-channel convergence $w'\in (-2r^{-1},-r^{-1})\cup(r^{-1},2r^{-1})$. This introduces an additional systematic error, because we are ignoring the integral of the discontinuity in the excised $w'$-region, but this error is naturally suppressed by the kernel $\KK_{J_*}$ and becomes increasingly smaller as one increases $J_*$. As a result, $\KK_{J_*}$ can be used to suppress at the same time both errors introduced by the range of integration and by the use of hard truncations of the OPE in the modeling of the discontinuity. 

With these caveats in place, it will be useful to flesh out the precise way in which we approximate the contribution of the discontinuities in Eq.\ \eqref{dispak} and the corresponding crossing equation \eqref{kms2ab}.

Along the right truncated branch cut ($z':=rw'\in (1,2)$) we use the KMS condition and the expressions in \eqref{appad} to write 
\bea
\label{kms2ba}
&&\disc[g(z',\bar z')]= \disc[g(1-z', 1- \bar z')] 
\nonumber\\
&&\simeq \sum_{\OO_{\Delta, J}~{\rm truncated}} a_{\Delta, J} \sum_{s=0}^J p_s(J)\disc[(1-z')^{h_\OO-\Delta_\phi+s}] (1-\bar z')^{\bar h_\OO-\Delta_\phi-s}
\\
&&=-2 \sum_{\OO_{\Delta, J}~{\rm truncated}} a_{\Delta, J} \sum_{s=0}^J p_s(J) \sin\left[ \pi \left(\frac{\tau_\OO}{2} -\Delta_\phi + s\right)\right] (z'-1)^{\frac{\tau_\OO}{2}-\Delta_\phi+s} (1-\bar z')^{\frac{ \tau_\OO}{2}-\Delta_\phi+J-s}
\nonumber
~.
\eea
Here we have approximated the discontinuity using a truncated OPE in the crossed channel, where the sum $\sum_{\OO_{\Delta, J}~{\rm truncated}}$ runs over a subset of leading-twist operators. In the third line we used the explicit expression for the discontinuity of a generic power and reformulated the quantities $h = \frac{\Delta-J}{2}$, $\bar h = \frac{\Delta + J}{2}$ in terms of the twist $\tau = \Delta-J$.
Similarly, for the integral along the left truncated branch cut $(rw'\in (-2,-1))$ we set
\bea
\label{kms2bb}
&&\disc[g(z',\bar z')]= \disc[g(1+z', 1+\bar z')] 
\nonumber\\
&&\simeq \sum_{\OO_{\Delta, J}~{\rm truncated}} a_{\Delta, J} \sum_{s=0}^J p_s(J)\disc[(1+z')^{h_\OO-\Delta_\phi+s}] (1+\bar z')^{\bar h_\OO-\Delta_\phi-s}
\\
&&=2 \sum_{\OO_{\Delta, J}~{\rm truncated}} a_{\Delta, J} \sum_{s=0}^J p_s(J) \sin\left[ \pi \left(\frac{\tau_\OO}{2} -\Delta_\phi + s\right)\right] (-z'-1)^{\frac{\tau_\OO}{2}-\Delta_\phi+s} (1+\bar z')^{\frac{ \tau_\OO}{2}-\Delta_\phi+J-s}
\nonumber
~.
\eea

It is straightforward to numerically check that the kernel $\KK_{J_*}$ suppresses the integrands along the left and right branch cuts away from the points $z'\simeq -1$ and $z'\simeq 1$. We can, therefore, see explicitly in Eqs.\ \eqref{kms2ba}, \eqref{kms2bb} how the contributions of higher-twist operators are suppressed.

In summary, after the implementation of the above approximations, we obtain an approximate version of $\IT_\disc[J_*,rw,rw^{-1}]$ in Eq.\ \eqref{kms2ac}, which will be denoted from now on as $\IT_\disc^{(\rm approx)}[J_*,\{a\};rw,rw^{-1}]$. This function involves integrals along the truncated branch cuts $(rw' \in (-2,-1)\cup(1,2))$ and the approximate discontinuity, which involves a subset of the unknown coefficients $a_\OO$. The notation $\{a\}$ refers collectively to this dependence.

\subsubsection{Capturing the tail functions with Neural Networks}

The second ingredient that we need to tackle efficiently in the crossing equation \eqref{dispak} is a flexible, generic way to model the multiple tail functions $A_{\Delta_*(J),J}(r)$. We propose the use of Neural Networks (NNs) for this purpose, because of their versatility in expressing generic functions and because modern Machine Learning libraries, like {\tt PyTorch} or {\tt TensorFlow}, allow us to run efficiently non-convex optimization problems involving such representations. In the applications below we represent the tail functions using Multi-Branch MLPs. These are Neural Network architectures with an initial layer connected to $\frac{J_*}{2}+1$ subnets\footnote{$J_*$ is an even integer in our setup.} each consisting of 2 hidden layers. Each subnet is learning a single corresponding tail function $A_J(r)$. A graphical depiction of the architecture and a detailed description of the technical characteristics of the NNs that we used can be found in Appendix \ref{nn}. Throughout our computations we employed networks that have tens of thousands of optimizable parameters. We will call collectively the vector of NN parameters $\vec{\theta}$ and denote the NN representations of the tail functions $A_{\Delta_*(J),J;{\theta}}(r)$.

We note in passing that the use of fully connected feedforward NNs with multiple output channels (instead of Multi-Branch MLPs) was much less efficient, yielding losses that were several orders of magnitude higher than the Multi-Branch MLPs.

\subsubsection{Features of the approximate KMS condition}

We are now in prime position to write down an approximate KMS condition that is amenable to numerical analysis and goes beyond the traditional hard truncation schemes or soft truncation schemes that are based on ad hoc tail approximations. The form of the new approximate KMS condition is
\bea
\label{kms2ca}
&&\sum_{J=0}^{J_*} \sum_{\Delta\leq \Delta_*(J)}  a_{\Delta,J}\, \Bigg[
C_J^{(\nu)}\left(\frac{1}{2}(w+w^{-1})\right) r^{\Delta-2\Delta_\phi} 
- C_J^{(\nu)}\left(\frac{1}{2}(\tilde w+ \tilde w^{-1})\right) \tilde r^{\Delta-2\Delta_\phi} \bigg]
\nonumber\\
&&+\sum_{J=0}^{J_*} \Bigg[ A_{\Delta_*(J),J;{\boldsymbol{\theta}}}(r) \, C_J^{(\nu)}\left(\frac{1}{2}(w+w^{-1})\right) 
- A_{\Delta_*(J),J;{\boldsymbol{\theta}}}(\tilde r) \, C_J^{(\nu)}\left(\frac{1}{2}(\tilde w+\tilde w^{-1})\right) 
\Bigg]
\\
&&+ \IT_\disc^{(\rm approx)}[J_*,\{a\};rw,rw^{-1}] - \IT_\disc^{(\rm approx)}[J_*,\{a\};\tilde r \tilde w, \tilde r \tilde w^{-1}] = 0
~.\nonumber
\eea
The coefficients $\{a\}$ that appear in $\IT_\disc^{(\rm approx)}$ may or may not be part of the truncated conformal block expansions in the first line. If they do not appear in the first line with spin $J\leq J_*$, then their contributions are necessarily part of the tail functions approximated by NNs on the second line.

It is useful to highlight the following features of Eq.\ \eqref{kms2ca}:
\begin{itemize}
    \item[(1)] In the exact KMS condition \eqref{kms2ab} one can freely tune the parameter $J_*$. The presence of the approximations in \eqref{kms2ca} can pose some limitations. For example, when $J_*=0$ only a single tail for scalar operators needs to be included, which makes the second line of \eqref{kms2ca} relatively simple with a single unknown function. However, this is also where the approximation in $\IT_\disc^{(\rm approx)}$ performs worst. As we increase $J_*$, the approximation in $\IT_\disc^{(\rm approx)}$ improves, but the second line involves an increasing number of unknown functions, which complicates the numerical analysis of \eqref{kms2ca}. This suggests that an intermediate, relatively low value of $J_*$ may be numerically optimal. For example, $J_*=10$ would involve the presence of 6 tail functions, which are amenable to efficient optimization. 
    
    \item[(2)] The scaling dimension cutoffs $\Delta_*(J)$, controlling the number of terms that are included explicitly in the first line of \eqref{kms2ca}, are free parameters. Their choice does not affect the accuracy of \eqref{kms2ca}. When an operator is included in the truncated OPE of the first line of \eqref{kms2ca}, we will say that {\it the operator is exposed}. 
    
    \item[(3)] The presence of the tail functions in our formulation allows us to bootstrap the contribution of an infinite number of operators in the thermal block expansion. As we remarked in the introduction, in any other truncation scheme based on a finite number of unknown parameters, the problem reduces to a linear regression problem with a unique solution. This immediately clashes with the fact that the KMS condition can have an infinite number of solutions and obstructs one's ability to discover them. By including unknown optimizable tail functions in our formulation, we obtain a more flexible framework that evades this problem. Our approach leads to non-convex optimization problems, typically involving a rich landscape of many different solutions. 
    
    \item[(4)] In guided searches for specific solutions, the discontinuity in \eqref{kms2ca} plays an important role. Since discontinuities capture a more fundamental part of the correlation function (compared to the full correlation function), one can view $\IT_\disc^{(\rm approx)}$ as a source for the exposed unknown coefficients $a_{\Delta, J}$ and the unknown tail functions $A_{\Delta_*(J),J;{\boldsymbol{\theta}}}(r)$.
\end{itemize}

We will revisit several specific aspects of points (3) and (4) when discussing holographic CFTs in Section \ref{holo2}. In that context, the KMS condition is known to have an infinite number of solutions corresponding to bottom-up holographic CFTs that are dual to arbitrary higher-derivative theories of gravity in AdS.

\subsubsection{KMS as a non-convex optimization problem}
\label{losses}

To solve Eq.\ \eqref{kms2ca} numerically, we discretize the values of $(z,\bar z)=(rw, rw^{-1})$ on a grid of finite points and evaluate the LHS of \eqref{kms2ca} on this grid to form a vector $\vec F(\vec a,\vec \theta)$ that depends algebraically on the (exposed) unknown parameters $a_{\Delta, J}$ and the NN parameters $\vec \theta$ that appear in \eqref{kms2ca}. Then, we choose a non-negative loss function $\LL(a,\vec \theta)$ and perform non-convex optimization to determine the configurations
\beq
\label{kms2da}
(\vec a_*, \vec \theta_*) = {\rm argmin}\,\LL(\vec a, \vec \theta)
\eeq
that minimize the loss in the KMS condition. We perform the optimization using {\tt Adam}, an optimizer based on Stochastic Gradient Descent that incorporates adaptive learning rates and momentum.

The choice of a point-wise evaluation of Eq.\ \eqref{kms2ca} is obviously one of the many possibilities one can employ. Another common choice in such problems includes the evaluation of derivatives at the crossing-symmetric point $z=\frac{1}{2}$. In this paper, we choose point-wise evaluation, because it is a natural approach when one tries to determine directly the tail functions $A_{\Delta_*(J),J}(r)$. We work with a uniform grid on the $z$-plane that covers most of the common convergence region of the two conformal block expansions in the crossing equation. The grid of 243 points that we used throughout the paper appears in Figure\ \ref{fig:grid243}. It covers the region $|z|<0.95$, $|1-z|<0.95$.

We implemented several different types of loss functions. We found that the following ones were the most promising.

\subsubsection*{Mean absolute loss $\LL_{\overline{\text{abs}}}$}

In this case, we define
\beq
\label{kms2db}
\LL_{\overline{\text{abs}}}(\vec F) := \frac{1}{N} \sum_{i=1}^N |F_i|
~,
\eeq
where $N$ is the number of grid points and $F_i$ the components of the KMS vector $\vec F$. We can apply this loss in generic situations. We optimize it varying all the unknown parameters $(\vec a,\vec \theta)$. One limitation of this loss function is that good minimum values cannot be determined a priori.

Another, related, common choice in this class of loss functions is the root mean square loss. We did not employ it extensively in this work.

\subsubsection*{Dot-product loss with no optimizable $a_{\Delta,J}$ coefficients $\LL_{\text{dot}(0)}$}

In some cases we may want to optimize over a set of tail functions without varying any $a_{\Delta,J}$ coefficients. This can happen, for example, because we insert into the KMS condition some information about $a_{\Delta,J}$ coefficients and there are no other exposed optimizable $a$-coefficients outside the tail functions. To denote the absence of optimizable $a$-coefficients we use the subscript $(0)$.

In such situations the KMS vector $\vec F$ takes the form
\beq
\label{kms2dc}
\vec F = \vec g_\theta - \vec h
~,
\eeq
where $\vec h$ is a fixed, known vector and $\vec g_\theta$ is the vector that contains the optimizable NN parameters $\vec \theta$. We define
\beq
\label{kms2dd}
\LL_{\text{dot}(0)}(\vec F(\vec\theta)) : = 1 - \frac{|\vec g_\theta \cdot \vec h|}{|\vec g_\theta||\vec h|} + \left|1-\frac{\vec g_\theta \cdot \vec h}{|\vec h|^2}\right|
~.
\eeq
The first term, $1-\frac{|\vec g_\theta \cdot \vec h|}{|\vec g_\theta||\vec h|}$, is 1 minus the absolute value of the cosine of the angle between the vectors $\vec g_\theta$ and $\vec h$. It is, therefore, manifestly non-negative. Its presence aims to suppress the component of $\vec g_\theta$ normal to $\vec h$. The square in the second term aims to equate the component of $\vec g_\theta$ parallel to $\vec h$ with $\vec h$. Indeed, in generic configurations of $\vec \theta$
\beq
\label{kms2de}
\vec g_\theta = c \vec h + \epsilon \hat h_\perp~, ~~ \vec h \cdot \hat h_\perp =0~, ~~ |\hat h_\perp|=1
\eeq
and
\beq
\label{kms2df}
\LL_{\text{dot}(0)}(\vec F(\vec\theta)) = \left( 1 - \frac{1}{\sqrt{1+\frac{\epsilon^2}{c^2 |\vec h|^2}}}\right) + \left| 1- c\right|
\underset{\epsilon\, \text{small}}{\simeq}
\frac{\epsilon^2}{2c^2 |\vec h|^2}+|1-c|
~.
\eeq
The goal of the optimization is to satisfy the KMS condition by minimizing $\frac{\epsilon}{c |\vec h|}$ simultaneously with $|1-c|$.

\subsubsection*{Dot-product loss with a single optimizable $a_{\Delta,J}$ coefficient $\LL_{\text{dot}(1)}$}

In other situations there will be a single exposed coefficient, call it $a$, to be  optimized together with a number of tail functions. In that case, the form of the KMS vector is 
\beq
\label{kms2dg}
\vec F(a,\vec \theta) = a \vec f + \vec g_\theta - \vec h
~,
\eeq
where $\vec f$ is the crossed thermal block for the exposed operator with thermal 1-point function proportional to $a$, $\vec g_\theta$ is the sum of the contributions that depend on the NN parameters $\vec \theta$ and $\vec h$ a fixed vector of known contributions. Instead of varying simultaneously $a$ and $\vec\theta$ in a loss function that contains both, we can minimize the loss function
\beq
\label{kms2di}
\LL_{\text{dot}(1)}(\vec F(\vec\theta)) : = 1 - \frac{|(\vec g_\theta -\vec h) \cdot \vec f|}{|\vec g_\theta-\vec h||\vec f|}
\eeq
with respect to the NN parameters $\vec{\theta}$ only, and at the very end set
\beq
\label{kms2dj}
a = - \frac{(\vec g_{\theta_*} - \vec h)\cdot \vec f}{|\vec f|^2}
~,
\eeq
where $\vec \theta_*={\rm argmin}\, \LL(\vec F(\vec\theta))$. The minimization of the loss \eqref{kms2di} orients the vector $\vec g_\theta -\vec h$ parallel to the known vector $\vec f$ and Eq.\ \eqref{kms2dj} fixes the coefficient $a$ to satisfy the KMS condition in the direction of $\vec f$. 

This prescription is an obvious modification of the previous approach with loss \eqref{kms2dd}. Compared to the absolute loss, here we do not vary over the unknown OPE coefficient $a$ during the optimization process. It is determined at the end of the computation by solving exactly one of the components of the KMS condition.

We used the subscript (1) in the notation of this loss to denote the presence of a single optimizable coefficient $a$. One can use this approach to determine multiple 1-point coefficients exposing them sequentially one by one, or by accordingly extending the prescription to handle multiple unknown 1-point coefficients simultaneously.

    \section{Generalized Free Fields}
\label{GFF}

In this section we consider the theory of Generalized Free Fields (GFFs), which can be solved analytically. As such, this theory provides a useful initial testing ground for the approach proposed above.

In the GFF CFT of a scalar primary field $\phi$ with scaling dimension $\Delta_\phi$ (in any spacetime dimension $d\geq 2$), the thermal 2-point function of $\phi$ can be expressed as a sum over images
\beq
\label{gffaa}
g(z,\bar z) = \sum_{m=-\infty}^\infty \frac{1}{\big[(m-z)(m-\bar z)\big]^{\Delta_\phi}}
~.
\eeq
The conformal block expansion of this expression receives contributions only from the identity operator $(\Delta=0,J=0)$ and the double-twist operators $[\phi\phi]_{n,J}$ (with scaling dimensions $\Delta_{n,J}=2\Delta_\phi + 2n +J$ and even spin $J=2\ell$, $\ell=0,1,\ldots$). The double-twist operators are conformal primary composites of two $\phi$'s with $J$ uncontracted spacetime derivatives and $n$ Laplacians $\Box=\d_\mu \d^\mu$. The conformal block expansion of \eqref{gffaa} reads
\beq
\label{gffab}
g(rw, rw^{-1}) = r^{-2\Delta_\phi} + \sum_{n=0}^\infty \sum_{\ell=0}^\infty a_{n,2\ell} \, C_{2\ell}^{(\nu)}\left( \frac{1}{2}(w+w^{-1}) \right) r^{2(n+\ell)}
\eeq
with 
\beq
\label{gffac}
a_{n,J} = 2\zeta(2\Delta_\phi+2n+J)\frac{(J+\nu)(\Delta_\phi)_{J+n}(\Delta_\phi-\nu)_n}{n!(\nu)_{J+n+1}}
\eeq
and the Pochhammer symbol $(a)_n = \frac{\Gamma(a+n)}{\Gamma(a)}$.

In this case, the discontinuity that arises from the crossed $t$-channel expansion is particularly simple: {\it only the identity operator contributes}. As a result, 
\bea
\label{gffad}
\IT_\disc^{({\rm approx})}[J_*;rw,rw^{-1}] &=& 4\sin(\pi \Delta_\phi) \bigg[
\int_{r^{-1}}^{2r^{-1}} dw'\, \KK_{J_*}(w,w') (rw'-1)^{-\Delta_\phi} (1-r w'^{-1})^{-\Delta_\phi}
\nonumber\\
&&-\int_{-2r^{-1}}^{-r^{-1}} dw'\, \KK_{J_*}(w,w') (-rw'-1)^{-\Delta_\phi} (1+rw'^{-1})^{-\Delta_\phi} \bigg]
\eea
is a fixed function of the spacetime coordinates that does not involve any non-trivial thermal 1-point functions. In Eq.\ \eqref{gffad} we assumed that the external scaling dimension $\Delta_\phi$ is not an integer. When $\Delta_\phi$ is an integer, the branch cut reduces to a pole and the discontinuity receives $\delta$-function contributions.

In what follows, we proceed to test the proposed approach for a specific randomly chosen case of a generalized free field with $\Delta_\phi=1.68$ in $d=4$ spacetime dimensions. We observed similar results for other values of $\Delta_\phi$, but do not report them explicitly here.

\subsection{Tests of the approximate KMS condition}
\label{testkms}

\begin{figure}[t!]
    \centering
    \begin{minipage}{0.4\textwidth}
        \centering
        \includegraphics[width=\textwidth]{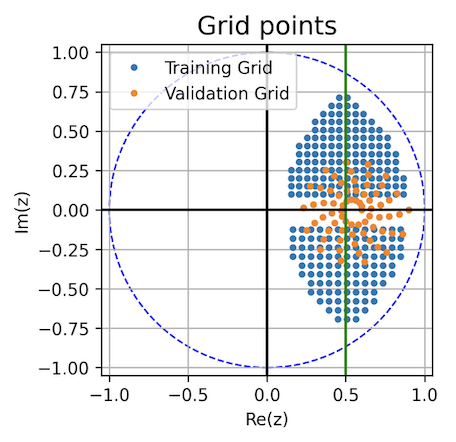} 
        \caption{\small{The blue points represent a grid of 243 points on the complex $z$-plane used for optimization during NN training. The grid avoids the real axis and covers points in the region $|z|<0.95$, $|1-z|<0.95$. The orange points represent the validation grid.
        \vspace{1.4cm}}}
        \label{fig:grid243}
    \end{minipage}
    \hfill
    \begin{minipage}{0.55\textwidth}
        \centering
        \begin{tabular}{|c|c|c|}
        \hline
        $J_*$ & $\LL_{\overline{\text{abs}}}$ & $\LL_{\text{dot}(0)}$   \\
        \hline\hline
        0 & 0.1177 & 0.0125  \\
        \hline
        2 & 0.0300 & $1.3\times 10^{-5}$  \\
        \hline
        4 & 0.0067 & $1.6 \times 10^{-7}$   \\
        \hline
        6 & 0.0013 & $3.3 \times 10^{-9}$   \\
        \hline
        8 & 0.0002 & $0.8 \times 10^{-10}$   \\
        \hline
    \end{tabular}
    \vspace{1.0cm}
        \captionof{table}{\small{The second column presents the mean absolute loss, Eq.\ \eqref{gffbb}, of the approximate KMS condition for the analytic solution with varying $J_*$ values. The third column presents the dot loss, Eq.\ \eqref{gffbc}, for the analytic solution. The spacetime dimension is $d=4$, the scaling dimension of the external operator is $\Delta_\phi=1.68$ and the grid used is depicted in Fig.\ \ref{fig:grid243}. 
        }}
        \label{tab:analytic_losses}
    \end{minipage}
    \vspace{-1.1cm}
\end{figure}

First, we want to examine how closely the approximate KMS condition captures the exact version as a function of $J_*$. For that purpose we substitute the exact GFF solution \eqref{gffac} into several versions of the approximate KMS condition and evaluate the loss functions we formulated above. Low loss values would demonstrate that our scheme is based on the `correct' equations. This is important. Recall that one of the major issues in generic truncation schemes has been the inability to control the systematic error introduced by the truncation.

\subsubsection{Loss on the analytic solution without exposed operators}
\label{analyticlossnoexposed}

In our first test, we consider the approximate KMS condition
\bea
\label{gffba}
&&r^{-2\Delta_\phi} - \tilde r^{-2\Delta_\phi}
+ \sum_{J=0}^{J_*} \Bigg[ A_{\Delta_*(J),J;{\boldsymbol{\theta}}}(r) \, C_J^{(\nu)}\left(\frac{1}{2}(w+w^{-1})\right) 
- A_{\Delta_*(J),J;{\boldsymbol{\theta}}}(\tilde r) \, C_J^{(\nu)}\left(\frac{1}{2}(\tilde w+\tilde w^{-1})\right) 
\Bigg]
\nonumber\\
&&+ \IT_\disc^{(\rm approx)}[J_*;rw,rw^{-1}] - \IT_\disc^{(\rm approx)}[J_*;\tilde r \tilde w, \tilde r \tilde w^{-1}] = 0
~.
\eea
We expose only the contribution of the identity in the conformal block expansion and keep everything else inside the tail functions $A_{\Delta_*(J),J;\boldsymbol{\theta}}(r)$ with a very low scaling dimension cutoff $\Delta_*(J)$ placed at the unitarity bound.

We evaluate Eq.\ \eqref{gffba} on the $(z,\bar z)$-grid of Figure\ \ref{fig:grid243} with 243 points,\footnote{In the grid of Figure \ref{fig:grid243}, we avoided on purpose points on the real axis, which correspond to 2-point functions at zero spatial separation $\sigma$. At those points, operators with different spin at the same scaling dimension make indistinguishable contributions to the 2-point function. Notice, however, that the region around the real axis, which is not sampled during training, is sampled via the orange points during validation. In addition, the grids above and below the real axis are chosen with a slight displacement that makes them inequivalent under the symmetry $\sigma \to -\sigma$.} producing a 243-dimensional vector $\vec F$ of algebraic equations. In addition, we formulate a loss function that quantifies how far this vector is from the zero-vector. Following the discussion of Section \ref{losses} we consider two options. The first is the mean absolute loss
\beq
\label{gffbb}
\LL_{\overline{\text{abs}}}(\vec F) = \frac{1}{N} \sum_{i=1}^N |F_i| 
~.
\eeq
The second is the dot-product loss $\LL_{\text{dot}(0)}$ of Eq.\ \eqref{kms2dd}
\beq
\label{gffbc}
\LL_{\text{dot}(0)}(\vec F) = 1 - \frac{|\vec g_\theta \cdot \vec h|}{|\vec g_\theta||\vec h|} + \left| 1 - \frac{\vec g_\theta \cdot \vec h}{|\vec h|^2} \right| 
~,
\eeq
where $\vec g_\theta$ is the vector of grid evaluations of the tail contributions in Eq.\ \eqref{gffba} and $\vec h$ the vector of grid evaluations of the function
\beq
\label{gffbd}
h(z,\bar z) = - \left\{ 
r^{-2\Delta_\phi} - \tilde r^{-2\Delta_\phi}
+ \IT_\disc^{(\rm approx)}[J_*;rw,rw^{-1}] - \IT_\disc^{(\rm approx)}[J_*;\tilde r \tilde w, \tilde r \tilde w^{-1}]
\right\}
~.
\eeq

Using the data of the exact solution \eqref{gffac} we can compute numerically the analytic values of all the tail functions. The values of the respective loss functions are presented in Table \ref{tab:analytic_losses} for even $J_*=0,2,4,6,8$. As expected, the result exhibits visible improvement with increasing $J_*$ both for $\LL_{\overline{\text{abs}}}$ and $\LL_{\text{dot}(0)}$. At $J_*=8$ we observe loss values at the order of $10^{-4}$ and $10^{-10}$ respectively.

\subsubsection{Loss on the analytic solutions with a single exposed operator}

It is also instructive to repeat the test of Section \ref{analyticlossnoexposed} by exposing a number of operators. For example, we can write an approximate KMS condition where the leading scalar operator is exposed and all other CFT data are inside the tail functions. The leading scalar operator cannot be the double-twist operator $[\phi\phi]_{0,0}$, because this operator contributes a constant in the KMS condition, which cancels out after crossing. Interestingly, the next leading scalar operator $[\phi\phi]_{1,0}$ alone is subtle for the following reason. 

The contribution of a double-twist operator $[\phi\phi]_{n,2\ell}$ to the crossed conformal block expansion in $d$ dimensions is
\beq
\label{gffbe}
a_{n,2\ell} \bigg[ r^{2n+J}C^{(\nu)}_J\left(\frac{1}{2}(w+w^{-1}) \right) - {\tilde r}^{2n+J} C^{(\nu)}_{J}\left( \frac{1}{2}(\tilde w + \tilde w^{-1}) \right) \bigg]
~.
\eeq
For $(n,2\ell) = (1,0),(0,2)$ we observe the peculiar identity\footnote{We have not observed similar identities for other operators at generic $(z,\bar z)$ points.}
\bea
\label{gffbf}
&&r^2 C^{(\nu)}_0\left(\frac{1}{2}(w+w^{-1}) \right) - {\tilde r}^{2} C^{(\nu)}_0\left( \frac{1}{2}(\tilde w + \tilde w^{-1}) \right)
\nonumber\\
&&=\frac{1}{\nu(2\nu+1)} \left[ r^2 C^{(\nu)}_2\left(\frac{1}{2}(w+w^{-1}) \right) - {\tilde r}^{2} C^{(\nu)}_2\left( \frac{1}{2}(\tilde w + \tilde w^{-1}) \right)\right]\nonumber
\\
&&=r(w+w^{-1})-1
~,
\eea
which implies that the scalar operator $[\phi\phi]_{1,0}$ comes together with the spin-2 $[\phi\phi]_{0,2}$ operator in the KMS condition to contribute the single term
\beq
\label{gffbg}
\big( a_{1,0} + \nu(2\nu+1) a_{0,2} \big) 
\big(r(w+w^{-1})-1 \big)
~.
\eeq
Notice that the combination
\beq
\label{gffbga}
a_{\Delta = 2\Delta_\phi+2} := a_{1,0} + \nu(2\nu+1) a_{0,2}
\eeq
is exactly the same type of combination of spin-dependent coefficients that appears in the zero-spatial-separation analysis of Ref.\ \cite{Marchetto:2023xap}. Accordingly, when we choose to expose the operators $[\phi\phi]_{1,0}, [\phi\phi]_{0,2}$, the approximate KMS condition (in $d=4$ dimensions) becomes
\bea
\label{gffbi}
&&r^{-2\Delta_\phi} - \tilde r^{-2\Delta_\phi}
+ \big( a_{1,0} + 3 a_{0,2} \big) 
\big(r(w+w^{-1})-1 \big)
\nonumber\\
&&+ \sum_{J=0}^{J_*} \Bigg[ A_{\Delta_*(J),J;{\boldsymbol{\theta}}}(r) \, C_J^{(\nu)}\left(\frac{1}{2}(w+w^{-1})\right) 
- A_{\Delta_*(J),J;{\boldsymbol{\theta}}}(\tilde r) \, C_J^{(\nu)}\left(\frac{1}{2}(\tilde w+\tilde w^{-1})\right) 
\Bigg]
\nonumber\\
&&+ \IT_\disc^{(\rm approx)}[J_*;rw,rw^{-1}] - \IT_\disc^{(\rm approx)}[J_*;\tilde r \tilde w, \tilde r \tilde w^{-1}] = 0
\eea
with $\Delta_*(J=0)=\Delta_*(J=2)=2\Delta_\phi + 2$ and all other $\Delta_*(J)$ at the unitarity bound.

As we discussed in subsection \ref{losses}, we can bootstrap this equation using the dot-loss $\LL_{\text{dot}(1)}$ of Eq.\ \eqref{kms2di}, where the optimizable parameters are only the NN parameters $\vec \theta$. The value of $a_{1,0} + 3 a_{0,2}$ is determined at the end of the computation by Eq.\ \eqref{kms2dj}. 
More specifically, in this context
\bea
\label{gffbj}
\LL_{\text{dot}(1)}(\vec F(\vec\theta)) : = 1 - \frac{|(\vec g_\theta -\vec h) \cdot \vec f|}{|\vec g_\theta-\vec h||\vec f|}
~,
\eea
where again $\vec g_\theta$ is the vector of grid evaluations of the tail contributions in Eq.\ \eqref{gffbi}, $\vec h$ the vector of grid evaluations of the function $h(z,\bar z)$ in Eq.\ \eqref{gffbd} and $\vec f$ the vector of grid evaluations of the function
\beq
\label{gffbk}
f(z,\bar z) = r(w+w^{-1})-1
~.
\eeq

In the second column of Table \ref{tab:a_analytic} we present the corresponding numerical values of $\LL_{\text{dot}(1)}$ on the grid of Figure\ \eqref{fig:grid243}, when we insert the analytic values of the tail functions for $J_*=2,4,6,8$. The choice $J_*=0$ is not viable here as we are exposing a spin-2 operator. In the third column of Table \ref{tab:a_analytic}, we list the values of $a_{1,0}+3a_{2,0}$ that are deduced from the analytic tail functions and Eq.\ \eqref{kms2dj}, which follows from the approximate KMS condition \eqref{gffbi}. Once again, it is satisfying to observe that the accuracy of the approximate KMS condition increases for increasing $J_*$ achieving $\LL_{\text{dot}(1)}$ of the order of $10^{-9}$ at $J_*=8$. We also notice a clear convergence of the approximate-KMS-derived value of $a_{1,0}+3a_{2,0}$ towards the analytic value 15.06013.

\begin{table}[t!]
    \centering
    \begin{tabular}{|c|c|c|}
        \hline
        $J_*$ & $\LL_{\text{dot}(1)}$ & $a_{1,0} + 3\, a_{2,0}$ from dot loss   \\
        \hline\hline
        2 & $9.5 \times 10^{-6}$ & 15.16582  \\
        \hline
        4 &  $1.4 \times 10^{-6}$ & 15.07614  \\
        \hline
        6 &  $7.8 \times 10^{-8}$ & 15.06252  \\
        \hline
        8 &  $3.5 \times 10^{-9}$ & 15.06049 \\
        \hline
        \multicolumn{2}{c}{} & \multicolumn{1}{|c|}{Exact value: 15.06013} \\
    \cline{3-3}
    \end{tabular}
    \caption{\small{The second column presents the values of the loss function $\LL_{\text{dot}(1)}$, Eq.\ \eqref{gffbj}, for the analytic values of the tails in the GFF theory. The third column presents the value of the $a_{1,0}+3a_{2,0}$ combination that follows from the approximate KMS condition according to Eq.\ \eqref{kms2dj}. As $J_*$ increases, there is clear convergence towards the analytic result listed at the bottom of the table.}}
    \label{tab:a_analytic}
\end{table}

\subsection{Tail bootstrap}
\label{kmstails}

In this section we assume no input of thermal data from the analytic solution of the GFF theory and try to recover it by bootstrapping the approximate KMS condition \eqref{gffba}. We set $J_*=6$, we do not expose any operators and bootstrap the four tail functions $A_0(r), A_2(r), A_4(r), A_6(r)$. However, before proceeding to the actual computation, we need to make a preparatory comment about the training grid and a related addition to the loss functions.

\subsubsection{Constraining the tail asymptotics with $\LL_{\rm BC}[A]$}

We are optimizing by evaluating the approximate KMS condition on the grid of Figure\ \ref{fig:grid243}. This grid covers most of the common region of convergence of the $s$- and $t$-channel expansions, but not all of it. The OPE-convergence region is defined by the inequalities $|z|<1$, $|1-z|<1$, whereas the grid of Figure\ \ref{fig:grid243} covers points in the region $|z|<0.95$, $|1-z|<0.95$. We excluded points near the boundary, because they were numerically more challenging. 

With this choice, initial experimentation shows that the landscape of minima for the original loss functions $\LL_{\overline{\rm abs}}$ and $\LL_{{\rm dot}(0)}$ is complicated with many low-loss configurations, some of which are clearly distinct from the analytic GFF solution. For example, we frequently observed low-loss configurations with tail functions exhibiting $r\to 1$ asymptotics that are inconsistent with the general analysis of Appendix\ \ref{apptails}, which is based on the KMS condition. This is a clear indication that by not training on the full OPE-convergence region we are failing to extract the complete information of the KMS condition, which in turn allows many more optimal configurations beyond the GFF solution.

To guide the search more efficiently, and avoid spurious minima, it is helpful to supplement our optimization problem with additional information. An alternative to enlarging the training grid is to add a condition that enforces the KMS constraint governing the asymptotic behavior of tail functions as $r\to 1$.

In Appendix \ref{apptails} we use the KMS condition to argue that the tail functions $A_{2\ell}(r)$ in thermal two-point functions of identical scalars exhibit a universal singularity structure in the limit $r\to 1^-$, when the external scalar scaling dimension $\Delta_\phi$ obeys the inequality $\Delta_\phi \geq \frac{d-1}{2}$. Focusing on this regime, we can use Eq.\ \eqref{asae} to approximate the numerical value of the optimizable tail functions $A_{2\ell}(r)$ at a value $r=r_0$ close to $r=1$ in a theory-independent way. In the numerical results reported below, we used $r_0=0.9999$. In order to recover accurately the universal asymptotics of Eq.\ \eqref{asae} we need to go exponentially close to $r=1$. Even at $r_0=0.9999$, the application of \eqref{asae} introduces a systematic error. For example, when $\Delta_\phi=1.68$ in $d=4$ dimensions, the exact GFF values of $A_J(0.9999)$ (for $J=0,2,4,6$) are $(83.044, 250.765, 413.139, 573.387)$ and the corresponding values predicted by Eq.\ \eqref{asae} are $(84.822, 254.467, 424.111, 593.756)$).

Practically, we enforced the asymptotic boundary condition at $r_0$ by adding to the loss functions a new term $\LL_{\rm BC}[A]$ 
\beq
\label{lbcaapre}
\LL \longrightarrow \LL + \LL_{\rm BC}[A]
~.
\eeq
We found through preliminary testing that $\LL_{\rm BC}[A]$ can be implemented most efficiently in the following manner. For starters, we set 
\beq
\label{lbcaa}
\AA_{J;\theta}(r) := {\rm arcsinh}\left( \frac{A_{J,\theta}(r)}{2} \right)~, ~~
\AA_{J;{\rm asymp}}(r) := {\rm arcsinh}\left( \frac{A_{J,{\rm asymp}}(r)}{2} \right)
~,
\eeq
where $A_{J,\theta}(r)$ are the optimizable tail functions, and $A_{J,{\rm asymp}}(r)$ are the functions obtained by the asymptotic formula \eqref{asae}. With these specifications, we defined
\bea
\label{lbcab}
\LL_{\rm BC}[A] &:=& \left(\frac{1}{\frac{J_*}{2} + 1} \sum_{\ell=0}^{\frac{J_*}{2}} \left( \AA_{2\ell;\theta}(r_0) - \AA_{2\ell;{\rm asymp}}(r_0)\right)^2\right)^2
\\
&&+ \frac{1}{\frac{J_*}{2} + 1} \sum_{\ell=0}^{\frac{J_*}{2}} \left( \AA_{2\ell;\theta}(r_0) - \AA_{2\ell;{\rm asymp}}(r_0)\right)^2
+\frac{1}{2} \sum_{\ell=0}^1 \left| \AA_{2\ell;\theta}(r_0) - \AA_{2\ell;{\rm asymp}}(r_0)\right|
~.\nonumber
\eea

In this expression, the role of the arcsinh transformation should be clear. It allows us to renormalize the large numbers that appear near the $r\to 1$ singularity, and as such it facilitates a more efficient optimization of the Neural Networks. 

The specific structure of the loss function in \eqref{lbcab} (with a quartic, a quadratic and a linear term) is less obvious. We observed that with single-power losses the optimization was still unstable towards different types of configurations with comparable low loss, despite the removal of the previous spurious minima with the wrong asymptotics. The variation was especially pronounced for the spin-0 and spin-2 tail functions.\footnote{That is also the origin of the last term in \eqref{lbcab} that involves only the spin-0 and spin-2 tail functions.} The specific form in \eqref{lbcab} was found to reduce this instability. We will argue momentarily that \eqref{lbcab} not only reduces the instability but also pushes the result toward the GFF configuration.\footnote{Unfortunately, \eqref{lbcab} does not solve completely the problem of spurious minima. We will return to this issue below.}

\begin{figure}[t!]
    \centering
    \includegraphics[width=1.0\linewidth]{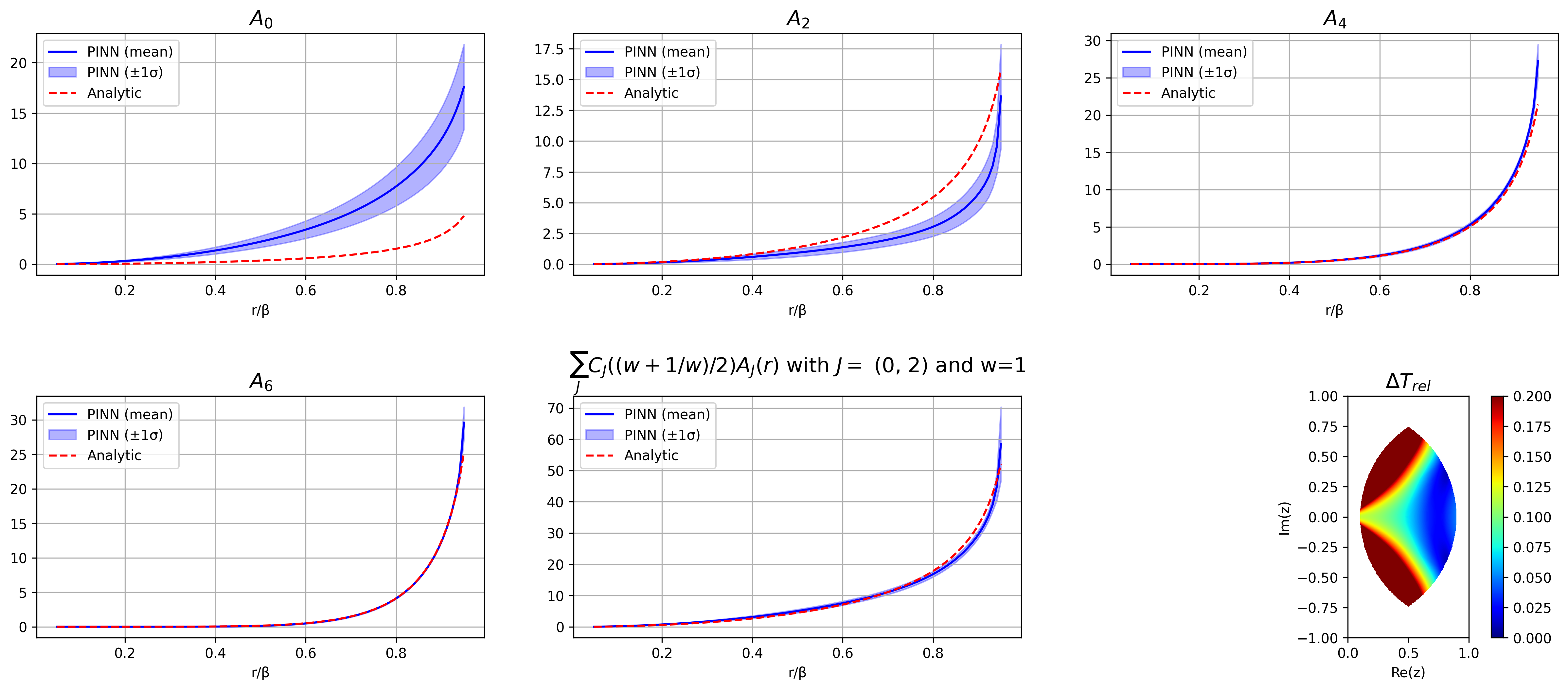}
    \caption{\small{Plots depicting the results obtained with $\LL_{\overline{\rm abs}}+\LL_{\rm BC}[A]$ in the $d=4$ GFF theory with $\Delta_\phi=1.68$ and $J_*=6$ without any exposed operators, and asymptotic boundary conditions for the tail functions at $r=0.9999$. The results are based on the 10 lowest-loss configurations within a pool of 1K independent training runs for 50K epochs. The mean loss of these configurations is $1.93 \times 10^{-3} \pm 1.06\times 10^{-4}$, whereas the absolute loss of the analytic GFF configuration is $1.29 \times 10^{-3}$. The first 4 plots depict (in blue) the mean and 1$\sigma$ deviation of the predicted tail functions. The dashed red curves represent the exact, analytic result of the GFF theory. The middle plot on the second line depicts the combined contribution of the two leading tail functions, $A_0$ and $A_2$, to the conformal block expansion of the thermal correlator for $w=1$ and its comparison against the analytic expression (in red). The final plot at the bottom right is a heatmap of the relative difference \eqref{tailsAbsab} between the predicted and analytic values in the training region.}}
    \label{fig:abs_no_exposed}
\end{figure}

\subsubsection{Bootstrap with $\LL_{\overline{{\rm abs}}}$}
\label{tailsAbs}

Using the loss function $\LL_{\overline{\rm abs}}(\vec F) + \LL_{\rm BC}[A]$ we obtained the results depicted in Figure\ \ref{fig:abs_no_exposed}. The plots in Figure \ref{fig:abs_no_exposed} present averages and standard deviations based on the 10 lowest-loss configurations that resulted after 50K epochs for 1K independent runs on the QMUL Apocrita cluster (with the same Neural Network hyperparameters). It is useful to highlight the following observations:

\paragraph{$(a)$} The optimization yields low-loss configurations with loss comparable to that of the analytic GFF solution ($1.93 \times 10^{-3} \pm 1.06\times 10^{-4}$ vs $1.29 \times 10^{-3}$ for GFF).

\paragraph{$(b)$} The leading low-loss configurations reproduce the analytic, higher-spin tail functions $A_4$, $A_6$ rather accurately and without significant variation. In contrast, there is significant variation in the profiles of the predicted functions $A_0$, $A_2$ with mean curves that are substantially far from the analytic GFF solution. Nevertheless, we observe that the obtained $A_0, A_2$ configurations have a total contribution to the crossing equation which is comparable to that of the analytic GFF solution. This is apparent from the plot of the combined conformal block contribution 
\beq
\label{tailsAbsaa}
T_{0+2}(r,w):=\sum_{J=0,2} A_J(r) C_J(\frac{1}{2}(w+w^{-1}))
\eeq
for $w=1$, and from the heatmap of the relative difference 
\beq
\label{tailsAbsab}
\Delta T_{rel} :=\frac{|T_{0+2;predicted}(r,w)-T_{0+2;analytic}(r,w)|}{|T_{0+2;predicted}(r,w)|+|T_{0+2;analytic}(r,w)|}
\eeq
between predicted and analytic tails inside the training region of Figure\ \ref{fig:grid243}. The $w=1$ curve in particular, shows that our setup has no trouble reproducing the spin-independent data $a_\Delta = \sum_J A_J(r) C_J(1)$ that appear in the bootstrap at zero spatial separation, \cite{Marchetto:2023xap,Buric:2025anb,Barrat:2025nvu}. 

The persistence of a landscape of many, roughly degenerate, low-loss $A_0, A_2$ configurations (despite the addition of \eqref{lbcab}) implies that we are still not extracting the full information of the KMS condition. We noted previously, in the expression \eqref{gffbg}, that there is an analytic ambiguity between the data $a_{1,0}$ and $a_{0,2}$ that cannot be resolved in the KMS condition that we are studying. We have not found, and do not expect, any similar, exact, analytic ambiguities that involve the full tail functions $A_0$ and $A_2$, so what we observe here is a numerical accident, whose origin lies in the form of the landscape of minima of the loss function. Indeed, the specific details of the loss function can influence such features and we will obtain further evidence of this statement in the next Section \ref{tailsDot}.

\paragraph{$(c)$} Although the mean predicted $A_0, A_2$ curves are far from the corresponding analytic GFF curves, some of the low-loss runs produced configurations that are close to the analytic GFF result. To illustrate this point, we present the predictions of the 10th in order-of-loss run in Figure\ \ref{fig:abs_no_exposed_example} in Appendix\ \ref{addplots}. This observation exhibits the fact that the optimization algorithm can indeed sample low-loss configurations in the vicinity of the GFF solution.

\paragraph{$(d)$} We have also checked that the low-loss configurations exhibit comparable low validation loss on the grid of orange points in Figure\ \ref{fig:grid243}. This implies that we can satisfy well the KMS condition in the heart of the region of OPE convergence (around $z=0.5$), even at points, which were not included in the optimization grid. In sharp contrast, the loss rises significantly outside the training region, towards the boundary $\{ |z|=1 \cup  |1-z|=1 \}$, for the optimal configurations that deviate from the GFF solution.

\subsubsection{Bootstrap with $\LL_{{\rm dot}(0)}$}
\label{tailsDot}

\begin{figure}[t!]
    \centering
    \includegraphics[width=1.0\linewidth]{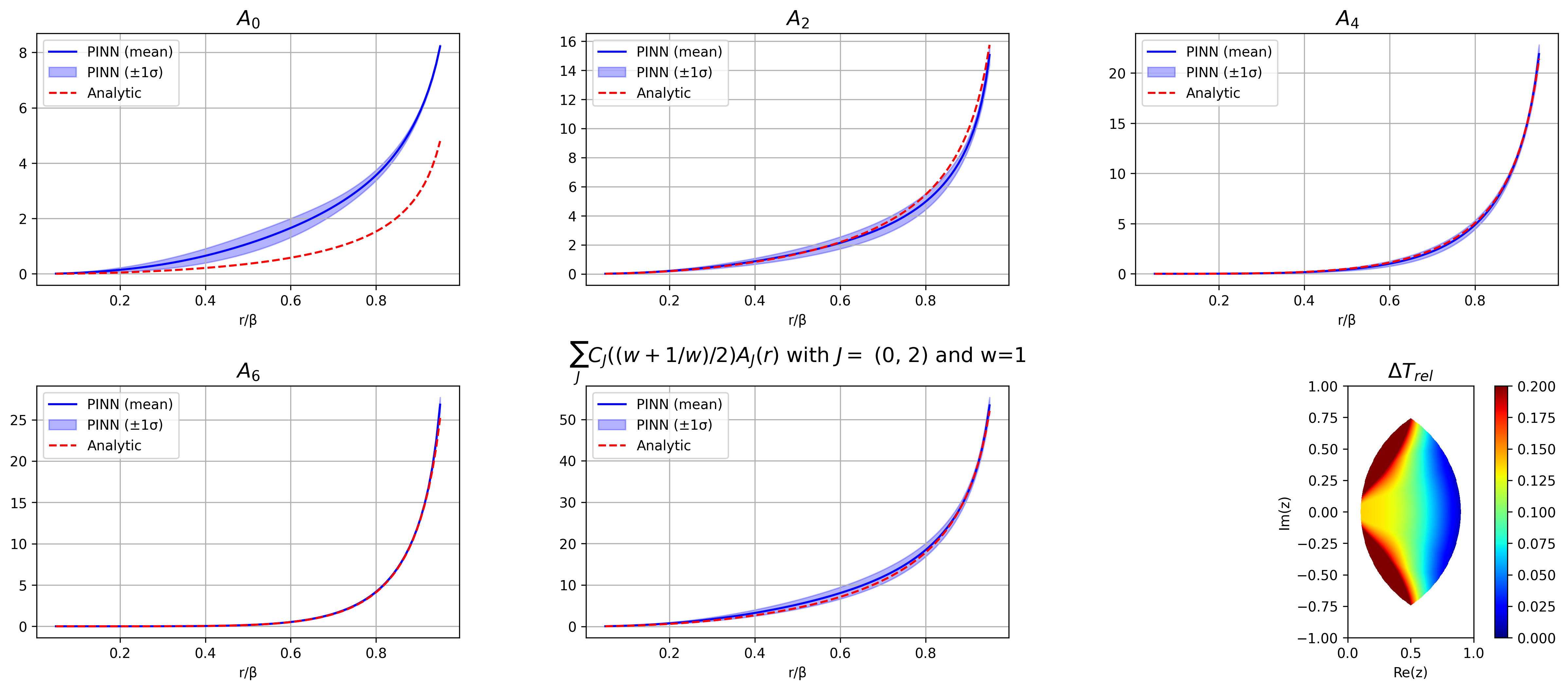}
    \caption{\small{Plots of the predicted tail functions obtained with $\LL_{{\rm dot}(0)}+\LL_{\rm BC}$ for $J_*=6$ in the $d=4$ GFF theory for $\Delta_\phi=1.68$. The obtained results are based on the 10 lowest-loss configurations for 1K independent runs of 50K epochs. The mean loss, $3.54\times 10^{-9} \pm 6.4 \times 10^{-10}$, should be compared to the loss of the analytic GFF solution $3.35 \times 10^{-9}$. The middle plot in the second line represents the combined contribution to the 2-point function of the $A_0$, $A_2$ tails at $w=1$. The heatmap depicts the relative difference \eqref{tailsAbsab} inside the training region.}}
    \label{fig:dot_no_exposed_no_input}
\end{figure}

Next we repeat the same exercise using the loss function $\LL_{{\rm dot}(0)}(\vec F)+\LL_{\rm BC}[A]$. The statistics collected from the 10 lowest-loss configurations from a pool of 1K independent runs with 50K epochs are depicted in Figure \ref{fig:dot_no_exposed_no_input}.

In this case, we observe that:

\paragraph{$(a)$} The optimization with the dot-loss \eqref{gffbc} is more efficient compared to the absolute loss optimization, producing visibly closer configurations to those of the analytic GFF solution. This is apparent when one compares the $A_0,A_2$ plots, and the corresponding heatmaps, in Figures\ \ref{fig:abs_no_exposed} and \ref{fig:dot_no_exposed_no_input}. We additionally observe that the dot-loss-optimized NNs achieve a very low loss with values $3.54\times 10^{-9} \pm 6.4 \times 10^{-10}$, which are close to the loss of the analytic GFF solution $3.35 \times 10^{-9}$.

\paragraph{$(b)$} The issue with the numerical ambiguity of the $A_0, A_2$ tail functions still remains, but it has improved compared to Figure\ \ref{fig:abs_no_exposed}. 

\paragraph{$(c)$} Similar to the case of the absolute loss, we can also verify here that the dot-loss optima satisfy  the KMS condition well in the orange validation grid of Figure\ \ref{fig:grid243} around $z=0.5$, but do poorly near the boundary of the OPE convergence region for configurations that deviate from the GFF solution. For example, the explicit validation of our configurations in a ring region $(\{0.975<|z|<0.99\}\cup\{0.975<|1-z|<0.99\})$ gave losses of the order of $10^{-3}$, when the corresponding loss of the analytic GFF solution was of the order of $10^{-7}$.

\vspace{0.5cm}
To obtain a better understanding of the situation, we explored how these results are affected when we incorporate additional information from the exact solution of the GFF theory---for example, a low-lying CFT datum, or the exact value of the tail functions inside the training region (instead of the approximate universal asymptotic value near $r=1$).

\begin{figure}[t!]
    \centering
    \includegraphics[width=1.0\linewidth]{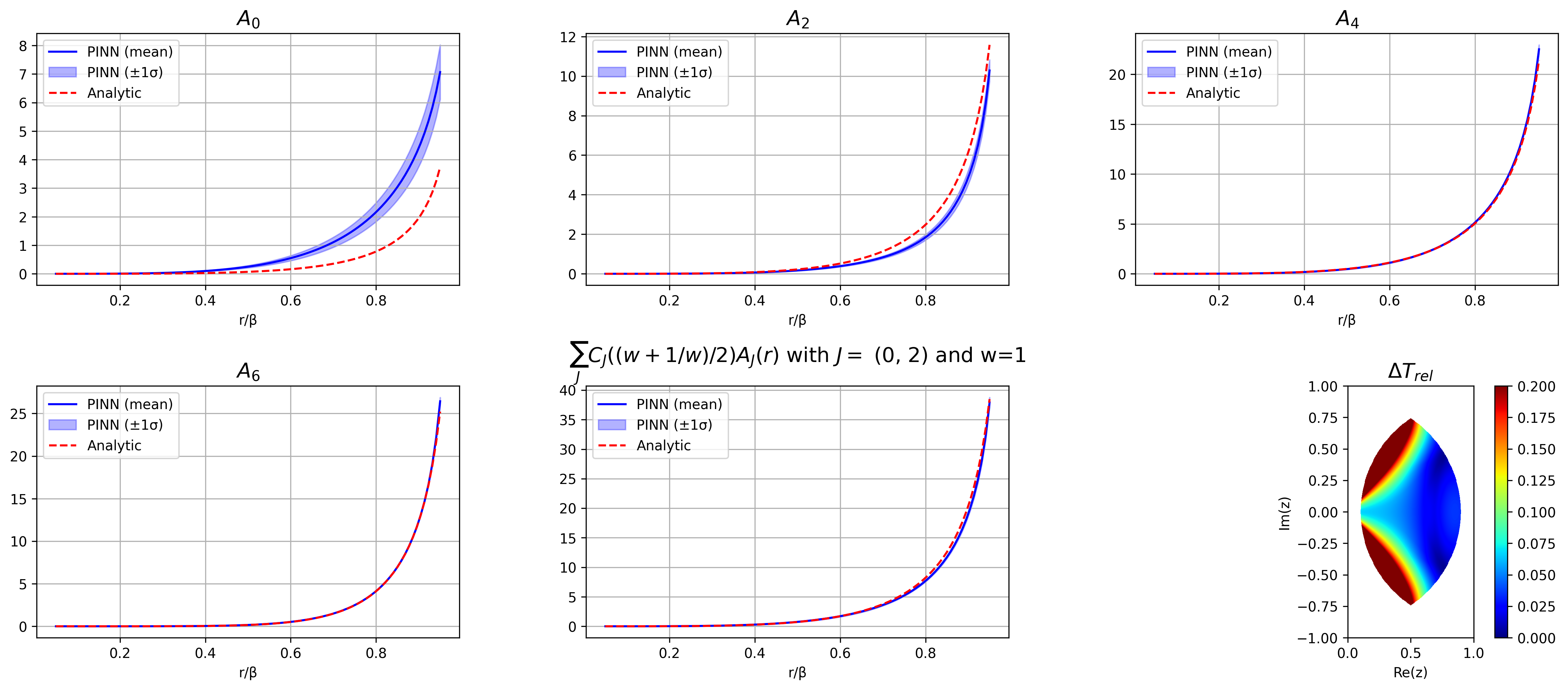}
    \caption{\small{Results obtained with the loss $\LL_{{\rm dot}(0)}+\LL_{\rm BC}$, performing the same runs as in Fig.\ \ref{fig:dot_no_exposed_no_input}, but with the additional input of the analytic value $a_{1,0}+3a_{0,2}\simeq 15.06$.}}
    \label{fig:dot_pinn_no_exposed_with_input}
\end{figure}

\subsubsection*{Supplying the exact value of a CFT datum}

The coefficient $a_{1,0}$ would be the leading scalar datum that contributes to the KMS condition, but as we noted in \eqref{gffbg}, it contributes together with $a_{0,2}$ through the combination $a_{1,0}+3a_{0,2}$ (in $d=4$). As a result, we fixed $a_{1,0}+3a_{0,2}$ to its analytic value $(\simeq 15.06)$ and repeated the exercise of Figure\ \ref{fig:dot_no_exposed_no_input} (with the same approximate asymptotic value for the tail functions at $r=0.9999$). The results are reported in Figure\ \ref{fig:dot_pinn_no_exposed_with_input}. When compared to Figure\ \ref{fig:dot_no_exposed_no_input} we observe further improvement towards the configurations of the analytic GFF solution, but part of the $A_0, A_2$ ambiguity and a visible discrepancy between the predicted $A_0$ and the analytic GFF $A_0$ still remain.

\subsubsection*{Supplying the value of the tail functions at an intermediate radius}

As an alternative, we replaced the input of the universal asymptotic behavior of the tail functions in the vicinity of $r=1$, with exact information about the value of the tail functions at some radius $r_i$ inside the training region. No further information from the exact solution was used (e.g.\ no analytic values of low-lying CFT data). This constraint was implemented with a simple quadratic loss and can be viewed as a set of sum-rules that are specific to the GFF theory. 

\begin{figure}[t!]
    \centering
    \includegraphics[width=1.0\linewidth]
    {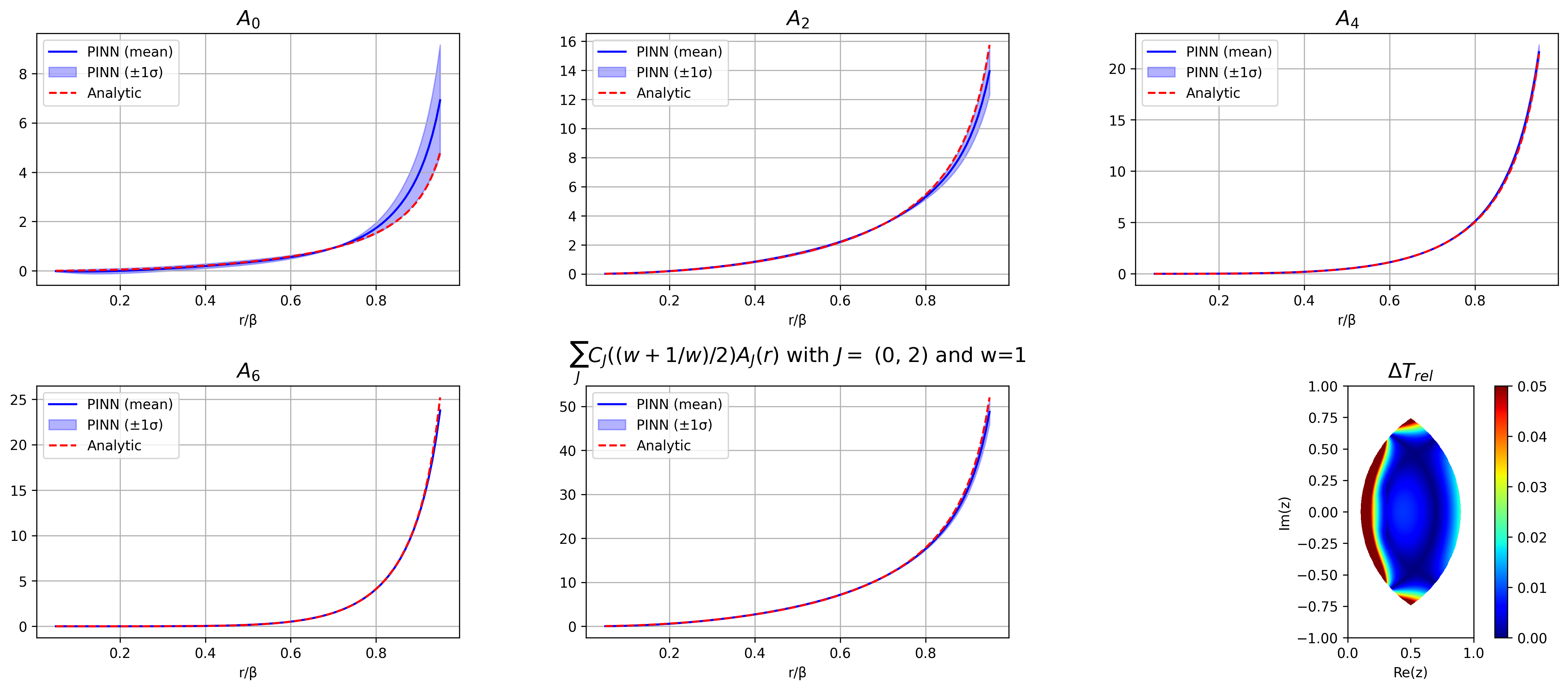}
    \caption{\small{Results obtained with the dot-loss $\LL_{{\rm dot}(0)}$, performing the same runs as in Fig.\ \ref{fig:dot_no_exposed_no_input} after replacing the condition on the asymptotic values of the tail functions at $r=0.9999$ with a condition that sets the values of the tail functions at $r_i=0.7$ to the analytic values of the GFF solution. Notice that the maximum of the color bar scale in the heatmap is now set to 0.05, compared to the higher value of 0.20 in the previous Figs.\ \ref{fig:dot_no_exposed_no_input}, \ref{fig:dot_pinn_no_exposed_with_input}.}}
    \label{fig:dot_pinn_no_exposed_finite_bc}
\end{figure}

The results obtained in this fashion (for $r_i=0.7$) are reported in Figure\ \ref{fig:dot_pinn_no_exposed_finite_bc}. In this case, the loss of the predicted configurations is $2.56 \times 10^{-8} \pm 1.21 \times 10^{-8}$ (vs $3.35 \times 10^{-9}$ for the analytic GFF solution). Although the dot-loss is slightly higher than the one observed in the runs of Figure\ \ref{fig:dot_no_exposed_no_input}, we notice that there is now much better agreement with the analytic tail functions, and the $A_0,A_2$ variation has been suppressed significantly without the need to perform any finetuning of the loss similar to the one reported in the context of Eq.\ \eqref{lbcab}.

For this implementation, the precise choice of the intermediate radius $r_i$ does not matter. As long as it is not too close to 0 or 1, e.g. when $r_i\in [0.3,0.7]$, we found comparable results in all the checks we performed. 

It is interesting to ask what happens if we impose generic, non-GFF values $A_J(r_i)$. We might expect that such generic input is problematic; for example, we might expect that it leads to incorrect asymptotic behavior near $r=1$, failing to produce 2-point functions that satisfy the KMS condition sufficiently well everywhere in the OPE-convergence region. We will explore this question in more detail in the next subsection. There, we will make the additional observation that non-GFF values $A_J(r_i)$ also lead to enhanced instability towards many low-loss configurations, in contrast to the relative stability of the configurations in Figure\ \ref{fig:dot_pinn_no_exposed_finite_bc}. This observation forms the basis of a potential strategy to recover the GFF $A_J(r_i)$ values from our analysis without any input from the exact GFF solution.

\begin{figure}[t!]
    \centering
    \includegraphics[width=1.0\linewidth]
    {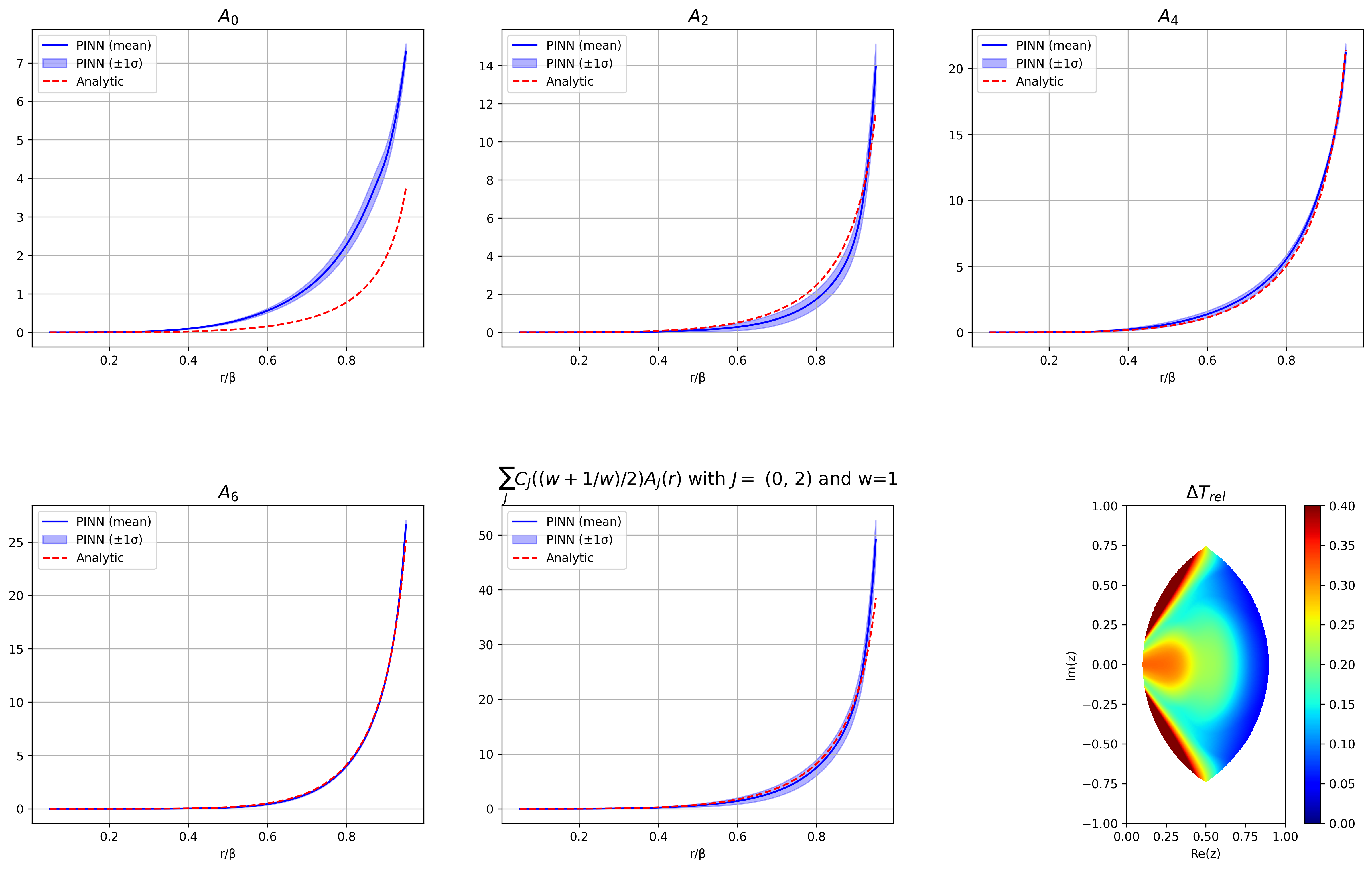}
    \caption{\small{Results obtained with the loss $\LL_{{\rm dot}(1)}+\LL_{\rm BC}$ and the CFT datum $a_{1,0}+3a_{0,2}$ exposed. $\LL_{\rm BC}$ in Eq.\ \eqref{lbcab} is set once again at $r=0.9999$ using information about the asymptotic behavior of the tails in the vicinity of $r=1$. The mean loss of the predicted functions is $6.12\times 10^{-7} \pm 1.26 \times 10^{-7}$ with the loss of the analytic solution at $7.78 \times 10^{-8}$. The predicted coefficient is $a_{1,0}+3a_{0,2}=13.29 \pm 2.82$ with the analytic value at $15.06013$. The first 4 plots (from top left to right) depict the predicted tail functions. The 5th plot and the heatmap provide information about the combined contribution of the tail functions $A_0, A_2$ to the conformal block expansion at $w=1$ and the full training region respectively.}}
    \label{fig:dot_pinn_exposed_asymptoticBC}
\end{figure}

\subsection{$a_{\Delta,J}$ bootstrap}
\label{kmsa}

Finally, we would like to expose one low-lying operator and use the dot-loss $\LL_{{\rm dot}(1)}$ to optimize the tail functions and recover the unknown exposed CFT datum from Eq.\ \eqref{kms2dj}. For the reasons explained above, we expose the operators $[\phi\phi]_{1,0}$, $[\phi\phi]_{0,2}$ and compute the combination of coefficients $a_{1,0}+3a_{0,2}$ that appears naturally in the KMS condition. Once again, we set $d=4$, $\Delta_\phi=1.68$ and collect the 10 lowest-loss configurations for the loss $\LL_{{\rm dot}(1)}+\LL_{\rm BC}$ in 1K independent runs with 50K epochs. For $\LL_{\rm BC}$ we continue to use Eq.\ \eqref{lbcab}.

Fixing the universal asymptotic behavior of the tail functions at $r=0.9999$ we obtained the results summarized in Figure\ \ref{fig:dot_pinn_exposed_asymptoticBC}. In analogy to the results of the previous subsection, the optimization in this setup struggles to accurately capture the leading tails, $A_0, A_2$, but better recovers  their combined contribution to the conformal block expansion. The analytic value of $a_{1,0}+3a_{0,2}= 15.06013$ is within the predicted 1$\sigma$ range $13.29 \pm 2.82$, but this prediction is clearly affected by the difficulty to recover the GFF profile of the tail functions $A_0, A_2$.

\begin{figure}[t!]
    \centering
    \includegraphics[width=1.0\linewidth]
    {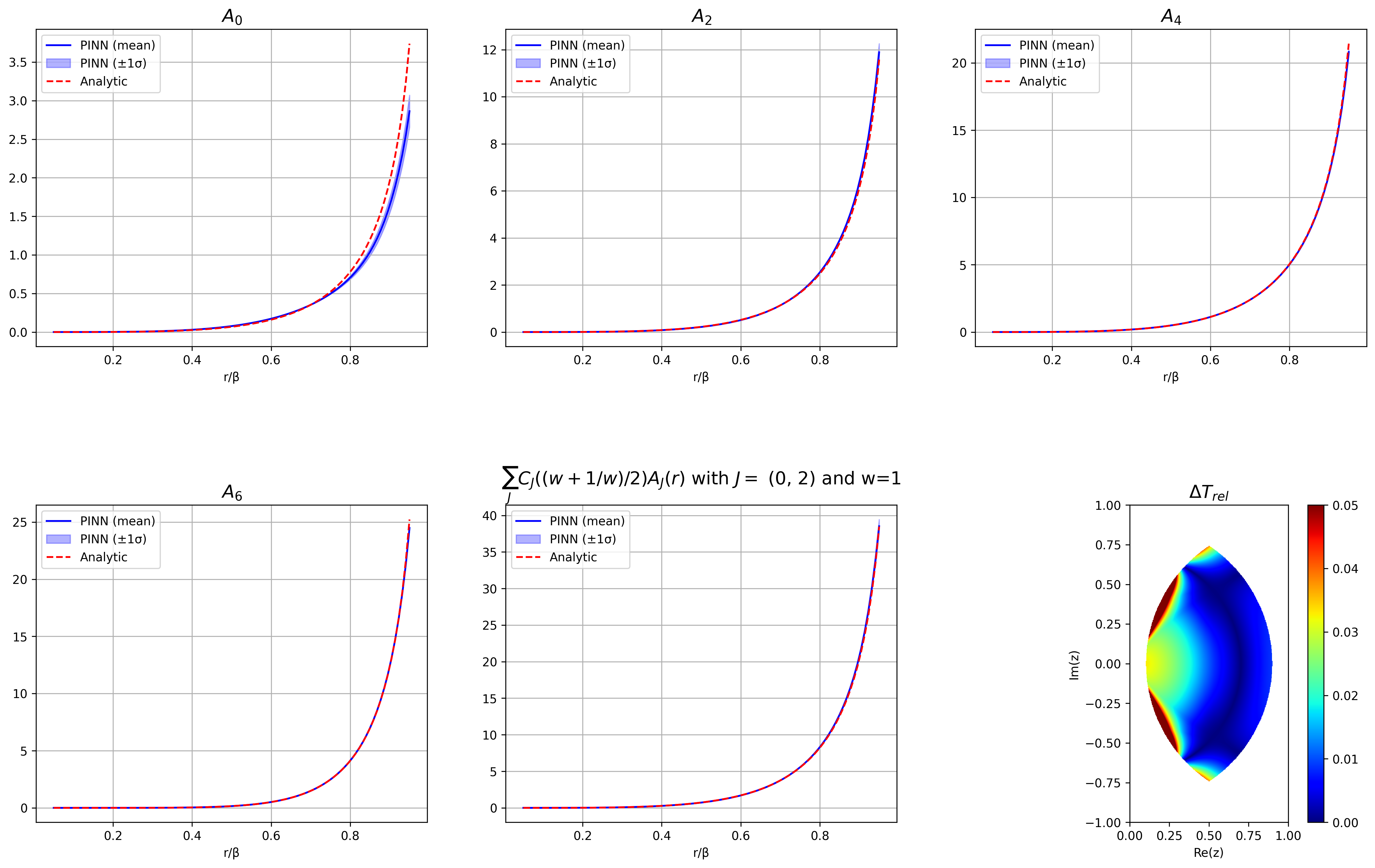}
    \caption{\small{Results obtained with the dot-loss $\LL_{{\rm dot}(1)}$ and the CFT datum $a_{1,0}+3a_{0,2}$ exposed. In this case, the conditions on the asymptotic values of the tail functions at $r=0.9999$ have been replaced by conditions that set the values of the tail functions at $r=0.7$ to the analytic values of the GFF solution. The mean loss of the predicted functions is $9.49 \times 10^{-9} \pm 6.27 \times 10^{-10}$ with the loss of the analytic solution at $7.78 \times 10^{-8}$. The predicted coefficient is $a_{1,0}+3a_{0,2}=15.0647 \pm 0.0291$ with the analytic value at $15.06013$. The first 4 plots (from top left to right) depict the predicted tail functions. The 5th plot and the heatmap provide information about the combined contribution of the tail functions $A_0, A_2$ to the conformal block expansion at $w=1$ and the training region respectively.}}
    \label{fig:dot_pinn_exposed_finite_bc}
\end{figure}

For reference, in Figure\ \ref{fig:dot_pinn_exposed_finite_bc} we also present the result of the same computation when the asymptotic value of the tail functions at $r=0.9999$ are replaced by their analytic values at $r=0.7$. Similar to the previous subsection, we notice that this information allows us to recover very accurately the analytic solution and the predicted value of $a_{1,0}+3a_{0,2}$ is now $15.0647 \pm 0.0291$, which is much closer to the analytic one at $15.06013$.

\begin{figure}[t!]
    \centering
    \includegraphics[width=1.0\linewidth]
    {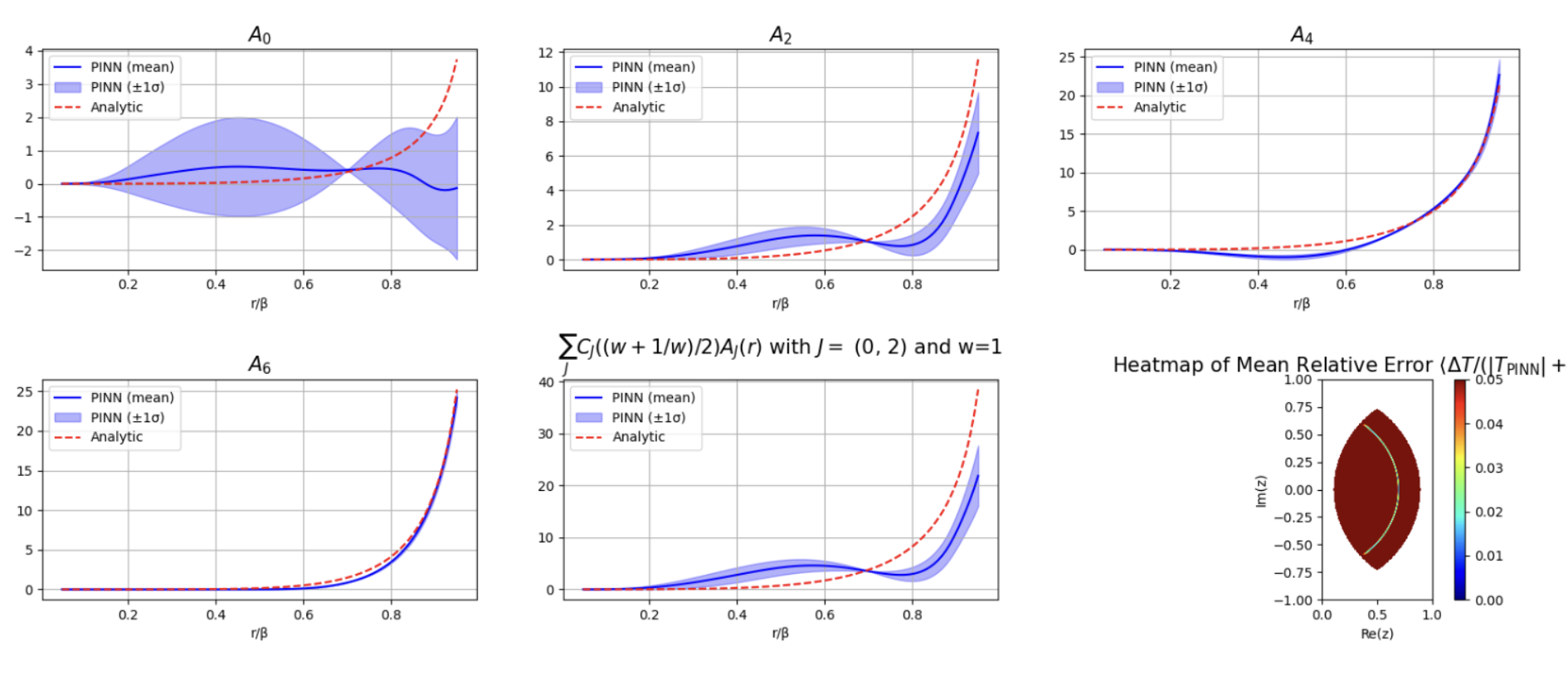}
        \caption{\small{Results obtained with the dot-loss $\LL_{{\rm dot}(1)}$ and the CFT datum $a_{1,0}+3a_{0,2}$ exposed. In this case, we impose conditions on the tail functions at $r=0.7$ (as in Figure\ \ref{fig:dot_pinn_exposed_asymptoticBC}), but to values that differ from the analytic values of the GFF solution. The mean loss of the predicted functions is $6.64 \times 10^{-9} \pm 8.67 \times 10^{-9}$ (to be compared against  the loss of the analytic solution at $7.78 \times 10^{-8}$). The predicted coefficient, $a_{1,0}+3a_{0,2}=28.784 \pm 4.083$, is now significantly farther away from the analytic value at $15.06013$. The first 4 plots (from top left to right) depict the predicted tail functions. The 5th plot and the heatmap provide information about the combined contribution of the tail functions $A_0, A_2$ to the conformal block expansion at $w=1$ and the training region respectively. The thin yellowish line within the heatmap represents the $r=0.7$ curve, where we impose the condition on $A_J(0.7)$. Since these values are relatively close to the GFF values, $\Delta T_{rel}$ is suppressed there.}}
    \label{fig:dot_pinn_exposed_wrong_finite_bc}
\end{figure}

Let us now examine what happens when, instead of the GFF values $\vec A_{\rm GFF} := \{A_J(0.7)\big|_{\rm GFF}\}$, we impose a random vector $\vec A$ of tail values at $r=0.7$. We performed several stochastic runs on the QMUL Apocrita cluster for $\vec A$ away from $\vec A_{\rm GFF}$, observing that we could still recover low-loss configurations inside the training region. The resulting configurations had two features: $(1)$ mean curves that deviate significantly from the GFF curves and $(2)$ considerable standard deviations. A typical example for $\vec A = \vec A_{\rm GFF} + \kappa\, |\vec A_{\rm GFF}|\, \vec v_{\rm random}$, with factor $\kappa = 0.5$ and random unit 4-vector $\vec v_{\rm random}=(0.327696, -0.176708, -0.382838, -0.845473)$ is presented in Figure\ \ref{fig:dot_pinn_exposed_wrong_finite_bc}. 

The accumulated evidence supports the picture that there is an island of 4-vectors $\vec A$ around $\vec A_{\rm GFF}$ with stochastic optimization results that exhibit small standard deviations. The size of the island correlates with the size of the chosen standard deviation cutoff. This picture opens up an exciting prospect: the possibility to extract an approximation of $\vec A_{\rm GFF}$ (and corresponding predictions like those of Figure\ \ref{fig:dot_pinn_exposed_finite_bc} that reproduce the GFF solution at high accuracy) from an analysis solely within the limited training region of Figure\ \ref{fig:grid243}, without having to address the numerical difficulties near the boundary of the OPE convergence region. We will not pursue a full implementation of this strategy in this section, but we will consider a closely related approach in the context of holographic CFTs in Section \ref{holo2}.

\subsection{Brief summary}
\label{5sum}

The above exercises in the context of the GFF theory provide a number of useful lessons about the approach we have introduced, including some of the potential challenges that arise during the implementation. Most notably, we can see explicitly in the above examples that tail functions can be bootstrapped, but the relevant optimization is subtle if the training grid is restricted and does not allow us to extract the full information of the KMS condition. 

To address this problem one has two options: $(1)$ extend the training grid to cover the OPE-convergence region as much as possible and explore how this affects the optimization outcome, or $(2)$ supplement the optimization of the KMS condition on a fixed restricted grid, like the one in Figure\ \ref{fig:grid243}, with additional assumptions. For any option that can extract the full information of the KMS condition (at fixed discontinuity), we would expect to recover a unique solution. In the context of the GFF theory in this section, apart from a constant corresponding to the OPE coefficient $a_{0,0}$ and the fact that the KMS condition cannot distinguish the coefficients $a_{1,0}$ and $a_{0,2}$, that solution should be the GFF solution. A similar statement, was shown at zero-spatial separation in the recent analyses of Refs\ \cite{Buric:2025anb,Barrat:2025nvu}.

In the present work, numerical limitations did not allow us to make significant progress with option (1), but we hope to revisit this point in the future. In the context of option (2), we explored several concrete avenues. We observed that universal information about the behavior of tail functions near $r=1$ can improve the stability of the optimization. When combined with proper loss functions, it nudges the result toward the GFF solution and recovers zero-spatial separation data, but leaves room for configurations that are not KMS-invariant outside the training region. At the same time, we observed that extra information inside the training region can be very restrictive when it  comes from the exact GFF solution. Reading this statement backwards, suggests the exciting possibility of recovering this extra information inside the training region by demanding that the optimization problem produces a stable, (numerically) unique minimum. We have numerical evidence for this proposal, but not a proper, complete argument that proves it.

In the following sections we develop these observations further in the context of holographic CFTs. This context is also interesting for another reason. One of the key simplifying features of the GFF theory is that it has a fixed universal contribution to the discontinuity (that comes only from the identity operator). In generic cases, $\IT_{\rm Disc}$ will not be a fixed function; it will be a function of in principle unknown CFT data. Holographic CFTs provide a setup where we can start discussing the challenges of theories with more complicated discontinuities.

\section{Holographic CFTs: setup}
\label{holo1}

In this section we set up the KMS condition for thermal 2-point functions of scalar operators, $\langle \phi\phi\rangle_\beta$, in holographic CFTs with a classical (super)gravity dual. We work at leading order in the large-$c$ limit and assume that the scaling dimension $\Delta_\phi$ of the operator $\phi$ is a positive, non-integer, real number above the unitarity bound $(\Delta_\phi \geq \frac{d-2}{2})$. We focus on situations where the thermal 1-point function of the operator $\phi$ vanishes, namely $\langle \phi\rangle_\beta = 0$.

In Appendix \ref{largec} we review for the convenience of the reader basic properties of the large-$c$ counting for CFTs on $S^1 \times \IR^{d-1}$, while in Appendix \ref{logs} we outline the subtleties of the case with integer $\Delta_\phi$. The latter will be useful in Section \ref{sym}, where we present a preliminary discussion of the $d=4$ $\NN=4$ SYM case in the context of our approach.

The interest in holographic CFTs is obvious, because of their connection to gravity and black holes. Beyond this connection, such theories also pose a useful testing ground for novel bootstrap methodologies at finite temperature, since 2-point functions computed holographically in {\it any} two- or higher-derivative classical theory of gravity have the correct analytic structure and automatically solve the KMS condition. As a result, it is interesting to ask how one could recover such solutions independently within a bootstrap method. As we discussed in the introduction, current numerical techniques based on hard or soft truncations of the spectrum in the KMS condition are not suitable for such studies due to well-known limitations. The approach proposed in this paper does not have those limitations, and it is therefore interesting to explore how far it can take us in this holographic CFT context.

\subsection{Spectra and discontinuities}
\label{holospectrum}

In holographic CFTs, and at leading order in the large-$c$ limit, scalar operators behave as generalized free fields coupled to the energy-momentum sector. The latter is intricately connected to the classical properties of the dual planar black hole solution. More specifically, the spectrum of operators that is expected to contribute to the conformal block expansion of a scalar 2-point function $\langle \phi\phi\rangle_\beta$ comprises:
\begin{itemize}
    \item The identity operator.
    \item The energy-momentum tensor $T_{\mu\nu}$. 
    \item The double-twist operators $[\phi\phi]_{n,J}$, with $J=2\ell$ for $\ell=0,1,2,\ldots$.
    \item The multi-trace energy-momentum tensor operators $[T^k]_J$. These are composite operators with $k$ insertions of the energy-momentum tensor and an arbitrary number of index contractions. Conformal primaries with derivative insertions are also possible but do not contribute at leading order in the large-$c$ limit on $S^1\times \IR^{d-1}$. We review the argument in Appendix\ \ref{largec}. 
\end{itemize}
The scaling dimensions, the spin and the notation for the corresponding $a_{\Delta,J}$ coefficients for each of the above contributions is summarized in Table \ref{tab:holoOps}.

\begin{table}[t!]
    \centering
    \begin{tabular}{|c|c|c|c|}
        \hline
        Operator & Scaling dimension $\Delta$ & Spin $J$ & $a_{\Delta, J}$ \\
        \hline\hline
        Identity & 0 & 0 & $a_{\boldsymbol{1}}=1$ \\
        \hline
        $T_{\mu\nu}$ & $d$ & 2 & $a_T$ \\
        \hline
        $[\phi\phi]_{n,J}$ & $2(\Delta_\phi + n + \ell)$ & $0 \leq 2\ell$ & $a_{n,J}$ \\
        \hline
        $[T^k]_J$ & $d\, k$ & $0\leq 2\ell\leq 2k$ & $a^{(k)}_J$ \\
        \hline
    \end{tabular}
    \caption{\small{List of the operators contributing to the conformal block expansion of the thermal 2-point function $\langle \phi \phi\rangle_\beta$ in holographic CFTs. The scaling dimension of the operator $\phi$ is $\Delta_\phi$.}}
    \label{tab:holoOps}
\end{table}

Similar to the GFF case, the double-twist operators have vanishing discontinuity. The approximate contributions of the remaining  operators to $\IT_\disc$  can be summarized as follows. For concreteness, and for more direct reference to the case analyzed in the next section, we fix the spacetime dimension to $d=4$. The {\bf identity} contribution is:
\bea
\label{holospecaa}
\IT_{\disc}^{({\rm approx})(\bf 1)}[J_*;rw, rw^{-1}] &=& 
4 \sin(\pi \Delta_\phi) \bigg[ 
\int^{2r^{-1}}_{r^{-1}} dw' \, 
\KK_{J_*}(w,w')\, (rw'-1)^{-\Delta_\phi} (1-rw'^{-1})^{-\Delta_\phi}
\nonumber\\
&&- \int_{-2r^{-1}}^{-r^{-1}}  
\KK_{J_*}(w,w')\, (-rw'-1)^{-\Delta_\phi} (1+rw'^{-1})^{-\Delta_\phi}
\bigg]
~.
\eea
The {\bf energy-momentum} contribution is:
\bea
\label{holospecab}
&&\IT_{\disc}^{({\rm approx})(T)}[J_*;rw, rw^{-1}] = 
4 a_T \sum_{s=0}^2 p_s(J) \sin(\pi(-\Delta_\phi +s)) 
\nonumber\\
&&\hspace{1.5cm} \bigg[ \int_{r^{-1}}^{2r^{-1}} dw'\, \KK_{J_*}(w,w') (rw'-1)^{1 - \Delta_\phi +s} (1-rw'^{-1})^{3 -\Delta_\phi -s} 
\nonumber\\
&&\hspace{1.5cm} - \int_{-2r^{-1}}^{-r^{-1}} dw'\, \KK_{J_*}(w,w') (-rw'-1)^{1 - \Delta_\phi +s} (1+rw'^{-1})^{3 -\Delta_\phi -s}
\bigg]
~.
\eea
The contribution of the {\bf multi-trace energy-momentum tensors} $[T^k]_J$ is:
\bea
\label{holospecac}
&&\IT_\disc^{({\rm approx})([T^k]_J)}[J_*;rw,rw^{-1}] = - 4 (-1)^{\frac{J}{2}} a^{(k)}_J \sum_{s=0}^J p_s(J) \sin\left(\pi\left( -\Delta_\phi +s \right) \right) 
\nonumber\\
&&\hspace{1.5cm} \bigg[ \int_{r^{-1}}^{2r^{-1}} dw'\, \KK_{J_*}(w,w') (rw'-1)^{2k-\frac{J}{2} - \Delta_\phi +s} (1-rw'^{-1})^{2k +\frac{J}{2} -\Delta_\phi -s} 
\nonumber\\
&&\hspace{1.5cm} - \int_{-2r^{-1}}^{-r^{-1}} dw'\, \KK_{J_*}(w,w') (-rw'-1)^{2k-\frac{J}{2} - \Delta_\phi +s} (1+rw'^{-1})^{2k + \frac{J}{2} -\Delta_\phi -s}
\bigg]
\, .
\eea

\subsection{Expectations from holography}
\label{holoexpect}

In the AdS/CFT correspondence, the finite-temperature CFT on $S^1\times \IR^{d-1}$ is captured by a planar, asymptotically AdS$_{d+1}$ black hole solution. A scalar conformal primary operator $\phi$ in the CFT corresponds in the dual gravitational description to a scalar field $\Phi$. By studying the bulk-boundary propagator of the field $\Phi$ one can extract the holographic 2-point function $\langle \phi \phi \rangle_\beta$ on the boundary. Different classical gravitational theories in the bulk (e.g.\ theories with arbitrary higher-derivative terms) are characterized by different black hole solutions that produce different holographic 2-point functions of scalar operators. In this context, holography becomes an apparatus that produces an infinite class of consistent solutions to the KMS condition.  

Holographic computations of this type have been performed by many authors in the past. Recent papers, related to the present discussion, include \cite{Fitzpatrick:2019zqz,Parisini:2023nbd,Karlsson:2019dbd,Li:2019tpf,Li:2019zba,Li:2020dqm,Esper:2023jeq,Ceplak:2024bja,Buric:2025anb}. The conformal-block expansion of the holographic correlators reveals interesting properties, some of which are worth summarizing here.

First, the contribution of the energy-momentum tensor comes with the coefficient $a_T$ that  is related to the thermal 1-point function of the energy-momentum tensor $b_T$ through the equation
\beq
\label{holoexpectaa}
a_T = - \frac{2\Delta_\phi}{(d-2)(d-1)} \frac{\Gamma\left(\frac{d}{2}\right)}{2\pi^{\frac{d}{2}}} \frac{b_T}{C_T}
~.
\eeq
Here $C_T$ is the 2-point function coefficient of the energy-momentum tensor, which is proportional to the $c$ Weyl anomaly coefficient of the CFT. At the same time, $b_T$ is connected to the entropy density $S$ of the $d$-dimensional CFT (and of the corresponding $(d+1)$-dimensional planar black hole) via the relation
\beq
\label{holoexpectab}
S = - b_T \, T^{d-1}
~.
\eeq

Additionally, the conformal block expansion of the holographic thermal 2-point function reveals contributions from the multi-traces $[T^k]_J$ and the double-twist operators $[\phi\phi]_{n,J}$. Interestingly, the multi-trace $[T^k]_J$ data $a^{(k)}_J$ can be deduced from a relatively straightforward analysis of the asymptotic behavior of the bulk-boundary propagator near the AdS boundary, \cite{Fitzpatrick:2019zqz}. In contrast, the double-twist data $a_{n,J}$ require the analysis of the bulk-boundary propagator in the full geometry, including information from the vicinity of the black hole horizon. This information is relatively harder to extract, see \cite{Parisini:2023nbd} for a recent relevant computation, and is not always readily available. As a result, it is very interesting to explore if it is possible to extract the double-twist data from the multi-trace energy-momentum data with an independent bootstrap computation.

Another noteworthy feature of the holographic 2-point functions is the universality of the lowest-twist multi-trace energy-momentum data $a^{(k)}_{2k}$, originally observed in \cite{Fitzpatrick:2019zqz}. As one varies the couplings in the bulk gravitational action, the data $a^{(k)}_J$ vary as well, but not all of them are independent. The data of the lowest-twist operators $[T^k]_{2k}$ (with twist $\tau_k = \Delta_k - J = 4k - 2k = 2k$) exhibit a fixed relation with $a_T$: once $a_T$ is fixed, so are all the lowest-twist multi-trace coefficients $a^{(k)}_{2k}$ for $k\geq 2$. For example, in $d=4$
\bea
\label{holoepectac}
&&a^{(2)}_4 = \frac{\Delta_\phi(7\Delta_\phi^2 + 6 \Delta_\phi + 4)}{201600(\Delta_\phi -2)} \left(\frac{120}{\Delta_\phi} a_T \right)^2
~,\nonumber\\
&&a^{(3)}_6 = \frac{\Delta_\phi(1001\, \Delta_\phi^4 + 3575\, \Delta_\phi^3 + 7310 \, \Delta_\phi^2 + 7500\, \Delta_\phi + 3024)}{10378368000(\Delta_\phi-3)(\Delta_\phi-2)} \left(\frac{120}{\Delta_\phi} a_T \right)^3
\eea
etc., irrespective of what the bulk gravitational action is. In Appendix \ref{holoresults}, we collect further explicit formulas for the lowest-twist data and some exact results for the non-lowest-twist data in the case of Einstein gravity.

\section{Holographic CFTs: an application}
\label{holo2}

In holographic CFTs we can perform two general types of exercises. We can either fix the discontinuity using input from independent gravitational computations and bootstrap the rest of the data, or we can limit the input from gravity and ask within the thermal bootstrap itself how to constrain the discontinuity as well. 

In the first case, which is more straightforward, we expect that fixing the discontinuity of a holographic CFT within our setup, will allow an approximate reconstruction of the thermal 2-point functions of the holographic theory. This amounts to a (numerical) derivation of double-twist data from (multi-trace) energy-momentum data. We will test this hypothesis in Section \ref{fixeddisc}.

The second question concerns the constraints we can derive about thermal data of generic holographic CFTs without fixing the discontinuity. This should also be possible within our approach, but is more demanding. We do not address this aspect fully in this work but provide some preliminary comments in Section \ref{dyndisc}.

\subsection{Fixed discontinuity}
\label{fixeddisc}

In this Section, we consider the 2-point function of a scalar operator with scaling dimension $\Delta_\phi=1.5$ in $d=4$ spacetime dimensions.\footnote{This particular, half-integer, value of $\Delta_\phi$ was chosen in order to have a direct comparison with double-twist data approximated with different methods in Ref.\ \cite{Buric:2025anb}. We note that we can obtain similar results in generic, non-half-integer, scaling dimensions.} We analyze the approximate KMS condition \eqref{kms2ca} setting $J_*=6$ (similar to the GFF analysis in previous sections).

\subsubsection{Discontinuity from Einstein gravity data}
\label{einsteindisc}

As a specific example, we will consider thermal data in a holographic CFT having an Einstein gravitational dual description. For such a theory, our first task is to find an approximation for the corresponding 2-point function discontinuity. 

From the previous Section \ref{holo1}, we recall that the discontinuity receives contributions only from the energy-momentum sector. There is an infinite number of such contributions, but for the purposes of this section we will model $\IT_{\disc}^{({\rm approx})}$ using only the contributions of the identity operator, the energy-momentum tensor and the multi-trace energy-momentum tensors
\beq
\label{q1aa}
[T^2]_0~, ~~ [T^2]_2~, ~~[T^2]_4~, ~~
[T^3]_4~, ~~ [T^3]_6~, ~~
[T^4]_8
\eeq
according to Eqs\ \eqref{holospecaa}-\eqref{holospecac}. In this list, we include all the operators that contribute to the discontinuity up to maximum twist $\tau_{\max}=8$; the twist is defined as $\tau:=\Delta-J$. This is an ad hoc truncation.\footnote{Unlike the truncations to the OPE expansion in common truncation schemes for crossing equations, here the situation is under better control, because (as we noted in Section \ref{approxcross}) the truncation occurs within the discontinuity and the kernel $\mathcal K_{J_*}$ naturally suppresses higher-twist contributions.} By including higher-twist operators one can probe the systematic error of this truncation, but we will not attempt to do so in this paper. Our primary goal is to fix some approximation of the discontinuity and examine the solutions of the corresponding approximate KMS condition.

To fix the discontinuity with these specifications we use the thermal OPE coefficients for the operators \eqref{q1aa}, as dictated by Einstein gravity in AdS/CFT through a holographic computation devised in \cite{Fitzpatrick:2019zqz} and described in Appendix \ref{holoresults} to find
\bea
\label{q1ab}
&&a_{T,{\rm GR}} = 1.21761  ,~
a_{0,{\rm GR}}^{(2)} = -1.37668 , ~
a_{2,{\rm GR}}^{(2)} = 1.58848 ,~
a_{4,{\rm GR}}^{(2)} = -4.05945 ,~
\nonumber\\
&&a_{4,{\rm GR}}^{(3)} = 1.77035 ,~
a_{6,{\rm GR}}^{(3)} = 8.52362 ,~
a_{8,{\rm GR}}^{(4)} = -15.9641 .
\eea

Using this information, we want to determine the double-twist datum $a_{1,0}+3a_{0,2}$, similar to the exercises we performed in the GFF theory in Section \ref{GFF}. Therefore, in the low-spin part of the approximate KMS condition \eqref{kms2ca} we expose only the energy-momentum contribution and the contributions of the double-twist operators $[\phi\phi]_{1,0}, [\phi\phi]_{0,2}$; the contribution of all other operators is included in the four tail functions $A_0,A_2,A_4,A_6$. 

Repeating the optimization of Section \ref{GFF}, with a loss of the form $\LL_{\rm KMS}+\LL_{\rm asymptotic\, BC}$, we encounter immediately the same technical issues we observed in the GFF case: a landscape of many configurations with comparable low loss. The problem arises for both types of KMS losses, $\LL_{\rm KMS}=\LL_{\overline{\rm abs}}$ or $\LL_{\rm dot(1)}$, and any type of loss $\LL_{\rm BC}$ for the implementation of the asymptotic boundary condition for the tails near $r=1$. In fact, in the present case we could not even identify an analog of \eqref{lbcab} to partially stabilize the outcome of the optimization. For those reasons, and motivated by the observations at the end of Section \ref{GFF}, we would like to implement a strategy that identifies a suitable vector of constraints on $A_J(r_i)$ (for $J=0,2,4,6$ at some intermediate radius $r_i$) leading to an almost unique minimum in our optimization. We will propose that this minimum is the true, unique solution of our approximate KMS condition everywhere in the OPE-convergence region.

We proceed to explain how to implement this strategy.

\subsubsection{Shooting method}
\label{shoot}

In Section \ref{kmsa} we observed that by specifying the 4-vector $\vec A:=\{A_J(r_i)\}$ we obtained many KMS minima with comparable low-loss, unless $\vec A$ was fixed at $\vec A_{\rm GFF}$ with the corresponding values of the tail functions of the exact GFF solution. In that case, the optimization yielded an almost unique minimum that approximated well the analytic GFF solution.

Using the same criterion of optimization stability, we would like to search for a similar vector $\vec A_*$ that replaces the GFF vector of Section \ref{kmsa} with one suitable for the holographic CFT corresponding to the prescribed approximate discontinuity. In principle, we can achieve this goal with a shooting method where we scan over optimizations with generic $\vec A$ vectors until we find values that yield tail curves with low standard deviation. Anticipating that $\vec A_*$ is in the vicinity of $\vec A_{\rm GFF}$, we can restrict the scan in a neighborhood of $\vec A_{\rm GFF}$. To assist this search, we can also use as another criterion the proximity of the mean asymptotic values of the tails $A_J(r)$ in the vicinity of $r=1$ to the expected universal values of Eq.\ \eqref{asae}. 

Unfortunately, such searches can be very expensive; they require a cluster run for each candidate $\vec A$, and $\vec A$ has to be sampled in a 4-dimensional domain (a $(\frac{J_*}{2}+1)$-dimensional domain, in general). Consequently, the results we obtained in this way were not very accurate, leaving a lot of room for technical improvement. That motivated us to look for a more efficient implementation of this approach.

\subsubsection{Soft intermediate-$r$ constraints with a ReLU loss}
\label{relu}

As an alternative, consider an optimization problem, where one attempts to satisfy at the same time the following three objectives: $(a)$ the approximate KMS condition with a fixed $\IT_{\disc}$ on the grid of Figure\ \ref{fig:grid243}, $(b)$ the universal asymptotics at $r=0.9999$, and $(c)$ a constraint that the values of $A_J(r_i)$ remain within a fixed neighborhood $U$ of a prescribed vector $\vec A_0$. We can select $\vec A_0$ at will, but for the rest of the discussion we set $\vec A_0 = \vec A_{\rm GFF}$. This is motivated by the fact that in the holographic case of interest, it is natural to start the search by exploring deviations of the tail functions at $r_i$ away from the GFF values $A_J(r_i)|_{\rm GFF}$. In this setup, we can perform a less expensive, more focused analysis of the effects of intermediate-$r$ constraints, and examine how the landscape of minima changes for neighborhoods $U$ of different sizes. Unlike the shooting method of the previous subsection, here $A_J(r_i)$ are not rigidly fixed throughout the optimization process. They are allowed to vary, and the algorithm can choose preferable values if these lead to configurations with lower KMS loss.

We can implement $(c)$ straightforwardly using a Rectified Linear Unit (ReLU) loss function in the following manner. Similar to \eqref{lbcaa}, it is convenient to impose conditions on the non-linear transformation of the tail functions 
\beq
\label{q2aa}
\AA_{J}(r) := {\rm arcsinh} \left( \frac{A_{J}(r)}{2} \right)
\eeq
instead of $A_J(r)$ directly to avoid having to train on large absolute values.
Then, we can constrain the allowed values of each $\AA_J(r_f)$ within a range of the GFF values using the mean ReLU loss\footnote{This choice is not unique. Higher powers of the ReLU function can also be employed. Although we do not expect the results to depend significantly on this choice, it would be useful to explore further the efficiency of alternative powers.}
\beq
\label{q2ab}
\LL_{\rm ReLU} = \frac{1}{\frac{J_*}{2}+1}\sum_{J} {\rm ReLU}\left( \bigg|\AA_{J}(r_i) - \AA_{J}(r_i)|_{\rm GFF} \bigg| - {\tt p} \times \bigg| \AA_{J}(r_i)|_{\rm GFF} \bigg|  \right)
~.
\eeq
The ReLU function is defined as
\beq
\label{q2ac}
{\rm ReLU}(x) = \max (0, x)
~.
\eeq
In this prescription, the factor $\tt p$ is an arbitrary real positive number that controls the size of the range of the region of allowed variation of $A_J(r_i)$ around the values of the central vector $A_J(r_i)|_{\rm GFF}$. ${\tt p}=0$ penalizes any variation away from the GFF values $A_J(r_i)|_{\rm GFF}$, but ${\tt p}>0$ penalizes only those configurations that deviate more than ${\tt p} \times \big| \AA_{J}(r_i)|_{\rm GFF}\big|$ from the GFF values. 

Putting everything together, the total loss function in the specific approach of this subsection has three components
\beq
\label{q2ad}
\LL = \LL_{\rm KMS} + \LL_{\rm ReLU} + \LL_{\rm BC}
~,
\eeq
where $\LL_{\rm KMS}$ is either the absolute or the dot-pinn loss, $\LL_{\rm ReLU}$ is the loss in Eq.\ \eqref{q2ab}, and $\LL_{\rm BC}$ is a loss that enforces the asymptotic behavior of the tails near $r=1$, e.g.\ $r=0.9999$. In this case, there is no need for convoluted functions like the one in Eq.\ \eqref{lbcab}. The simple quadratic loss
\beq
\label{q2ae}
\LL_{\rm BC} = \frac{1}{\frac{J_*}{2}+1}\sum_{J} \bigg(\AA_{J}(r_f) - \AA_{J}(r_f)|_{\rm GFF} \bigg)^2 
~
\eeq
works efficiently.

By analogy to the shooting method, where the goal was to find parameters yielding a stable and (numerically) unique optimization outcome, our main goal here is to likewise identify a special value ${\tt p}_*$ where the outcome of the optimization is similarly stable with small standard deviations. It is not immediately obvious that such a value exists, but here is what we can expect based on the analysis so far.

In the holographic case, the solution  we are searching for is different from GFF and, therefore, fixing $A_J(r_i)$ to the GFF values is not expected to yield low-loss configurations with small standard deviations. In fact, the loss of these configurations is expected to be relatively high because we are forcing unnatural conditions at $r_i$ together with the universal boundary conditions near $r=1$. As we increase ${\tt p}$, we allow the values $A_J(r_i)$ to vary and the algorithm to explore a larger set of lower-loss configurations. With increasing ${\tt p}$, the loss is expected to continue to decrease, but eventually it will saturate and settle towards the value we would recover by doing an unconstrained optimization in the absence of the ReLU loss. Again, in this regime of large enough ${\tt p}$, the outcome of the optimization includes many low-loss configurations and the standard deviations of statistical runs will be accordingly large. At intermediate values of ${\tt p}$, however, the standard deviations could be smaller and there could be a picture where they achieve a minimum value for a certain value of ${\tt p}$. Mirroring the discussion of the previous section, we would like to conjecture that this most stable configuration captures the true solution of the complete KMS condition. Could such a point exist? Possibly yes, because as we increase ${\tt p}$ away from ${\tt p}=0$, there will be a point where the neighborhood $U$ becomes sufficiently large to include the values $A_J(r_i)$ corresponding to the solution we are looking for. For those values the standard deviations are expected to be minimal and it is natural to expect that the corresponding configurations will dominate the ReLU optimization. Explicit computation in the next section shows that this picture is indeed correct and that allows us to identify a critical value ${\tt p}_*$ of stable optimization, which is conjectured to represent an approximation of the unique solution to the complete KMS condition.

\begin{figure}[t!]
    \centering
    \includegraphics[width=0.7\linewidth]
    {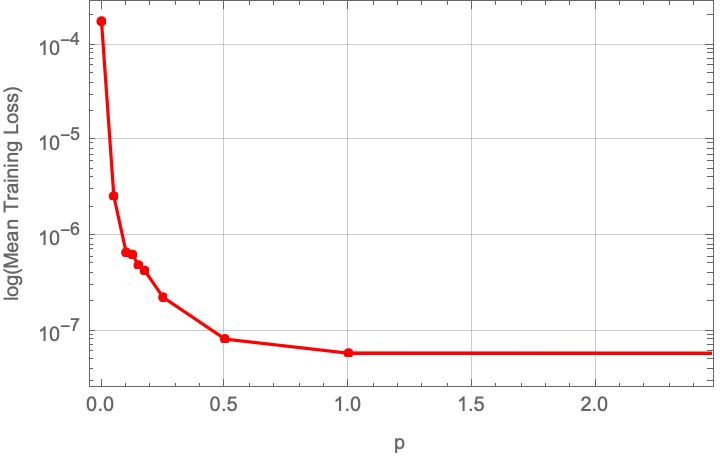}
    \caption{\small{Mean training loss for independent ReLU runs at different values of the factor ${\tt p}$. The mean is computed for the 10 lowest-loss configurations of each cluster run. The $y$-axis scale is logarithmic. There is a concentration of data points near the value ${\tt p}=0.15$, where we observed outcomes with the smallest statistical variation.}}
    \label{fig:lossvsp}
\end{figure}

\subsubsection{Preliminary results on a double-twist datum}
\label{doubletwist}

We are now in position to return to the analysis of the KMS condition in four-dimensional holographic CFTs. We fix the discontinuity using the (multi-trace) energy-momentum data \eqref{q1ab}, and set up the approximate KMS condition with $J_*=6$ to compute the double-twist datum $a_{1,0}+3a_{0,2}$. We use the dot-loss in the KMS optimization and the ReLU approach to determine a critical factor ${\tt p}_*$.

Let us examine what happens to the outcome of the optimization as we vary the factor ${\tt p}$. 1K independent runs were performed on the QMUL Apocrita cluster for 50K epochs with different values of ${\tt p}$ from 0 to 10. The mean training loss of the 10 lowest-loss configurations as a function of ${\tt p}$ is presented in Figure\ \ref{fig:lossvsp}. At ${\tt p}=0$ the loss takes high values at the order of $10^{-4}$ and very quickly drops (in accordance with the expectations at the end of the previous subsection) by three orders of magnitude to an almost constant value.

\begin{figure}[t!]
    \centering
    {\tiny ${\tt p}=0$}\\
    \includegraphics[width=0.73\linewidth,height=5.7cm]
    {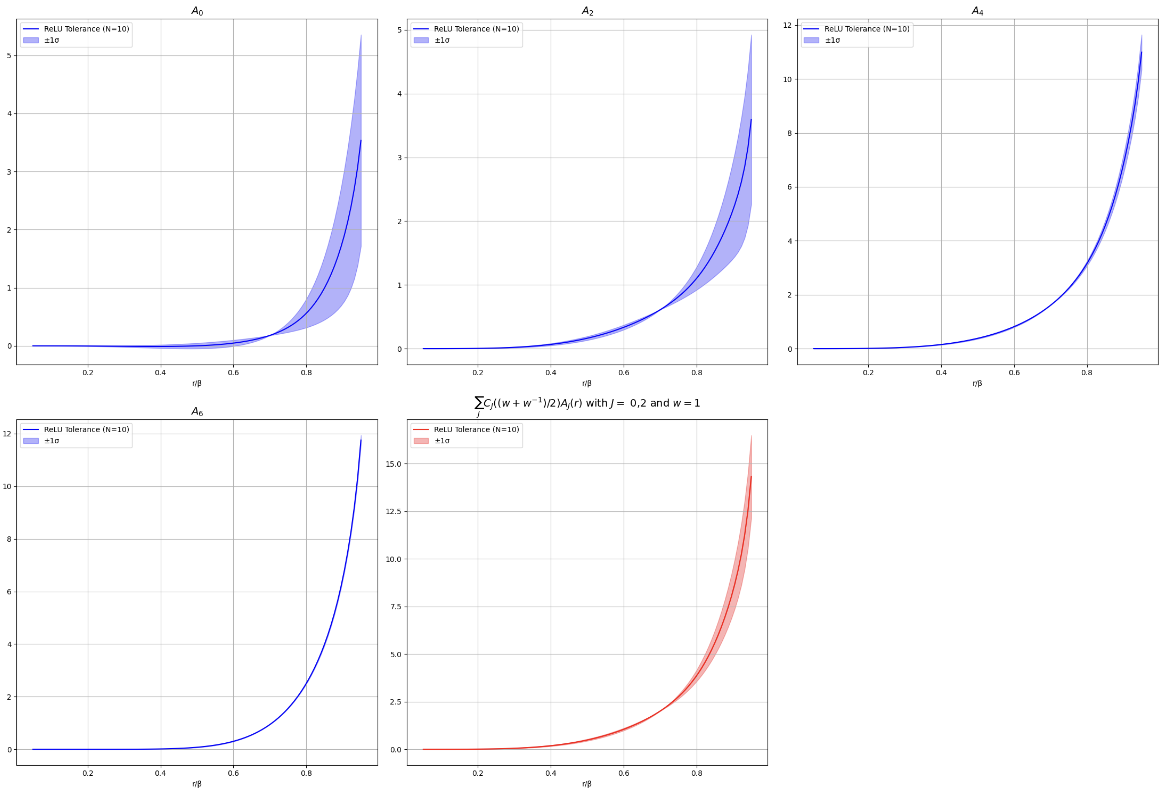}\\
    {\tiny ${\tt p}=0.15$}\\
    \includegraphics[width=0.75\linewidth,height=5.7cm]
    {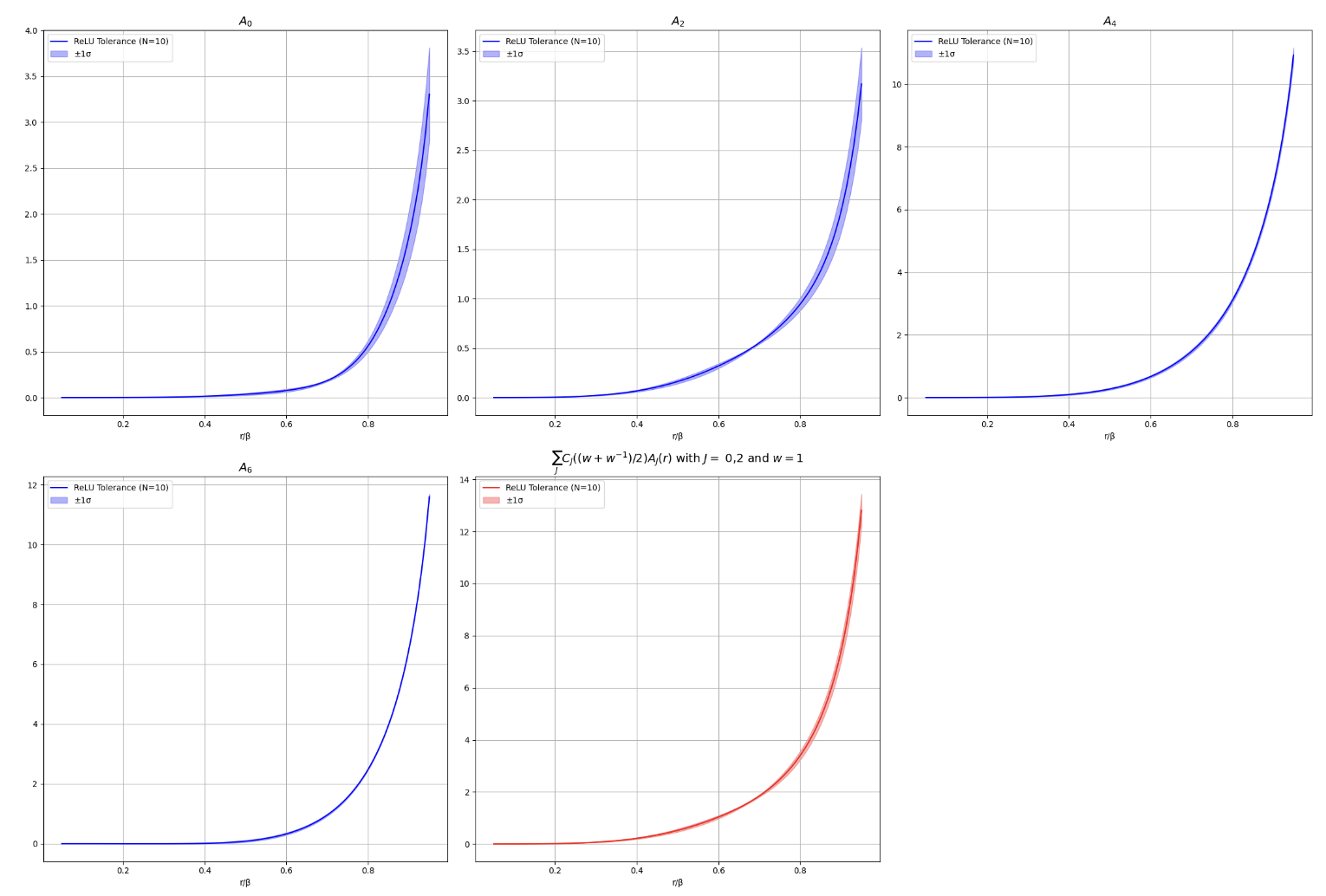}\\
    {\tiny ${\tt p}=10$}\\
    \includegraphics[width=0.72\linewidth,height=5.7cm]
    {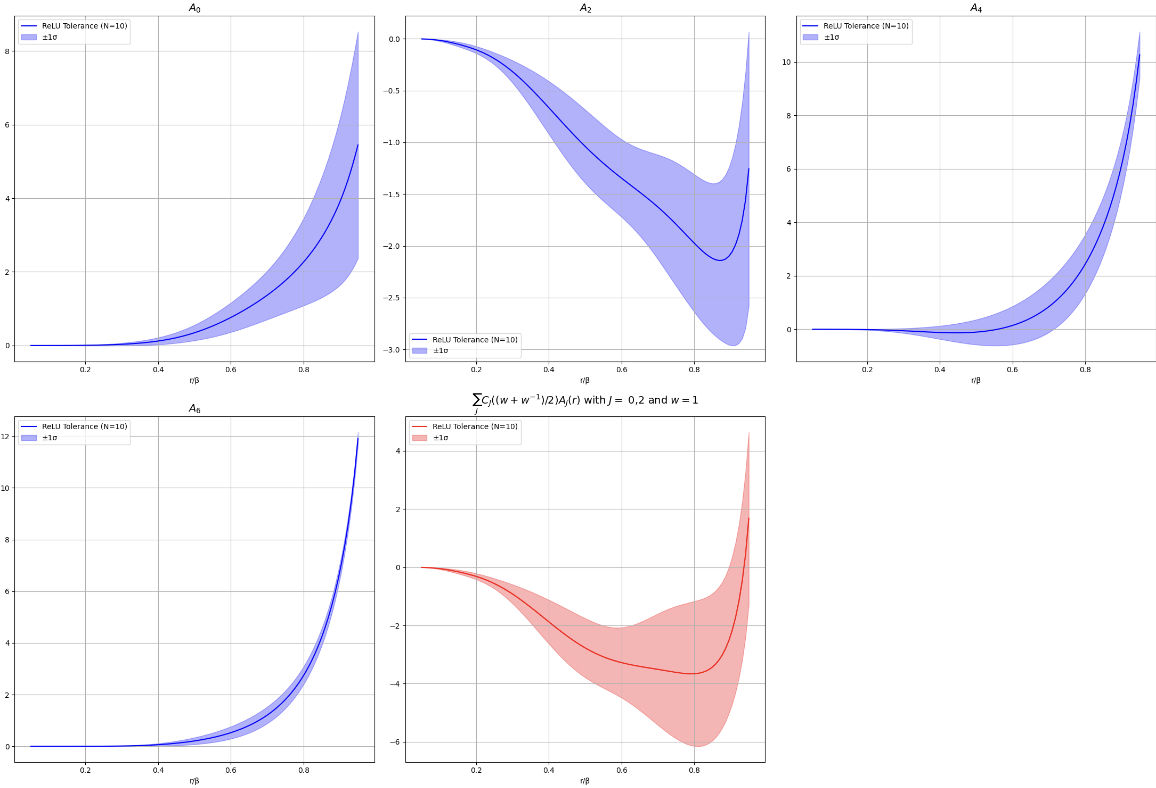}
    \caption{\small{The output of the ReLU optimization for three different values of ${\tt p} \in \{0,0.15,10\}$. In each case the first four plots represent the functions $A_0, A_2,A_4,A_6$. The last plot in red presents the combined contribution of $A_0$ and $A_2$ to the conformal block expansion at $w=1$.}}
    \label{fig:relu}
\end{figure}

Figure\ \ref{fig:relu} presents the explicit output of the tails for three different values of ${\tt p} \in \{0,0.15,10\}$. We observe the features anticipated in the previous subsection. There is visible variation of the output at ${\tt p}=0$, which then decreases as ${\tt p}$ increases up to some point around ${\tt p}=0.15$. Then the variation of the output increases again producing significantly deformed curves. We performed multiple cluster runs around the value ${\tt p}=0.15$ in order to determine the factor ${\tt p}$ of minimum instability, which was identified in the vicinity of ${\tt p}=0.15$.

Once ${\tt p}$ is fixed we can read off the corresponding value of the double-twist parameter $a_{10}+3a_{02}$ using Eq.\ \eqref{kms2dj}. Specifically, we found:
\bea
\label{doubletwistaa}
{\tt p} = 0.125 ~&:&~ a_{10}+3a_{02} = 9.81 \pm 0.35
~,\\
\label{doubletwistab}
{\tt p} = 0.150 ~&:&~ a_{10}+3a_{02} = 10.09 \pm 0.37
~,\\
\label{doubletwistac}
{\tt p} = 0.175 ~&:&~ a_{10}+3a_{02} = 10.58 \pm 0.46
~.
\eea
Adopting the conjecture that the configuration at ${\tt p}_*\sim 0.15$ is a sensible approximation of the sought-after solution of the KMS condition for this holographic CFT, we can use \eqref{doubletwistab} to make a preliminary estimate of the double-twist datum $a_{10}+3a_{02}$. 

We note in passing that another way to perform the computation is without enforcing an asymptotic condition on the tail function near $r=1$. That removes a potential systematic error from the explicit, subtle specification of the tail functions in that region. In that case, we remove the $\LL_{\rm BC}$ contribution to the loss function $\LL$ in \eqref{q2ad} and set
\beq
\label{dtada}
\LL = \LL_{\rm KMS} + \LL_{\rm ReLU}
~.
\eeq
Repeating the exercise with this loss we found that we could still observe the same qualitative features as in Fig.\ \ref{fig:relu}, namely the existence of a value of $\tt p_*$ with increased stability. That value was approximately ${\tt p_*} \simeq 0.20$ yielding 
\bea
\label{doubletwistaaExtra}
{\tt p}_* = 0.20 ~&:&~ a_{10}+3a_{02} = 9.37 \pm 0.44
~.
\eea

It is worth noting here that the asymptotic model of Ref.\ \cite{Buric:2025anb} predicts the following value for $a_{10}+3a_{02}$ (see formula (2.29) in that paper for $m=1$)
\beq
\label{dtad}
\big[ a_{10}+3a_{02} \big]_{\hspace{-0.1cm}\text{ \cite{Buric:2025anb}}} = 7.54
~.
\eeq
In the more recent work, \cite{Buric:2025fye}, the predicted value is 
\beq
\label{dtada}
\big[ a_{10}+3a_{02} \big]_{\hspace{-0.1cm}\text{ \cite{Buric:2025fye}}} = 7.686
~.
\eeq
There is significant difference between these results and Eq.\ \eqref{doubletwistac} (at the order of 20-30\%), but this is not completely unexpected. The estimates \eqref{doubletwistab}, or \eqref{doubletwistaaExtra}, contain several systematic errors, whose effect has not been properly analyzed at this stage. These include: 
\begin{itemize}
    \item The errors in the approximation of the discontinuity. These could be estimated by increasing $J_*$ and the number of multi-trace energy-momentum tensor contributions.
    \item The errors in the asymptotic boundary condition for the tails near $r=1$ using Eq.\ \eqref{asae}. These are not present in \eqref{dtada}, which is visibly closer to the estimates of Refs.\ \cite{Buric:2025anb,Buric:2025fye}.
    \item The imprecise determination of ${\tt p}$. In this context, it would also be useful to check the efficiency of the optimization using another variant of the ReLU loss, \eqref{q2ab}, e.g.\  a quadratic (ReLU)$^2$ loss. 
\end{itemize}

It is also worth noting here that an independent estimation of $a_{10}+3a_{02}$ directly from the numerical solution of the PDE in AdS in Ref.\ \cite{Buric:2025anb} yields
\beq
\label{dtadb}
\big[ a_{10}+3a_{02} \big]_{\rm holographic-PDE\, \text{\cite{Buric:2025anb}}} \in [8.84, 10.97]
~.
\eeq
Both our estimates, \eqref{doubletwistab}, \eqref{doubletwistaaExtra}, are comparable to this result.

We again stress that the estimates \eqref{doubletwistab}, \eqref{doubletwistaaExtra} are based on the conjecture that the ${\tt p}_*$-configurations approximate the solution of the complete KMS condition. This remains a conjecture that needs to be better understood. For example, it would be interesting to establish whether a suitable ${\tt p}_*$ value always exists in problems with fixed discontinuity.

\subsection{Dynamic discontinuity}
\label{dyndisc}

So far we have been dealing exclusively with situations where the discontinuity is fixed to a specific form and does not involve any unknown thermal OPE coefficients. In generic problems the discontinuity contains unknown thermal OPE coefficients that need to be included and take active part in the optimization. Some of these coefficients can either be exposed or can be treated as implicit parts of the tail functions. At the end of the day, the optimization should constrain these data and accordingly modify the allowed form of the discontinuity.

In the holographic CFTs discussed in this and the previous section, it is the data of the energy-momentum sector that contribute to the discontinuity. We know that the KMS condition cannot completely fix these data, but there are certain restrictions they have to obey, most notably the universality relations of the lowest-twist multi-trace energy-momentum tensors observed in \cite{Fitzpatrick:2019zqz}. As a preliminary study in this context, we attempted to investigate if fixed discontinuities with arbitrarily fixed (multi-trace) energy-momentum data lead, after optimization, to configurations with lowest-losses that depend on the details of the discontinuity. We did not detect any such dependence and therefore no signs that discriminate between the discontinuities that obey the universality relations and those that do not. For this check we used the absolute loss combined with a loss for the universal asymptotic boundary condition. As we remarked above, this approach allows for a lot of spurious minima, which could have contaminated our analysis. It would be useful to repeat this analysis with a more proper treatment, e.g.\ the ReLU loss prescription, that eliminates the spurious minima. It is also important to further develop a version of the ReLU loss approach when the data entering the approximation of the discontinuity are optimizable. We intend to return to this aspect in future work.

On a more general note, it is possible that in some theories the spectrum is such that the KMS condition fixes the discontinuity and the corresponding 2-point functions uniquely. The results of Ref.\ \cite{Barrat:2025wbi} suggest this may be the case for the $O(N)$ vector models in $d = 3$  dimensions. It would be interesting to explore this case as well with the methods introduced in this paper.

\section{Preliminary discussion of $\NN=4$ SYM theory in the supergravity regime}
\label{sym}

As explained in Appendix \ref{logs}, thermal 2-point functions of (scalar) operators with integer scaling dimensions in holographic CFTs are subtle, because of the presence of logarithmic contributions to the conformal block expansion. The source of the logarithms are poles in certain 3-point function coefficients, which are canceled between the respective contributions of degenerate multi-trace energy-momentum operators and double-twist operators. In this section, we discuss the case of integer scaling dimensions in some detail, laying out the form of the OPEs, the contributions of the discontinuity and the corresponding KMS condition.

The primary example that we have in mind are thermal 2-point functions of half-BPS superconformal primary operators $\phi = S_p$ (with scaling dimension $\Delta_\phi =p$, $p=2,3,\ldots$) in the supergravity limit of 4d $\NN=4$ SYM theory. The aim of the present discussion is to set the stage for future work in this direction.

Similar to Section \ref{holo1}, we assume that the contributing operators to the thermal 2-point function $\langle \phi \phi \rangle_\beta$ are the identity, the energy-momentum tensor, the double-twist operators $[\phi\phi]_{n,J}$ and the multi-trace energy-momentum tensor operators $[T^k]_J$ for $k\geq 2$ and $J=2\ell\leq 2k$. 

In Appendix \ref{logs} we recall that when a multi-trace energy-momentum tensor $[T^k]_J$ is degenerate with a double-twist operator $[\phi\phi]_{n,J}$, namely when
\beq
\label{symaa}
2k = p + n + \frac{J}{2}
~,
\eeq
its contribution to the thermal OPE gets modified to
\beq
\label{symab}
g(rw,rw^{-1}) \supset
\bigg( a'_{[T^k]_J} \log (z\bar z) + \big( a_{[T^k]_J} + a_{[\phi\phi]_{n,J}} \big) \bigg) C_{J}^{(\nu)}\left(\frac{1}{2}(w+w^{-1})\right) r^{4k-2p}
~.
\eeq
The meaning of the coefficient $a'_{[T^k]_J}$ is explained in Eq.\ \eqref{logsaia}. Clearly, in this case the KMS condition cannot track the coefficients $a_{[T^k]_J}$ and $a_{[\phi\phi]_{n,J}}$ independently, but it can in principle track the combinations $a'_{[T^k]_J}$ and $a_{[T^k]_J} + a_{[\phi\phi]_{n,J}}$.

\subsection{Discontinuities}
\label{symdisc}

When the scaling dimension $\Delta_\phi$ of the external operator is an integer $p$, many of the discontinuities simplify and receive only $\delta$-function contributions. The double-twist operators have vanishing discontinuity. As a result, the only potential contributions arise from the identity operator and the energy-momentum sector. Here we list the explicit form of these contributions. 

\paragraph{Identity.} 
The identity operator contributes the following discontinuity:
\bea
\label{symdiscaa}
\IT^{({\bf 1})}_{\disc}[J_*; rw, rw^{-1} ] &=& \frac{4\pi r^{-p}}{\Gamma (p)} \bigg\{ (-1)^{p-1} \d_{w'}^{p-1}
\left[ \KK_{J_*}(w,w') (1-rw'^{-1})^{-p} \right] \bigg |_{w'=r^{-1}} 
\nonumber\\ 
&&\hspace{1.5cm} - \d_{w'}^{p-1} \left[ \KK_{J_*}(w,w') (1+rw'^{-1})^{-p} \right] \bigg |_{w'=-r^{-1}}\bigg\}
~.
\eea

\paragraph{The energy-momentum tensor.}
The contribution to the discontinuity from the energy-momentum tensor, with thermal OPE coefficient $a_T$, is:
\bea
\label{symbdiscba}
&&\IT^{(T)}_{\disc}[J_*; rw, rw^{-1} ] =
\nonumber\\
&&= 4 \pi a_T \sum_{s=0}^{2} \frac{r^{-p+s+1}}{\Gamma (p-s-1)} 
\left\{ (-1)^{p+s} \partial_{w'}^{p-s-2} [\KK_{J_*}(w,w')(1-rw'^{-1})^{3-p-s}]\bigg |_{w'=r^{-1}} \right.
\nonumber\\
&&\hspace{5.0cm} -\left. \partial_{w'}^{p-s-2} [\KK_{J_*}(w,w')(1+rw'^{-1})^{3-p-s}]\bigg |_{w'=-r^{-1}} \right\}
~.
\eea
When the $\Gamma$-functions in the denominator have poles, the corresponding term does not contribute.

\paragraph{Multi-trace energy-momentum tensors.}
There are two different types of discontinuities in this case. We noted in Eq.\ \eqref{appad} that we can recast the conformal block expansion as a series in powers of $z$ and $\bar z$. In this form, multi-trace energy-momentum operators that are not degenerate with double-twist operators contribute to the conformal block expansion in the following way:
\beq
\label{symdiscca}
g(z,\bar z) \supset \sum_{[T^k]_{2\ell} \in \phi\times \phi} a_{2\ell}^{(k)}
\sum_{s=0}^{2\ell} p_s(2\ell) z^{2k-\ell-p+s} \bar z^{2k+\ell -p -s}
~.
\eeq
On the other hand, multi-trace energy-momentum operators that are degenerate with double-twist operators have a modified conformal block expansion, see Eq.\ \eqref{symab}, which yields contributions of the form:
\beq
\label{symdisccb}
g(z,\bar z) \supset \sum_{[T^k]_{2\ell}}
\bigg( {a'}_{2\ell}^{(k)} \log (z\bar z) + \big( a_{2\ell}^{(k)} + a_{n,2\ell} \big) \bigg) \sum_{s=0}^{2\ell} p_s(2\ell) z^{2k-\ell-p+s} \bar z^{2k+\ell -p -s}\bigg |_{2k=p+n+\ell}
~.
\eeq
Here for the thermal OPE coefficients we are using the notation of Table \ref{tab:holoOps}.

The first type of contributions in Eq.\ \eqref{symdiscca} exhibit a non-vanishing discontinuity when a power of $z$ is negative, namely when
\beq
\label{symdisccc}
2k - p - \ell \leq - 1 - s \leq - 1
~.
\eeq
Since $J=2\ell \leq 2k \Leftrightarrow \ell \leq k$, this implies
\beq
\label{symdisccd}
k \leq k + (k-\ell) + s \leq p-1
~.
\eeq
For example, this condition excludes contributions from the multi-trace energy-momentum operators for $p=2$, where $k \leq 1$ cannot be satisfied. For $p=3$, on the other hand, there is a contribution from the double-trace operator $[T^2]_4$, which is equal to
\bea
\label{symdisccda}
\IT^{([T^2]_4)}[J_*;rw,rw^{-1}]
=4\pi a_{4}^{(2)} \frac{(1-r^2)^3}{r} \bigg[ \KK_{J_*}(w,r^{-1}) - \KK_{J_*}(w,-r^{-1}) \bigg]
~.
\eea

The second type of contributions, in Eq.\ \eqref{symdisccb}, have a non-vanishing discontinuity that arises from logarithmic terms. Notice that \eqref{symdisccb} requires
\beq
\label{symdiscce}
2k-p-\ell = n \geq 0
~,
\eeq
which is complementary to \eqref{symdisccc}. The total approximate contribution to the discontinuity from such operators is
\bea
\label{symdisccea}
&&\IT^{({\rm approx})(\log)}_\disc[J_*;rw,rw^{-1}] = 
-4\pi \sum_{k=2}^{\infty}\sum_{\ell=0}^{k} \sum_{s=0}^{2\ell} {a'}^{(k)}_{2\ell}\, p_s(2\ell) \times
\nonumber\\
&&\times\bigg\{ -\int_{-2r^{-1}}^{-r^{-1}} dw'\, \KK_{J_*}(w,w') (1+rw')^{2k-\ell-p+s} (1+rw'^{-1})^{2k+\ell-p-s} 
\nonumber\\
&&\hspace{0.5cm} + \int_{r^{-1}}^{2r^{-1}} dw'\, \KK_{J_*}(w,w') \, (1-rw')^{2k-\ell-p+s} (1-rw'^{-1})^{2k+\ell-p-s} \bigg\}_{2k-p-\ell \geq 0}
~.
\eea

In conclusion, some multi-trace energy-momentum operators contribute with inverse power discontinuities via \eqref{symdiscca} and the rest with log discontinuities via \eqref{symdisccb}. As we increase the integer scaling dimension $p$ of the external operator, more and more operators exchange logarithmic discontinuities with inverse power discontinuities.

\subsection{KMS condition}
\label{symkms}

An (approximate) KMS condition along the lines of Eq.\ \eqref{kms2ca} can now be formulated for any thermal 2-point function of half-BPS superconformal primary operators $S_p$. For example, for $p=2$, and only with the contributions of the identity and energy-momentum tensor exposed, we obtain
\bea
\label{symkmsaa}
&&r^{-4} 
+ a_T\, C_2^{(1)}\left( \frac{1}{2}(w + w^{-1} ) \right) 
+ \sum_{\ell = 0}^{\frac{J_*}{2}} A_{\Delta_*(2\ell), 2\ell;{\boldsymbol{\theta}}}(r)\, C_{2\ell}^{(1)}\left( \frac{1}{2}(w + w^{-1} ) 
\right) 
\nonumber\\
&&+ \IT^{(\boldsymbol{1})}_{\disc,p=2}[J_*;rw, rw^{-1}]
+ \IT^{(\log)}_{\disc, p=2}[J_*; rw, rw^{-1} ]
+\IT^{(T)}_{\disc,p=2}[J_*;rw, rw^{-1}]
\nonumber\\
&&=\tilde r^{-4} 
+ a_T\, C_2^{(1)}\left( \frac{1}{2}(\tilde w + \tilde w^{-1} ) \right) 
+ \sum_{\ell = 0}^{\frac{J_*}{2}} A_{\Delta_*(2\ell), 2\ell;{\boldsymbol{\theta}}}(\tilde r)\, C_{2\ell}^{(1)}\left( \frac{1}{2}(\tilde w +\tilde w^{-1} ) \right)
\nonumber\\
&&+ \IT^{(\boldsymbol{1})}_{\disc,p=2}[J_*;\tilde r \tilde w, \tilde r \tilde w^{-1}]
+ \IT^{(\log)}_{\disc, p=2}[J_*; \tilde r \tilde w, \tilde r \tilde w^{-1} ]
+ \IT^{(T)}_{\disc,p=2}[J_*;\tilde r \tilde w, \tilde r \tilde w^{-1}]
~.
\eea
Higher $p$ cases can be treated in a similar fashion using the formulas of the previous subsection.

We intend to return to a detailed analysis of these equations in future work.

\section{Outlook}
\label{outlook}

In this work, we provided a new framework for the bootstrap of crossing equations in CFTs focusing, as a concrete example, on the KMS condition at finite temperature. The proposed approach combines two complementary elements: $(i)$ modeling of the high-spin contributions to the thermal OPE with discontinuities, in a controlled approximation that employs thermal dispersion relations, and $(ii)$ capturing low-spin, but arbitrarily high-scaling-dimension, contributions via tail functions represented by Neural Networks. After testing and calibrating this approach on Generalized Free Fields, we applied it to holographic CFTs and provided a preliminary prediction for a thermal 1-point coefficient of a simple double-twist operator.

Moving forward, we would like to better understand how to reliably extract the complete information of the KMS condition from everywhere within the simultaneous $s$ and $t$ OPE convergence region. Since it is numerically challenging to set up an optimization scheme near the boundary of this region, it is useful to obtain a better understanding of the properties of the tail functions in the bulk of the OPE convergence region, and how these are related to features of the optimization. The analysis of the GFF case led us to conjecture that the analytic solution (that obeys the KMS condition everywhere in the OPE convergence region) can be recovered from the KMS optimization within a subregion of the OPE convergence, when the evaluation of the tail functions $A_J(r_i)$ at a generic bulk point $r_i$ is constrained to take a (numerically) unique value that stabilizes the output of the optimization. This conjectured relation between special values of $A_J(r_i)$, the stability of the optimization and the exact KMS solutions needs to be better understood. The existence of such special values was also observed in the case of holographic CFTs, where we used the above conjecture as a guiding principle to identify approximations of physical thermal 2-point functions. 

In addition, when $\Delta_\phi \geq \frac{d-1}{2}$, we derived a universal constraint on the $r\to 1$ asymptotics of the tail functions using the KMS condition. Imposing this constraint as an asymptotic boundary condition---lying outside of the bulk convergence region used for NN training---had a positive effect in restricting the space of low-loss configurations. It would be interesting to understand what happens when the above inequality is not satisfied.

On a more conceptual level, the proposed approach is shifting the focus away from individual CFT data to suitable tail functions and discontinuities. It would be very interesting to study the general properties of tail functions (as formulated in this paper) and use them to reformulate CFT constraints, or use theory-specific constraints that are imposed on the tails to guide the bootstrap search towards specific CFTs. 

As an immediate next step to the analysis presented in this paper, one could explore in further detail the more general bootstrap with a fully dynamical discontinuity. It would be interesting to study holographic CFTs in this manner and demonstrate the universality relations of \cite{Fitzpatrick:2019zqz}. Further natural directions include going beyond the (super)gravity limit, and applying our methods to other CFTs, e.g.\ $O(N)$ vector models.

Our approach can be extended beyond the KMS bootstrap for CFTs at finite-temperature. In fact, the bootstrap program in many different directions can be reformulated and studied anew. For instance, in the 4-point bootstrap for zero-temperature CFTs, our prescription can supplement standard methods based on positivity constraints \cite{NPS}. There are many physically interesting setups where positivity constraints are absent (or hard to identify), where modeling the crossing equations with tails and discontinuities would be a source of new information. Interesting setups of this type beyond finite-temperature physics include the physics of defects, the 5-point bootstrap \cite{Poland:2023vpn, Poland:2023bny, Poland:2025ide}, and CFTs in the presence of boundaries. We intend to return and address many of the above open problems in the near future. 

\section*{Acknowledgments}
We would like to thank Kyriakos Papadodimas, Andrei Parnachev and Andreas Stergiou for useful discussions.
The work of CP was partially supported by the Science and Technology Facilities Council (STFC) Consolidated Grant ST/X00063X/1 “Amplitudes, Strings \& Duality”. MW was funded by a STFC
studentship. This research utilised Queen Mary's Apocrita HPC facility, supported by QMUL Research-IT. \href{http://doi.org/10.5281/zenodo.438045}{http://doi.org/10.5281/zenodo.438045}.

\begin{appendix}

\section{Comments on thermal dispersion relations}
\label{dispApp}

The material in this Appendix collects useful facts about dispersion relations for scalar 2-point functions in thermal CFTs. This is a partial review of the MSc thesis \cite{Strat} of one of the authors (AS). Similar results were previously obtained in Ref.\ \cite{Alday:2020eua}.

\subsection{Dispersion relations from Cauchy's theorem}
\label{cauchy}

Consider the thermal 2-point function $g(z,\bar z)=\langle \phi(x) \phi(0)\rangle_\beta$ of a single scalar primary $\phi$ (in the same notation as in Section \ref{kms1}, where $z=r w$). In what follows, we set for convenience $\beta=1$. Let $G(z,\bar z)$ denote a `subtracted' version of $g$ that improves the large-$w$ behavior of the function, e.g.\ $G(z,\bar z) = f(z,\bar z)g(z,\bar z)$ or $G(z,\bar z)=g(z,\bar z)-f(z,\bar z)$, for a suitably chosen function $f$.\footnote{The function $f(z,\bar z)$ is symmetric under $z \leftrightarrow \bar z$ but not necessarily under $z\to-z,~\bar z \to-\bar z$} The subtraction allows us to drop arc contributions in the following discussion and streamline the presentation.

By using Cauchy's theorem 
\beq
\label{cauchyaa}
G(rw, rw^{-1}) = \frac{1}{2\pi i} \oint_{C_w} dw' \frac{1}{w'-w} G(rw', rw'^{-1})
~,
\eeq
one can derive the dispersion relation 
\bea
\label{cauchyad}
G(rw, rw^{-1}) &=&    \int_{-r}^{r} dw' \, K(w,w')\,
\disc\left[ G(rw', r{w'}^{-1} ) \right]
\nonumber\\
&+&   \int_{-\infty}^{-r^{-1}} dw' \, K(w,w')\,
\disc \left[ G(rw', r{w'}^{-1} ) \right]
\\
&+&   \int^{\infty}_{r^{-1}} dw' \, K(w,w')\,
\disc \left[ G(rw', r{w'}^{-1} ) \right]
~,\nonumber
\eea
which involves three contributions from integrals around the branch cuts $(-r,r)$, $(-\infty, -r^{-1})$ and $(r^{-1},\infty)$ respectively. By default, $r<1$. All three terms involve a common kernel in this expression
\beq
\label{cauchyae}
K(w,w') :=\frac{1}{4\pi} \left( \frac{1}{w'-w} + \frac{1}{w'-w^{-1}} \right)
~.
\eeq
We assumed that, after a suitable choice of subtraction, potential arc contributions from infinity do not contribute in the contour-deformation arguments that led to this result.

Equivalently, changing variables $w' \to w'^{-1}$ in the inner branch-cut integral along $(-r,r)$, and assuming the invariance $G(rw',rw'^{-1})=G(rw'^{-1},rw')$, we can reformulate \eqref{cauchyad} as an integral solely over the `outer' branch cuts as
\beq
\label{cauchyaf}
G(rw,rw') = \left( \int_{-\infty}^{-r^{-1}} + \int_{r^{-1}}^{\infty} \right)dw' \; \KK(w,w') \disc[G(rw',rw'^{-1})]~~
~,
\eeq
with kernel
\beq
\label{cauchyag}
\KK(w,w') := \frac{1}{2\pi w'}\frac{w'^2-1}{(w'-w)(w'-w^{-1})}
~.
\eeq
In this derivation of the dispersion relation, the kernel is trivially spacetime-dimension independent.

The dispersion relation \eqref{cauchyaf} can be generalized as follows. Consider a meromorphic function $\FF(r,w,w')$ that has $N$ poles at $w'=w_n$ ($n=1,\dots, N$) anywhere in the $w'$-plane except at $w'=w$. A variant of Eq.\ \eqref{cauchyaa} is
\beq
\label{cauchyAA}
G(rw, rw^{-1}) = \frac{1}{2\pi i} \oint_{C_w} dw' \left( \frac{1}{w'-w} + \FF(r,w,w') \right) G(rw', rw'^{-1})
~.
\eeq
Repeating the previous contour argument, one can easily derive the generalized dispersion relation
\bea
\label{cauchyAB}
&&G(rw,rw^{-1}) = - \sum_{n=1}^N {\rm Res}_{w' \to w_n} \left\{ \FF(r,w,w') G(rw', rw'^{-1}) \right\}
\nonumber\\
&&\hspace{3cm} + \oint_{arc} \frac{dw'}{2\pi i}\left [ \left(\frac{1}{w'-w} + \FF(r,w,w') \right) G(rw', rw'^{-1}) \right]
\\
&&+\left( \int_{-\infty}^{-r^{-1}} + \int_{r^{-1}}^{\infty} \right) dw' \bigg\{ \left( \KK(w,w') +\frac{1}{2\pi} \FF(r,w,w') - \frac{1}{2\pi w'^2}\FF(r,w,w'^{-1})\right) 
\nonumber\\
&&\hspace{5cm}\times\disc[G(rw',rw'^{-1})] \bigg\}
~.\nonumber
\eea
For completeness, in this case we have explicitly included a potential arc contribution at $|w'| \to \infty$ in the second line of this expression. If $G$ is suitably subtracted so that the $\frac{1}{w'-w} G(rw',rw'^{-1})$ term does not contribute, and $\FF$ is reasonably well behaved at infinity, we can safely set the arc contribution to zero. Alternatively, the arc contribution can be made to vanish if the auxiliary function $\FF$ is chosen such that $\FF\sim - \frac{1}{w'}$ at large $|w'|$.

\paragraph{Special cases.} We note two special cases:
\begin{itemize}
    \item For $\FF=0$ we recover trivially \eqref{cauchyaf}.

    \item Ref.\ \cite{Alday:2020eua} chose
    \beq
    \label{cauchyAC}
    \FF(r,w,w') = -\frac{1}{2(w'-w)} + \frac{1}{2(w'+w)} - \frac{1}{w'}
    ~.
    \eeq
\end{itemize}

\subsection{Dispersion relations from the Lorentzian OPE inversion formula}
\label{lor}

The dispersion relations \eqref{cauchyAB} can also be derived from the Lorentzian OPE inversion formula \cite{Iliesiu:2018fao}. We note here that this is unlike the corresponding situation for the dispersion relations in zero-temperature 4-point correlation functions, where the direct application of Cauchy's theorem yields the Bissi-Dey-Hansen dispersion relation \cite{Bissi:2019kkx} in terms of the single-discontinuity, whereas a derivation via the Lorentzian OPE inversion formula \cite{Caron-Huot:2017vep} yields the Carmi-Caron Huot dispersion relation \cite{Carmi:2019cub}. The latter is expressed in terms of the double-discontinuity of the 4-point correlation function.

Here we summarize the derivation of a dispersion relation for the subtracted thermal 2-point function from the Lorentzian OPE inversion formula following \cite{Alday:2020eua}. This is the main dispersion relation used in the main text. The resulting formula does not follow from a direct application of \eqref{cauchyAB}.  

First, let us recall the spectral sum decomposition of the 2-point function $g(z,\bar z)$ and the Lorentzian OPE inversion formula for $g(z,\bar z)$, \cite{Iliesiu:2018fao},
\beq
\label{subda}
g(rw, rw^{-1}) = \sum_{J=0}^\infty \oint_{-\epsilon-i\infty}^{-\epsilon + i \infty} \frac{d\Delta}{2\pi i}\, a(\Delta,J)\, C_J^{(\nu)}\left( \frac{1}{2}(w+w^{-1}) \right) r^{\Delta-2\Delta_\phi}
~,
\eeq
\bea
\label{subdb}
a(\Delta,J) &=& K_J \int_0^1 \frac{dr}{r}\, r^{2\Delta_\phi -\Delta} 
\bigg\{ (-1)^{d+1} \int_{-r}^{r} \frac{dw}{w} (w-w^{-1})^{2\nu} F_J(w) {\rm Disc}\left[ g(rw, rw^{-1} ) \right]
\nonumber\\
&& +  \int^{-r^{-1}}_{-\infty} \frac{dw}{w} (w-w^{-1})^{2\nu} F_J(w^{-1}) {\rm Disc}\left[ g(rw, rw^{-1} ) \right]
\\
&&+  \int_{r^{-1}}^{\infty} \frac{dw}{w} (w-w^{-1})^{2\nu} F_J(w^{-1}) {\rm Disc}\left[ g(rw, rw^{-1} ) \right] 
\bigg\}
\nonumber\\
&& + \theta(J_0 - J) a_{\rm arc}(\Delta, J)
~.\nonumber
\eea
The coefficient $K_J$ is
\beq
\label{subdbda}
K_J = \frac{\Gamma(J+1)\Gamma(\nu)}{4\pi \Gamma(J+\nu)}
~,
\eeq
precisely as in Eq.\ \eqref{dispae} in the main text. In the last line of \eqref{subdb} we included the arc contributions for $J\leq J_0$. $J_0$ has a value related to the Regge behavior of the discontinuity, \cite{Iliesiu:2018fao}. 

Following \cite{Alday:2020eua} we rewrite \eqref{subda} by separating out the contribution of a truncated $s$-channel expansion
\bea
\label{subdc}
&&g(rw,rw^{-1}) = \sum_{J=0}^{J_*} \sum_{\Delta}  a_{\Delta,J}\, C_J^{(\nu)}\left(\frac{1}{2}(w+w^{-1})\right) r^{\Delta-2\Delta_\phi} 
\nonumber\\
&&+\sum_{J>J_*}^{\infty} \int_{-\epsilon-i\infty}^{-\epsilon+i\infty} \frac{d\Delta}{2\pi i}\, a(\Delta,J)\, C_J^{(\nu)}\left( \frac{1}{2}(w+w^{-1}) \right) r^{\Delta-2\Delta_\phi}
%\nonumber
~.
\eea
Notice that the separated first term is truncated only in spin. It still includes an infinite sum over scaling dimensions. $J_*$ is an arbitrary cutoff on the spin (it can be less, equal or larger than $J_0$). We want to insert $a(\Delta, J)$ from the Lorentzian OPE inversion formula \eqref{subdb} into the last term of \eqref{subdc} and obtain an expression in terms of the discontinuity of $g$. It is clear without computation that this manipulation will produce a term that involves an integral of $\disc[g]$ multiplied by a modified kernel, which should be contrasted with the expression in \eqref{cauchyAB} that involves the fixed kernel $\KK(w,w')$ in Eq.\ \eqref{cauchyag}.

The last term in \eqref{subdc} is
\beq
\label{subde}
\II = \sum_{J>J_*}^{\infty} \int_{-\epsilon-i\infty}^{-\epsilon+i\infty} \frac{d\Delta}{2\pi i}\, a(\Delta,J)\, C_J^{(\nu)}\left( \frac{1}{2}(w+w^{-1}) \right) r^{\Delta-2\Delta_\phi} 
~.
\eeq
Substituting $a(\Delta,J)$ in the form provided by the Lorentzian OPE inversion formula in \eqref{subdb}, exchanging the $J$-sum and $\Delta$-integral with the $r$, $w$-integrals in \eqref{subdb}, yields \cite{Iliesiu:2018fao} the expression
\beq
\label{subdf}
\II = 2 \left( \int_{-\infty}^{-r^{-1}} +  \int^{\infty}_{r^{-1}} \right)
dw' \, \KK_{J_*}(w,w')\,
{\rm Disc}\left[ g(rw', r{w'}^{-1} ) \right]
+ \II_{\rm arc}
~,
\eeq
where
\bea
\label{subdg}
&&\KK_{J_*}(w,w') =  w'^{-1} (w'-w'^{-1})^{2\nu} 
\left[\sum_{J=J_*+1}^\infty K_J C_J^{(\nu)} \left(\frac{1}{2}(w+w^{-1})\right) F_J(w'^{-1}) \right]
\nonumber\\
&&=\frac{1}{2}\KK(w,w')
-  w'^{-1} (w'-w'^{-1})^{2\nu} 
\left[\sum_{J=0}^{J_*} K_J C_J^{(\nu)} \left(\frac{1}{2}(w+w^{-1})\right) F_J(w'^{-1}) \right]
~.
\eea
$\KK(w,w')$ is the kernel that appears in the standard thermal dispersion relation \eqref{cauchyag}. We note that this derivation requires the mathematical identity
\beq
\label{subdga}
\KK(w,w') = 2w'^{-1} (w'-w'^{-1})^{2\nu} 
\left[\sum_{J=0}^\infty K_J C_J^{(\nu)} \left(\frac{1}{2}(w+w^{-1})\right) F_J(w'^{-1}) \right]
~,
\eeq
which is valid in any spacetime dimension.\footnote{Notice that the LHS of \eqref{subdga} is manifestly spacetime-dimension-independent, but the RHS involves a series over spacetime-dimension-dependent terms.} The cases $d=2$ and $d=4$ provide a simple verification of these expressions. 

$\II_{\rm arc}$ is a potential contribution from $a_{\rm arc}$ in \eqref{subdb}
\beq
\label{subdiaa}
\II_{\rm arc} = \sum_{J>J_*}^{\infty} \int_{-\epsilon-i\infty}^{-\epsilon+i\infty} \frac{d\Delta}{2\pi i}\, \theta(J_0-J) a_{\rm arc}(\Delta,J)\, C_J^{(\nu)}\left( \frac{1}{2}(w+w^{-1}) \right) r^{\Delta-2\Delta_\phi} 
~.
\eeq
This contribution requires $J\leq J_0$ to be non-vanishing. Therefore, if $J_*<J_0$
\beq
\label{subdiab}
\II_{\rm arc} = \sum_{J=J_*+1}^{J_0} \int_{-\epsilon-i\infty}^{-\epsilon+i\infty} \frac{d\Delta}{2\pi i}\, a_{\rm arc}(\Delta,J)\, C_J^{(\nu)}\left( \frac{1}{2}(w+w^{-1}) \right) r^{\Delta-2\Delta_\phi} 
~,
\eeq
but whenever $J_*\geq J_0$, it vanishes automatically. In the main text we present formulas with the implicit assumption $J_*\geq J_0$.

To summarize, in this subsection we have shown that
\bea
\label{subdia}
&&g(rw,rw^{-1}) = \sum_{J=0}^{J_*}  \sum_{\Delta} a_{\Delta,J}\, C_J^{(\nu)}\left(\frac{1}{2}(w+w^{-1})\right) r^{\Delta-2\Delta_\phi} 
\nonumber\\
&&+ 2 \left( \int_{-\infty}^{-r^{-1}} +  \int^{\infty}_{r^{-1}} \right)
dw' \, \KK_{J_*}(w,w')\,
{\rm Disc}\left[ g(rw', r{w'}^{-1} ) \right]
+ \II_{\rm arc}
~.
\eea

\section{Universal features of the tail functions in the vicinity of $r=1$}
\label{apptails}

In this Appendix we argue that, under certain conditions, the tail functions $A_J(r)$ exhibit a universal behavior in the limit $r\to 1$ that follows from the KMS condition. Information about this universal behavior can be used to facilitate better optimization results when we vary tail functions in search of KMS solutions.

The conformal block expansion of a thermal 2-point function of identical scalar operators is
\beq
\label{asaa}
g(rw,rw^{-1}) = r^{-2\Delta_\phi} + \sum_{\ell=0}^\infty A_{2\ell}(r) C_{2\ell}^{(\nu)}\left(\xi(w)\right)
~,
\eeq
where $\xi(\omega)=\frac{1}{2}(w+w^{-1})$. Using the orthogonality of the Gegenbauer polynomials
\beq
\label{asab}
\int_{-1}^1 d\xi \, (1-\xi^2)^{\nu-\frac{1}{2}} C_n^{(\nu)}(\xi) C_m^{(\nu)}(\xi) = h_n^{(\nu)} \delta_{nm}
~, ~~~ h_n^{(\nu)} = \frac{2^{1-2\nu}\pi \Gamma(n+2\nu)}{n!(n+\nu)\Gamma(\nu)^2}
\eeq
we obtain
\beq
\label{asac}
A_{2\ell}(r) = \frac{1}{h^{(\nu)}_{2\ell}} \int_{-1}^1 d\xi\, (1-\xi^2)^{\nu-\frac{1}{2}} C_{2\ell}^{(\nu)}(\xi)\, 
\bigg[ g(r w(\xi),r w(\xi)^{-1}) - r^{-2\Delta_\phi}\bigg]
~.
\eeq
The function $g(z,\bar z)$ diverges at $z=-1,1$. The divergence is captured by the KMS condition through the identity contribution in the crossed channel
\beq
\label{asad}
g(z,\bar z) \sim \frac{1}{|1-z|^{2\Delta_\phi}} = \frac{1}{(1-2r\xi + r^2)^{\Delta_\phi}}~~~\text{for }z\to 1
\eeq
and
\beq
\label{asad}
g(z,\bar z) \sim \frac{1}{|1+z|^{2\Delta_\phi}} = \frac{1}{(1+2r\xi + r^2)^{\Delta_\phi}}~~~\text{for }z\to -1
~.
\eeq
This behavior introduces a potential $r\to 1$ divergence in the functions $A_{2\ell}(r)$ in \eqref{asac} from the regions of the $\xi$-integral near $\xi=1$ and $\xi=-1$ as $r\to 1$. 
If there is such a divergence, we expect the functions
\bea
\label{asae}
&&\tilde A_{2\ell}(r) = 
\frac{1}{h^{(\nu)}_{2\ell}} \int_0^1 d\xi\, (1-\xi^2)^{\nu-\frac{1}{2}} C_{2\ell}^{(\nu)}(\xi)\, \frac{1}{(1-2r\xi + r^2)^{\Delta_\phi}}
\nonumber\\
&&\hspace{1.5cm}+\frac{1}{h^{(\nu)}_{2\ell}} \int_{-1}^0 d\xi\, (1-\xi^2)^{\nu-\frac{1}{2}} C_{2\ell}^{(\nu)}(\xi)\, \frac{1}{(1+2r\xi + r^2)^{\Delta_\phi}}
\eea
to exhibit the same behavior near the corresponding point as $A_{2\ell}(r)$.

In fact, we can maintain the correct form of the divergence and simplify things even further by noticing that in the vicinity of $\xi\sim \pm 1$
\beq
\label{asaf}
C_{2\ell}^{(\nu)}(\xi) \sim f_{2\ell}^{(\nu)} := \frac{2^{1-2\nu}\sqrt{\pi}\,\Gamma(2\ell+2\nu)}{(2\ell)!\Gamma(\nu)\Gamma(\nu+\frac{1}{2})}
~.
\eeq
The quantity 
\bea
\label{asag}
\tilde{\tilde A}_{2\ell}(r) &=& 
\frac{f_{2\ell}^{(\nu)}}{h^{(\nu)}_{2\ell}} \left( \int_0^1 d\xi\,   \frac{(1-\xi^2)^{\nu-\frac{1}{2}}}{(1-2r\xi + r^2)^{\Delta_\phi}}
+ \int_{-1}^0 d\xi\, \frac{(1-\xi^2)^{\nu-\frac{1}{2}}}{(1+2r\xi + r^2)^{\Delta_\phi}} \right)
\nonumber\\
&=&\frac{2 (2\ell+\nu) \Gamma(\nu)}{\sqrt{\pi}\Gamma(\frac{1}{2}+\nu)} \int_0^1 d\xi\,   \frac{(1-\xi^2)^{\nu-\frac{1}{2}}}{(1-2r\xi + r^2)^{\Delta_\phi}}
\eea
is also expected to have the same singularity structure as $A_{2\ell}(r)$.

The expression \eqref{asag} implies that $A_{2\ell}(r)$ is divergent at $r=1$ when 
\beq
\label{asak}
\nu-\frac{1}{2}-\Delta_\phi \leq -1 ~\Leftrightarrow~ \Delta_\phi \geq \nu +\frac{1}{2} 
~.
\eeq
In the narrow window $\nu\leq \Delta_\phi<\nu+\frac{1}{2}$, it implies instead that the tail functions $A_{2\ell}(r)$ are regular at $r=1$, but this is not in immediate contradiction with the fact that $g(z,\bar z)$ diverges at $z\to \pm 1$. Indeed, at $z=\pm 1$ the series
\beq
\label{asam}
g(1,1) = 1+ \sum_{\ell=0}^\infty A_{2\ell}(1) C_{2\ell}^{(\nu)}(1) 
= 1+ \sum_{\ell=0}^\infty \frac{\Gamma(2\ell+2\nu)}{(2\ell)!\Gamma(2\nu)} A_{2\ell}(1)
\eeq
can still diverge for finite $A_{2\ell}(1)$.

When the tail functions are divergent, the above argument also implies that the divergence is universal, independent of the details of the CFT, and directly related to the identity contribution in the crossed channel. We can verify the condition for the divergence, \eqref{asak}, and determine its type in the GFF case as follows.

For GFFs
\beq
\label{asan}
A_{2\ell}(r) = r^{2\ell} \sum_{n=0}^\infty a_{n,2\ell}\, r^{2n}
~,
\eeq
with
\beq
\label{asao}
a_{n,J} = 2\zeta(2\Delta_\phi+2n+J)\frac{(J+\nu)(\Delta_\phi)_{J+n}(\Delta_\phi-\nu)_n}{n!(\nu)_{J+n+1}}
~.
\eeq
At leading order in the large-$n$ limit (at fixed $J$)
\beq
\label{asap}
a_{n,J} \sim \frac{2(J+\nu)\Gamma(\nu)}{\Gamma(\Delta_\phi)\Gamma(\Delta_\phi-\nu)} (n(n+J))^{\Delta_\phi - \nu -1}
\eeq
and the series
\beq
\label{asaq}
A_{2\ell}(1) = \sum_{n=0}^\infty a_{n,2\ell}
\eeq
diverges when 
\beq
\label{asar}
2\Delta_\phi -2\nu - 2 \geq -1 \Leftrightarrow \Delta_\phi - \nu - \frac{1}{2} \geq 0 
~,
\eeq
reproducing the inequality in \eqref{asak}.

Moreover, the asymptotic behavior of the $a$-coefficients in \eqref{asap} suggests that in GFF the functions $A_{2\ell}(r)$ diverge in the following way as $r\to 1^-$
\beq
\label{asas}
A_{2\ell}(r) \sim  \frac{2(2\ell+\nu)\Gamma(\nu)}{\Gamma(\Delta_\phi)\Gamma(\Delta_\phi-\nu)} \frac{\Gamma(2\Delta_\phi-2\nu - 1)}{(1-r^2)^{2\Delta_\phi - 2\nu - 1}} r^{2\ell}
~.
\eeq

\section{Neural networks}
\label{nn}

As we mentioned in the main text, the tail functions $A_{\Delta_*(J), J}(r)$ are modeled by an artificial neural network. In this appendix, we provide more details regarding the architecture we employed, as well as the reasoning behind it.

\subsection{GFF}
The functions we are trying to capture are defined in Eq.\ \eqref{dispaj}. For double twist operators, they reduce to

\begin{equation}\label{NNa}
    A(r)_{n_*(J),\,J} = \sum_{n=n_*(J)+1}^{\infty} a_{n,J} \;r^{2n+J} ~~.
\end{equation}
Optimizing a model to directly capture \eqref{NNa} can be challenging. For example, the model could learn the wrong small $r$ behavior, i.e.\ the wrong starting power, as that information resides only within Eq. \eqref{NNa} and not within any of the loss functions we use. Similarly, the model could also fail to capture the $r\to 1$ behavior that is implied by the KMS condition, 

\begin{equation}\label{NNb}
    A(r)_{n_*(J),\,J} \propto (1-r^2)^{1+2\nu -2\Delta_\phi} ~~,
\end{equation}
as described in Appendix \ref{apptails}. In addition, the values of the tail functions close to $r\to 1$ can be several orders of magnitude larger than those at other points, which can cause severe numerical instabilities.

For all these reasons, we define our neural network model as,

\begin{equation}\label{NNc}
    A(r)_{n_*(J),\,J} = 2\,r^{2n_*(J)+J+2} \, (1-r^2)^{1+2\nu-2\Delta_\phi}\, \sinh\left\{{\rm NN}(r^2) \right\} ~~,
\end{equation}
where ${\rm NN}(r)$ is the trainable neural network. In this expression, the prefactor implements the correct low and high $r$ behavior, as well as the boundary condition, $A(0)_{n_*(J),\,J} = 0$. The power of $r$ in the prefactor is chosen to be such that the desired ${\rm NN} (r^2)$ behaves as a constant at low $r$. The hyperbolic sine function is introduced for regularization purposes. For small output values, it simply behaves like ${\rm NN}(r^2)$ whereas for larger outputs it behaves like $\exp{[{\rm NN}(r^2)]}$. In that way, the huge output variation needed to capture $A(r)_{n_*(J),\,J}$ is achieved without a huge variation of the trainable model's output. Finally, the model's input is always squared in order to enforce the $r\to-r$ symmetry that the GFF $A(r)_{n_*(J),\,J}$ have by construction.

\begin{figure}[h]
    \centering
    \includegraphics[width=0.8\linewidth]{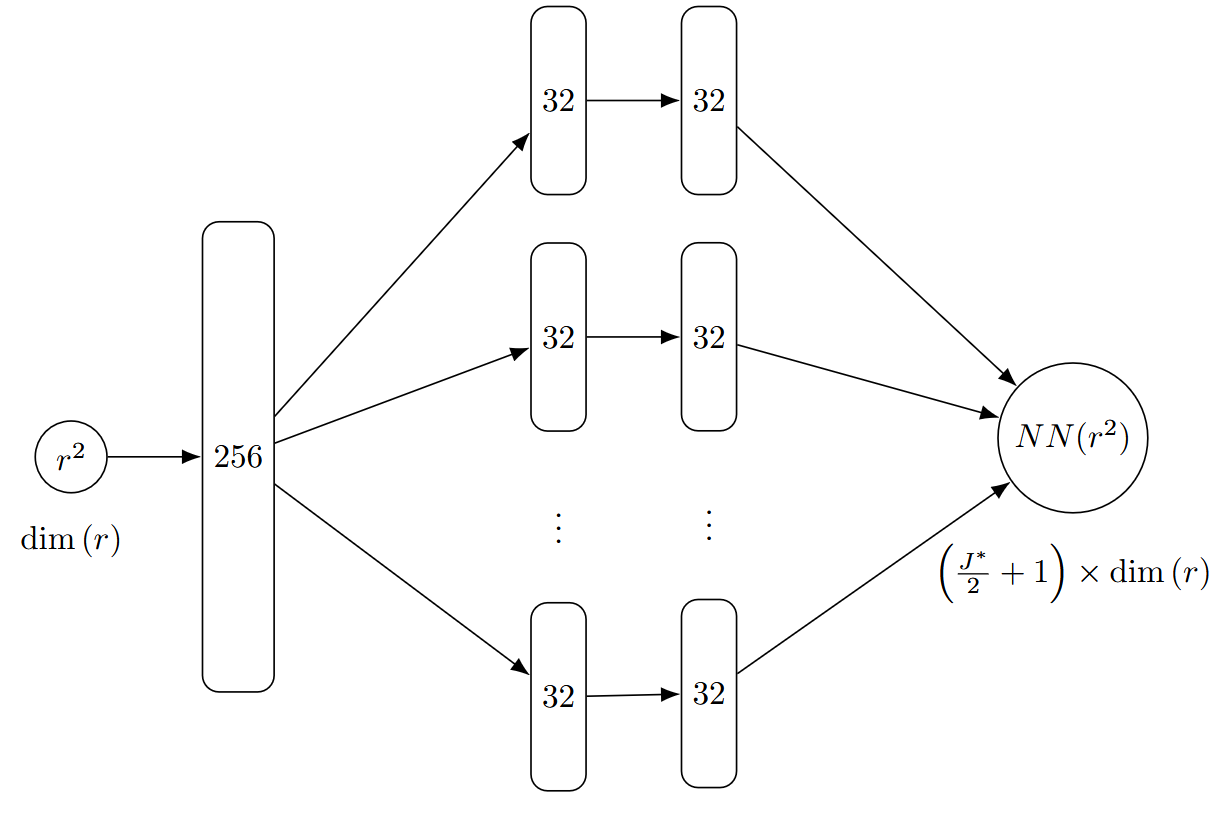}
    \caption{\small{The Neural Network model used for the GFF runs. This architecture sums a total of $512 + 
    \left(\frac{J_*}{2} +1 \right)9313$ trainable parameters.}}
    \label{fig:NN}
\end{figure}

As for the trainable ${\rm NN}(r^2)$, the model consists of two main parts: the backbone and the sub-networks. The backbone is a fully connected Deep Neural Network (DNN) that performs the first processing of the input. Consequently, the output is passed to $\frac{J_*}{2}+1$ sub-networks. These subnets aim to capture the contribution of each spin, i.e.\ to represent the sums \eqref{NNa} for different spins.  Between layers, on both backbone and sub-nets, the inputs are passed through hyperbolic tangent activation functions. Finally, each sub-net yields a single value leading to the full model outputting a $\frac{J_*}{2}+1$ sized vector. A visual representation of such an architecture is given in Figure \ref{fig:NN}.

Another point worth highlighting is the number of NN trainable parameters. A model with our proposed architecture involves a total of 
\begin{equation}
    \label{NNd}
    \mathcal{P} = 2 n_b +  (\mathcal{B}-1)n_b(n_b+1) + \left(\frac{J_*}{2}+1\right)\left( n_s (n_b+1) + (\mathcal{S}-1) n_s(n_s+1) + n_s+1\right)
\end{equation}
parameters, where
\begin{itemize}
    \item $n_b$ is the number of nodes per backbone layer,
    \item $\mathcal{B}>0$ is the number of backbone layers,
    \item $n_s$ is the number of nodes per subnet layer,
    \item $\mathcal{S}>0$ is the number of subnet layers.
\end{itemize}
In our applications, we observed that we were able to efficiently bootstrap the KMS condition with relatively small models (for GFF: $n_b = 256$, $\mathcal B = 1$, $n_s = 32$, $\mathcal S = 2$). In most cases, our models had a few tens of thousands parameters and individual runs could be performed on a laptop with execution times of about 45 minutes for 50K epochs. It is possible that larger models might be able to yield better results at the cost of requiring greater computational power to train.

\subsection{Holographic CFTs}
In the case of holographic CFTs, we used a very similar model, with some minor modifications. Here, we want to model sums of the form,
\begin{equation}\label{NNe}
    A_{J}(r) = \sum_{n=0}^{\infty} a_{n,J} \;r^{2n+J} + \sum_{k=\frac{J}{2}}^{\infty} a_{J}^{(k)} r^{kd-2\Delta_\phi}
\end{equation}
for an some appropriate lower boundary truncation characterized by some $\Delta_{*}(J)$.

The first difference from the GFF model comes from the existence of the energy-momentum data in the spectrum. These terms break the $r\to-r$ symmetry of the GFF case, and therefore we should not square the input before passing it to the neural network. Second, some changes have to be made to the starting powers in the prefactor. The purpose here is to once again factor out the leading powers for the tails at small $r$. This is not as straightforward as in the GFF case, since the leading power for each $J$ can change as a function of $\Delta_\phi$, the space-time dimension $d$ and the exposed operators. In general, the model can be defined as,

\begin{equation}\label{NNf}
    A(r)_{\Delta_*(J),\,J} = 2\,r^{Q(\Delta_{*},J,\Delta_{\phi},d)} \, (1-r^2)^{1+2\nu-2\Delta_\phi}\, \sinh\left\{{\rm NN}(r) \right\} ~~,
\end{equation}
where the power,

\begin{equation}
    \label{NNg}
    Q(\Delta_{*}(J), J, \Delta_\phi,d) = \min\{
    2(n_{*}+1)+J,~d(k_{*}+1) -2\Delta_\phi \}
\end{equation}
with $n_*, k_*$ being the maximum $n,k$ of exposed operators for a given $\Delta_*(J)$.
\noindent
The rest of the model specifications, such as the architecture and the number of trainable parameters, was the same as in the GFF case.

\section{Large-$c$ scaling of correlation functions at finite temperature}
\label{largec}

The organization of the large-$c$ expansion of correlation functions in CFT is considerably different at finite temperature compared to the zero-temperature case. Although this is well-known, we are not aware of a detailed discussion of the relevant properties in the literature. Therefore, for the convenience of the reader, we include here a short review of pertinent material. 

Motivated by large-$N$ gauge theories, we set $c:=N^2$ and use a common terminology in terms of single- and multi-trace operators (which refers to operators with specific large-$N$ scaling properties).

We consider the thermal 2-point function $\langle \phi \phi\rangle_\beta$ of a single-trace operator $\phi$ and write schematically its conformal block expansion as
\beq
\label{largeNba}
\langle \phi(x_1) \phi(x_2) \rangle_\beta = \sum_\OO \frac{\langle \phi \phi \OO\rangle_{\beta=0}}{\langle \OO \OO \rangle_{\beta=0}} \langle \OO(x_2)\rangle_\beta\;,
\eeq
with an appropriate sum over single- and multi-trace conformal primaries $\OO$.

\subsubsection*{Case I: $\langle \phi \rangle_\beta = 0$}

In this case, $\langle \phi(x_1) \phi(x_2) \rangle_\beta$ has no disconnected contribution and scales at leading order in the large-$N$ limit as $N^0$. We proceed to discuss in detail the leading large-$N$ scaling of each potential contribution on the RHS of \eqref{largeNba}. To make the notation lighter, we will denote zero-temperature correlators by dropping the $\beta$-subscript 
$$\langle \cdots \rangle := \langle \cdots \rangle_{\beta=0}~.$$

\paragraph{$\boxed{\OO=\phi^2}$} We have
\beq
\label{largeNbb}
\langle \phi \phi \phi^2 \rangle \sim N^0~, ~~
\langle \phi^2 \phi^2 \rangle \sim N^0
~.
\eeq
Since $\phi$ has no thermal vev, the leading $\OO(N^2)$ contribution from large-$N$ factorization vanishes yielding
\beq
\label{largeNbc}
\langle \phi^2 \rangle_\beta \sim N^0
~.
\eeq
Consequently, 
\beq
\label{largeNbd}
\frac{\langle \phi \phi \phi^2 \rangle}{\langle \phi^2 \phi^2 \rangle} \langle \phi^2(x_2)\rangle_\beta \sim N^0
~.
\eeq

\paragraph{$\boxed{\OO_{k,n}=\phi \d^k \Box^n \phi}$} For the rest of the double-twist operators we have similarly
\beq
\label{largeNbe}
\langle \phi \phi \OO_{k,n} \rangle \sim N^0~, ~~
\langle \OO_{k,n} \OO_{k,n} \rangle \sim N^0~, ~~
\langle \OO_{k,n} \rangle_\beta \sim N^0
~,
\eeq
yielding
\beq
\label{largeNbf}
\frac{\langle \phi \phi \OO_{k,n} \rangle}{\langle \OO_{k,n} \OO_{k,n} \rangle} \langle \OO_{k,n} (x_2)\rangle_\beta \sim N^0
~.
\eeq

\paragraph{$\boxed{\OO = \phi^n_\d$ for $n\geq 3}$} By $\phi^n_\d$ we denote compactly any $n$-trace (conformal primary) operator with or without derivatives. Since the leading-order large-$N$ scaling of the (connected) correlation $\langle \phi \phi \phi^n_\d\rangle$ is $N^{-n}$ (the same as that of the connected $(n+2)$-point function $\langle \phi(x_1) \phi(x_2) \prod_{i=1}^n \phi(y_i)\rangle$), we obtain
\beq
\label{largeNbg}
\langle \phi \phi \phi^n_\d \rangle \sim N^{-n}~, ~~
\langle \phi^n_\d \phi^n_\d \rangle \sim N^0~, ~~
\langle \phi^n_\d \rangle_\beta \sim N^\alpha~, ~~ \alpha < n
~,
\eeq
yielding
\beq
\label{largeNbi}
\frac{\langle \phi \phi \phi^n_\d \rangle}{\langle \phi^n_\d \phi^n_\d \rangle} \langle \phi^n_\d (x_2)\rangle_\beta \sim N^{\alpha-n}
~,
\eeq
which is subleading to $N^0$. Hence, $n$-trace operators with $n\geq 3$ (beyond the double-twist operators) {\bf cannot appear} at leading order in the large-$N$ limit in the conformal block expansion \eqref{largeNba}.

\paragraph{$\boxed{\OO = T_{\mu\nu}}$} In this case,
\beq
\label{largeNbj}
\langle \phi \phi T \rangle \sim N^{-1}~, ~~
\langle T T \rangle \sim N^0~, ~~
\langle T \rangle_\beta \sim N 
~,
\eeq
yielding
\beq
\label{largeNbk}
\frac{\langle \phi \phi T \rangle}{\langle TT \rangle} \langle T (x_2)\rangle_\beta \sim N^{0}
~.
\eeq
Similar behavior is exhibited by any single-trace operator that has a non-vanishing thermal vev.

\paragraph{$\boxed{\OO = T^n}$} In this case,
\beq
\label{largeNbl}
\langle \phi \phi T^n \rangle \sim N^{-n}~, ~~
\langle T^n T^n \rangle \sim N^0~, ~~
\langle T^n \rangle_\beta \sim N^n 
~,
\eeq
yielding
\beq
\label{largeNbm}
\frac{\langle \phi \phi T^n \rangle}{\langle T^n T^n \rangle} \langle T^n (x_2)\rangle_\beta \sim N^{0}
~.
\eeq

\paragraph{$\boxed{\OO = T^n \phi^k~{\rm with~or~without~derivatives}}$} In this case the leading large-$N$ factorization contribution to the thermal vev of the operator vanishes yielding (as in similar cases above) a contribution to the conformal block expansion that is subleading. Hence, such operators do not contribute to the leading order. A special case are multi-trace energy-momentum operators with derivatives, the contributions of which  we dropped in the main text.

\subsubsection*{Case II: $\langle \phi \rangle_\beta \neq 0$}

In this case, $\langle \phi(x_1) \phi(x_2) \rangle_\beta$ has a disconnected contribution that scales at leading order in the large-$N$ limit as $N^2$. We now discuss in detail how to reproduce both the disconnected $\OO(N^2)$ contribution and the leading $\OO(N^0)$ connected contributions in the conformal block expansion. 

\paragraph{$\boxed{\OO=\phi^2}$} We have
\beq
\label{largeNca}
\langle \phi \phi \phi^2 \rangle \sim N^0~, ~~
\langle \phi^2 \phi^2 \rangle \sim N^0
~.
\eeq
Since $\phi$ has a non-vanishing thermal vev, 
\beq
\label{largeNcb}
\langle \phi^2 \rangle_\beta \sim N^2
~,
\eeq
and the leading $\OO(N^2)$ contribution from large-$N$ factorization is reproduced as expected from the term
\beq
\label{largeNcc}
\frac{\langle \phi \phi \phi^2 \rangle}{\langle \phi^2 \phi^2 \rangle} \langle \phi^2(x_2)\rangle_\beta \sim N^2
~.
\eeq

\paragraph{$\boxed{\OO_{k,n}=\phi \d^k \Box^n \phi}$} Since the leading $\OO(N^2)$ contribution to the large-$N$ factorization of $\langle \OO_{k,n}\rangle_\beta$ is annihilated by the derivatives, for the rest of the double-twist operators we have 
\beq
\label{largeNcd}
\langle \phi \phi \OO_{k,n} \rangle \sim N^0~, ~~
\langle \OO_{k,n} \OO_{k,n} \rangle \sim N^0~, ~~
\langle \OO_{k,n} \rangle_\beta \sim N^0
~,
\eeq
yielding
\beq
\label{largeNce}
\frac{\langle \phi \phi \OO_{k,n} \rangle}{\langle \OO_{k,n} \OO_{k,n} \rangle} \langle \OO_{k,n} (x_2)\rangle_\beta \sim N^0
~.
\eeq

\paragraph{$\boxed{\OO = \phi^n$ for $n\geq 3}$} Here we do not include any derivatives. We obtain
\beq
\label{largeNcf}
\langle \phi \phi \phi^n \rangle \sim N^{-n}~, ~~
\langle \phi^n \phi^n \rangle \sim N^0~, ~~
\langle \phi^n \rangle_\beta \sim N^n
~,
\eeq
yielding
\beq
\label{largeNcg}
\frac{\langle \phi \phi \phi^n \rangle}{\langle \phi^n \phi^n \rangle} \langle \phi^n (x_2)\rangle_\beta \sim N^{0}
~,
\eeq
which contributes to the leading $\OO(N^0)$ result. In contrast, $n$-trace products of $\phi$ with derivatives will not contribute at leading order because the leading $\OO(N^n)$ contribution to their vev vanishes.

\paragraph{$\boxed{\OO = T_{\mu\nu}}$} In this case,
\beq
\label{largeNci}
\langle \phi \phi T \rangle \sim N^{-1}~, ~~
\langle T T \rangle \sim N^0~, ~~
\langle T \rangle_\beta \sim N 
~,
\eeq
yielding
\beq
\label{largeNcj}
\frac{\langle \phi \phi T \rangle}{\langle TT \rangle} \langle T (x_2)\rangle_\beta \sim N^{0}
~.
\eeq
Similar behavior is exhibited by any single-trace operator that has a non-vanishing thermal vev.

\paragraph{$\boxed{\OO = T^n}$} In this case,
\beq
\label{largeNck}
\langle \phi \phi T^n \rangle \sim N^{-n}~, ~~
\langle T^n T^n \rangle \sim N^0~, ~~
\langle T^n \rangle_\beta \sim N^n 
~,
\eeq
yielding
\beq
\label{largeNcl}
\frac{\langle \phi \phi T^n \rangle}{\langle T^n T^n \rangle} \langle T^n (x_2)\rangle_\beta \sim N^{0}
~.
\eeq

\paragraph{$\boxed{\OO = T^n \phi^k~{\rm with~or~without~derivatives}}$} Similar to the previous subsection, when there are derivatives the leading large-$N$ factorization contribution to the thermal vev of the operator vanishes yielding a contribution to the conformal block expansion that is subleading. However, operators of the schematic form $\OO = T^n \phi^k$ without derivatives, can contribute to leading order since
\beq
\label{largeNcm}
\langle \phi \phi [T^n \phi^k] \rangle \sim N^{-n-k}~, ~~
\langle [T^n\phi^k] [T^n\phi^k] \rangle \sim N^0~, ~~
\langle [T^n\phi^k] \rangle_\beta \sim N^{n+k} 
~,
\eeq
yielding
\beq
\label{largeNcn}
\frac{\langle \phi \phi [T^n\phi^k] \rangle}{\langle [T^n\phi^k] [T^n\phi^k] \rangle} \langle [T^n\phi^k] (x_2)\rangle_\beta \sim N^{0}
~.
\eeq

\section{Integer external scaling dimensions in holographic CFTs}
\label{logs}

In this section we review the appearance of logarithms in the conformal block expansions of thermal 2-point functions $\langle \phi\phi\rangle$ in holographic CFTs, when the scaling dimension of the external operator $\Delta_\phi$ is integer. The source of the logarithms are poles in the 3-point function coefficients $C_{\phi\phi [T^k]_J}$ when $\Delta_\phi$ is an integer. These poles are canceled by corresponding poles in the 3-point function coefficients $C_{\phi\phi [\phi^2]_{n,J}}$ with double-twist operators. We review the mechanism behind this cancellation and how it leaves behind logarithmic contributions.

When $\Delta_\phi =p\in \IN$, some energy-momentum multi-trace operators can be degenerate with double-twist operators. For example, let us assume that the energy-momentum multi-trace operator $\OO_T := [T^k]_{J}$ is degenerate with the double-twist operator $\OO_\phi :=[\phi\phi]_{n,J}$. This implies
\bea
\label{logsaa}
&&\Delta_{[T^k]_{J}} = 4k = 2p + 2n + J = \Delta_{[\phi\phi]_{n,J}}
\nonumber\\
&&~~~~~~ \Leftrightarrow ~~ 2k = p + n + \frac{J}{2}
~.
\eea
Such operators contribute to the thermal conformal block expansion with terms of the form (up to a constant factor $\beta^{-\Delta}\frac{J!}{2^J (\nu)_J}$)
\beq
\label{logsab}
\bigg( f_{\phi\phi \OO_T} g^{\OO_T \OO_T}  b_{\OO_T}  
+ f_{\phi\phi \OO_\phi} g^{\OO_\phi \OO_\phi} b_{\OO_\phi}
\bigg) G^{\Delta_\phi}_{\Delta, J}(z,\bar z)
~,
\eeq
where $G^{\Delta_\phi}_{\Delta, J}(z,\bar z)=C_J^{(\nu)}\left(\frac{1}{2}(w+w^{-1})\right)r^{\Delta-2\Delta_\phi}$ is the standard thermal conformal block. In \eqref{logsab} $\Delta = \Delta_{\OO_\phi} = \Delta_{\OO_T}$. 

The effect we want to discuss follows from the fact that the corresponding 3-point function coefficients share a similar pole structure (see e.g.\ \cite{Esper:2023jeq} Eqs\ (B.4)-(B.6), \cite{Fitzpatrick:2019zqz} and Appendix \ref{holoresults} for concrete examples) of the form
\begin{align}
\label{logsac}
f_{\phi\phi \OO_T} &\propto \frac{\lambda_{\rm sing}^{(\OO_T)}}{\Delta_\phi - p} + \lambda_{\phi\phi \OO_T}\;,
\\
f_{\phi\phi \OO_\phi} &\propto \frac{\lambda_{\rm sing}^{(\OO_\phi)}}{\Delta_\phi - p} + \lambda_{\phi\phi \OO_\phi}
~,
\end{align}
with simple poles at integer values of $\Delta_\phi$. 

To understand what happens when $\Delta_\phi=p$ we set $\varepsilon := \Delta_\phi - p$ and discuss the limit $\varepsilon \to 0$. For starters, let us
isolate any of the terms $z^{\frac{\Delta - J}{2}-\Delta_\phi + s}\, \bar z^{\frac{\Delta+J}{2}-\Delta_\phi-s}$ in the power expansion of 
\beq
\label{logsad}
G^{\Delta_\phi}_{\Delta,J}(z,\bar z) = \sum_{s=0}^J p_s(J)\, z^{\frac{\Delta - J}{2}-\Delta_\phi + s} \bar z^{\frac{\Delta+J}{2}-\Delta_\phi-s}
~.
\eeq
When $\varepsilon \ll 1$, we can expand in the following manner 
\bea
\label{logsae}
&&z^{\frac{\Delta_{\OO_T} - J}{2}-\Delta_\phi + s} \bar z^{\frac{\Delta_{\OO_T} + J}{2}-\Delta_\phi - s}
= z^{2k - \frac{J}{2} - p + s - \varepsilon}\, 
\bar z^{2k+\frac{J}{2}-p-s - \varepsilon}
\nonumber\\
&&\hspace{4cm} \simeq z^{2k-\frac{J}{2} - p + s}\, (1 - \varepsilon \log z) ~
\bar z^{2k+\frac{J}{2}-p-s}\, (1 - \varepsilon \log \bar z) +\ldots
~,
\nonumber\\
&& z^{\frac{\Delta_{\OO_\phi} - J}{2}-\Delta_\phi + s} 
\bar z^{\frac{\Delta_{\OO_\phi} + J}{2}-\Delta_\phi - s}
= z^{n + s} \bar z^{n+J-s} = z^{2k-\frac{J}{2} - p + s} \bar z^{2k-p+\frac{J}{2}-s} +\ldots
~,
\eea
the dots indicate terms with higher powers in $\varepsilon$. In the last line, and in the second equality, we used \eqref{logsaa}. Then, as we take the limit $\varepsilon \to 0$, the corresponding term in the expression \eqref{logsab} behaves as
\bea
\label{logsaf}
&&\left( \frac{\lambda_{\rm sing}^{(\OO_T)}}{\varepsilon}  + \lambda_{\phi\phi \OO_T} \right)g^{\OO_T \OO_T} b_{\OO_T} (1 - \varepsilon \log z) z^{2k-\frac{J}{2} - p + s}
(1 - \varepsilon \log \bar z) \bar z^{2k-p+\frac{J}{2}-s}
\nonumber\\
&&+\left(\frac{\lambda_{\rm sing}^{(\OO_\phi)}}{\varepsilon}  + \lambda_{\phi\phi \OO_\phi} \right) g^{\OO_\phi \OO_\phi}b_{\OO_\phi} z^{2k-\frac{J}{2} - p + s}  \bar z^{2k-p+\frac{J}{2}-s}
~.
\eea
The cancellation of the divergent $\varepsilon^{-1}$ terms requires
\beq
\label{logsag}
\lambda_{\rm sing}^{(\OO_T)} g^{\OO_T \OO_T} b_{\OO_T} = - \lambda_{\rm sing}^{(\OO_\phi)} g^{\OO_\phi \OO_\phi} b_{\OO_\phi}
~.
\eeq
After this cancellation, and the limit $\varepsilon \to 0$, the expression \eqref{logsaf} becomes
\bea
\label{logsai}
&&(- \lambda_{\rm sing}^{(\OO_T)} \, \log (z\bar z) + \lambda_{\phi\phi \OO_T})\, g^{\OO_T \OO_T}\, b_{\OO_T}\, z^{2k-\frac{J}{2} - p + s} \bar z^{2k - p +\frac{J}{2} - s} 
\nonumber\\
&&+ \lambda_{\phi\phi \OO_\phi}\, g^{\OO_\phi \OO_\phi}\, b_{\OO_\phi} z^{2k-\frac{J}{2} - p + s} \bar z^{2k - p + \frac{J}{2} - s}
\\
&&=(- \lambda_{\rm sing}^{(\OO_T)} \, \log (z\bar z) + \lambda_{\phi\phi \OO_T})\, g^{\OO_T \OO_T}\, b_{\OO_T}\,  z^{n + s} \bar z^{n+J-s}
+ \lambda_{\phi\phi \OO_\phi}\, g^{\OO_\phi \OO_\phi}\, b_{\OO_\phi}  z^{n + s} \bar z^{n+J-s}\;,
\nonumber
\eea
where in the second line we used \eqref{logsaa} to re-express the result in terms of the positive integer $n$. Notice that only the log term contributes to the discontinuity. Let us denote its coefficient as
\beq
\label{logsaia}
a'_{\OO_T} :=  - \lambda_{\rm sing}^{(\OO_T)}\, g^{\OO_T \OO_T}\, b_{\OO_T} \frac{J!}{2^J(\nu)_J}
~.
\eeq
Re-assembling the full contribution of the degenerate operators $[T^k]_J$, $[\phi\phi]_{n,J}$ to the conformal block expansion we find that it takes the form
\beq
\label{logsaj}
g(z,\bar z) \supset
\bigg( a'_{[T^k]_J} \log (z\bar z) + \big( a_{[T^k]_J} + a_{[\phi\phi]_{n,J}} \big) \bigg) G^{\Delta_\phi}_{\Delta,J}(z,\bar z)
~.
\eeq

\section{Holographic multi-trace energy-momentum coefficients $a^{(k)}_J$}
\label{holoresults}
One way to compute thermal one-point coefficients $a_J^{(k)}$ for the energy-momentum sector of a holographic theory is through the study of boundary `heavy-light' four-point correlators of the form
\begin{align}
    \langle \phi_H(0)\phi_L(z,\bar{z})\phi_L(1)\phi_H(\infty)\rangle
\end{align}
at zero temperature, where the light operators $\phi_L$ have $\Delta_L\ll C_T$ and the heavy operators $\phi_H$ have $\Delta_H\sim C_T$. At large $C_T$, this correlator is argued to be equivalent to a bulk two-point function $\langle\Phi_L(x_1)\Phi_L(x_2)\rangle_{\text{BH}}$ of the dual light scalar in a black hole background created by the heavy operator. This is then dual to precisely the kind of boundary thermal two-point function $\langle\phi_L(x_1)\phi_L(x_2)\rangle_\beta$ studied in this work.
%\par

Refs.\ \cite{Fitzpatrick:2019zqz}, \cite{Karlsson:2019dbd} use this relation to compute the thermal one-point unknowns $a_{\mathcal{O}}$ for the energy-momentum sector with $\mathcal{O}= [T^k]_J$, which we denote $a_J^{(k)}$. These coefficients are found in terms of the parameters appearing in a static, spherically-symmetric solution to a general higher-derivative gravity theory. Taking $d=4$ henceforth, such a solution has the form
\begin{align}
    ds^2=r^2f(r)dt^2+\frac{dr^2}{r^2h(r)}+r^2\sum_{i=1}^3dx_i^2,
\end{align}
with functions $f(r),h(r)$ that have a near-boundary ($r\rightarrow\infty$) expansion
\begin{align}
    f(r)=&\;1-\frac{1}{r^4}\sum_{i=0}^\infty \frac{f_{4i}}{r^{4i}}, \\
    h(r)=&\;1-\frac{1}{r^4}\sum_{i=0}^\infty \frac{h_{4i}}{r^{4i}}.
\end{align} 
When these expansions are truncated to the $r^{-4}$ term and we impose $f_0=h_0$, we reduce to Einstein gravity in asymptotically $AdS_4$ spacetime. Ref.~\cite{Fitzpatrick:2019zqz} demonstrates that one needs to impose $f_0=h_0$ for consistency with the boundary `heavy-light' conformal block decomposition and moreover, that leading-twist coefficients $a_{2k}^{(k)}$ universally depend only on $f_0$, i.e. they are agnostic to higher-derivative effects and depend only on Einstein gravity. We note, however, that additional non-minimal gravitational interaction terms in the $AdS$ effective action will induce non-universal corrections to leading-twist coefficients, as studied in e.g. \cite{Fitzpatrick:2020yjb,Huang:2023ikg,Huang:2024wbq}.
\par
While \cite{Fitzpatrick:2019zqz} and \cite{Esper:2023jeq} report all $a_J^{(k)}$ up to $k=2$ and leading twist unknowns $a_{2k}^{(k)}$ up to $k=4$, our study uses certain sub-leading twist unknowns at $k=3$. Using the calculation scheme described in \cite{Fitzpatrick:2019zqz}, we find:
\begin{align}
    a_6^{(3)}=&\;\frac{\Delta  (\Delta  (\Delta  (143 \Delta  (7 \Delta +25)+7310)+7500)+3024)}{10378368000 (\Delta -3) (\Delta -2)}f_0^3, \label{eq:a63}\\
    a_4^{(3)}=&\;\frac{\Delta  (\Delta  (\Delta  (\Delta  (143 \Delta  (7 \Delta -15)-2760)-2390)+2244)+2160)}{5189184000 (\Delta -4) (\Delta -3) (\Delta -2)}f_0^3 + \nonumber\\
    &\;+\frac{\Delta  (\Delta +1) (\Delta +2) (\Delta  (143 \Delta +427)+540)}{43243200 (\Delta -4) (\Delta -3) (\Delta -2)}f_0 f_4,\label{eq:a43}\\
    a_2^{(3)}=&\;\frac{\Delta  (\Delta  (\Delta  (\Delta  (\Delta  (143 \Delta  (7 \Delta -48)+12615)-3980)-6156)-11736)-1440)}{3459456000 (\Delta -5) (\Delta -4) (\Delta -3) (\Delta -2)}f_0^3 \nonumber \\
    &\;+\frac{\Delta  (\Delta +1) (\Delta  (11 \Delta  (\Delta  (26 \Delta -43)+107)+4446)+6120)}{43243200 (\Delta -5) (\Delta -4) (\Delta -3) (\Delta -2)} f_0 f_4 \nonumber \\
    &\;-\frac{\Delta  (\Delta +1) (\Delta  (\Delta  (\Delta  (143 \Delta +101)+6200)+15912)+15840)}{86486400 (\Delta -5) (\Delta -4) (\Delta -3) (\Delta -2)} f_0 h_4 \nonumber \\
    &\;+\frac{\Delta  (\Delta +1) (\Delta +2) (\Delta +3) (\Delta +4)}{48048 (\Delta -5) (\Delta -4) (\Delta -3) (\Delta -2)}f_8, \\
    a_0^{(3)}=&\;\frac{(\Delta -8) \Delta  (\Delta  (\Delta  (\Delta  (\Delta  (143 \Delta  (7 \Delta -48)+17115)-10460)+1584)+21384)+12960)}{10378368000 (\Delta -6) (\Delta -5) (\Delta -4) (\Delta -3) (\Delta -2)}f_0^3 \nonumber \\
    &\;+\frac{(\Delta -8) \Delta  (\Delta +1) (\Delta +2) (\Delta  (\Delta  (143 \Delta -185)+1422)+3240)}{43243200 (\Delta -6) (\Delta -5) (\Delta -4) (\Delta -3) (\Delta -2)}f_0 f_4 \nonumber \\
    &\;+ \frac{\Delta  (\Delta +1) (\Delta +2) (\Delta +3) ((\Delta -4) \Delta +72)}{144144 (\Delta -6) (\Delta -5) (\Delta -4) (\Delta -3) (\Delta -2)}(3f_8-h_8).
\end{align}
Equations \eqref{eq:a63} and \eqref{eq:a63} also appeared in \cite{Li:2019zba}.

\section{Additional plots}
\label{addplots}

\begin{figure}[t!]
    \centering
    \includegraphics[width=1.0\linewidth]{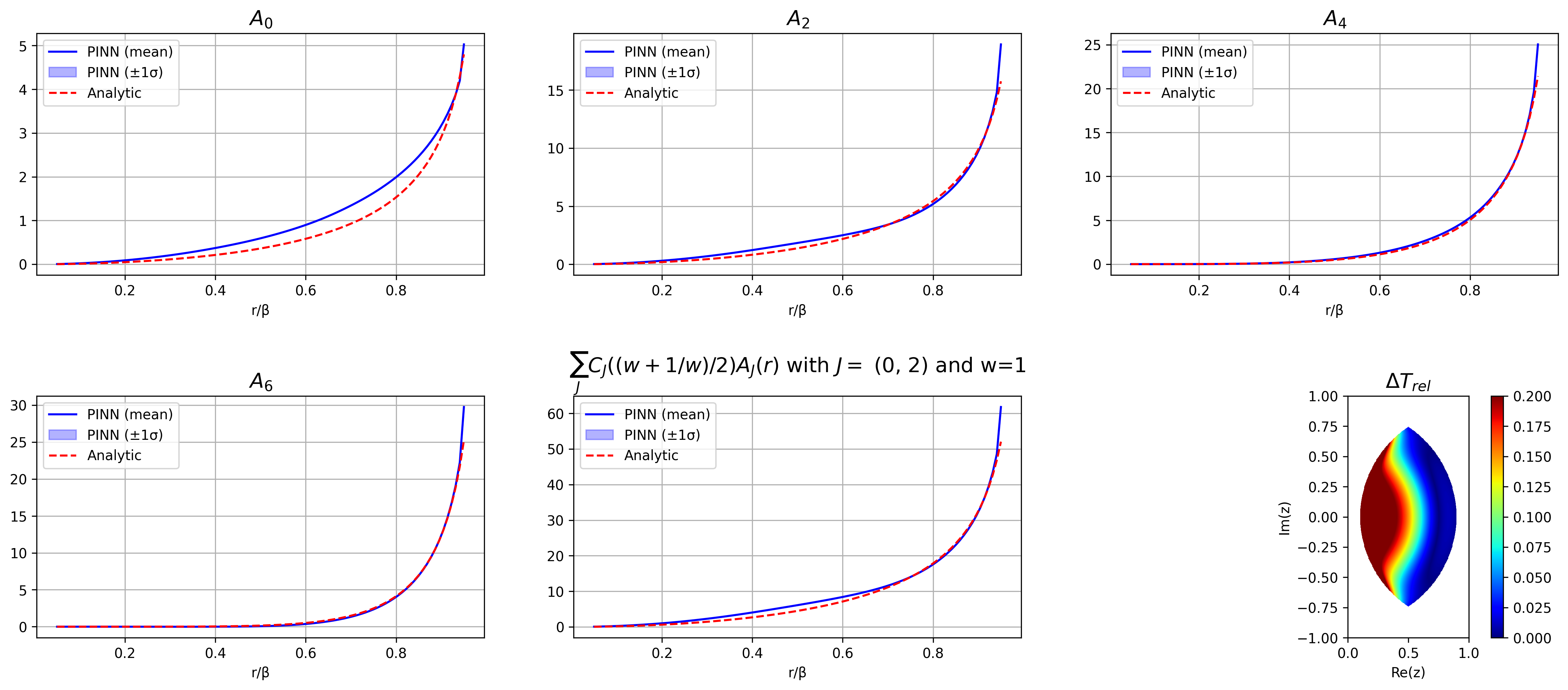}
    \caption{\small{Plots summarizing the results obtained in the $d=4$ GFF theory for the 10th in order-of-loss run in a pool of 1K independent training runs with 50K epochs. We use $\Delta_\phi=1.68$ without any exposed operators and asymptotic boundary conditions for the tail functions at $r=0.9999$.}}
    \label{fig:abs_no_exposed_example}
\end{figure}

In this appendix we include a few additional plots (Figure\ \ref{fig:abs_no_exposed_example}) that flesh out
an example of a low-loss configuration obtained by optimizing $\LL_{\overline{\rm abs}}$ in the context of the GFF analysis of Section \ref{tailsAbs}. In contrast to Figure\ \ref{fig:abs_no_exposed}, where many optima exhibit $A_0, A_2$ functions that are displaced relative to the corresponding GFF analytic curves, the result in Figure\ \ref{fig:abs_no_exposed_example} is much closer to the analytic curves, demonstrating that the analytic GFF solution is part of the minima of $\LL_{\overline{\rm abs}}$ and the optimization can access it.

\end{appendix}

%\newpage

%% Bibliography

\bibliography{finiteT}

\end{document}